\documentclass[10pt]{iopart}

\usepackage{graphicx}
\usepackage{url}
\usepackage{esint}
\usepackage{amsfonts}

\usepackage[russian,ngerman,english]{babel}

\newcommand{\muas}[0]{\hbox{\rm $\mu$as}}

\newcommand{\ve}[1]{\mbox{\boldmath$#1$}}

\let\oldbibitem\bibitem
\renewcommand\bibitem[2][]{\oldbibitem{#2}}

\begin{document}

\title[Light propagation in 2PN approximation in the field of one moving monopole]
{Light propagation in 2PN approximation in the field of one moving monopole II. Boundary value problem}

\author{Sven Zschocke}

\address{Institute of Planetary Geodesy - Lohrmann Observatory, Dresden Technical University,
Helmholtzstrasse 10, D-01069 Dresden, Germany}

\ead{sven.zschocke@tu-dresden.de}

\begin{abstract}
In  this investigation the boundary value problem of light propagation in the gravitational field of one arbitrarily moving body 
with monopole structure is considered in the second post-Newtonian approximation.  
The solution of the boundary value problem comprises a set of altogether three transformations: 
$\ve{k} \rightarrow \ve{\sigma}$, $\ve{\sigma} \rightarrow \ve{n}$, and $\ve{k} \rightarrow \ve{n}$.  
Analytical solutions of these transformations are given and the upper limit of each individual term is determined. 
Based on these results, simplified transformations are obtained by keeping only those terms relevant for  
the given goal accuracy of $1$\,nano-arcsecond in light deflection.   
Like in case of light propagation in the gravitational field of one body at rest, there are so-called enhanced terms  
which are of second post-Newtonian order but contain one and the same typical large numerical factor. 
Finally, the impact of enhanced terms beyond 2PN approximation is considered. It is found that enhanced 3PN terms are relevant 
for astrometry on the level of $1$\,nano-arcsecond in light deflection, while enhanced 4PN terms are negligible, except for 
grazing rays at the Sun.  
\end{abstract}

\pacs{95.10.Jk, 95.10.Ce, 95.30.Sf, 04.25.Nx, 04.80.Cc}


\section{Introduction}\label{Section0}

\subsection{The new era of space-based astrometry}
 
While advancement in astrometry has always been benefited from ground-based telescope improvements, the new era of space-based  
astrometry missions has initiated unprecedented accuracies in positional measurements of celestial objects, like Solar System  
objects, stars, galaxies, and quasars \cite{History_Astrometry1,History_Astrometry2,Kovalevsky}. 
Most notably, the astrometry missions Hipparcos and Gaia of the European Space Agency (ESA) have opened this new age in astronomy.  
These missions have (i) adapted from wide-field astrometry realized by optical instruments which are designed to measure large angles 
on the sky simultaneously, (ii) utilized the most modern technologies in the optical design of scanning satellite, 
(iii) taken advantage of appreciable developments in theoretical astrometry and applied gravitational physics. 

Approved by ESA in 1980 and launched on 8 August 1989, Hipparcos was the first ever astrometric satellite to precisely measure the positions  
and proper motions of stars in the vicinity of the Sun. The completion of the Hipparcos mission has led to the creation  
of three highly accurate catalogues of stellar positions, namely the star catalogues Hipparcos and Tycho in 1997 \cite{Hipparcos,Hipparcos1} 
and Tycho-2 in 2000 \cite{Hipparcos2}.  
In particular, the Hipparcos final catalogue \cite{Hipparcos} provides astrometric positions and stellar motions up to $1$ milli-arcsecond  
(${\rm mas}$) in angular accuracies for about $120$ thousand stars. The catalogues Tycho \cite{Hipparcos1} and Tycho-2 \cite{Hipparcos2}  
contain positions of about $1\,{\rm million}$ and $2.5\,{\rm million}$ stars, respectively,  
with an accuracy of up to $20\,{\rm mas}$ in angular resolution which still represents an unprecedented accuracy  
at that time, also in view of such huge number of individual stars. These catalogues set the precedent on stellar positions and 
are continuously used in space science research and for spacecraft navigation.  

Gaia is the second space-based mission ever and will provide fundamental data for many fields of astronomy.  
The Gaia mission was approved in 2000 by ESA as cornerstone mission and is aiming at precisions up to a few micro-arcseconds  
($\mu{\rm as}$) in determining positions and proper motions of stellar objects \cite{GAIA}, which is about $200$ times 
more accurately than the predecessor Hipparcos. Launched on 19 December 2013, the Gaia's main goal is to create an 
extraordinarily precise three-dimensional map of more than $1300$ million stars of our galaxy, in order to determine the structure 
and dynamics of the Milky way. The observational data of Gaia comprise not only astrometry but also spectro-photometry. 
For the brightest subset of targets, spectra will be acquired to obtain radial velocities of stellar objects by means of the Doppler effect 
which is essential for the understanding of the kinematics of our Galaxy \cite{GAIA1}.   

The highly precise measurements of the astrometry mission Gaia are
of fundamental importance to all the other fields of astronomy, specifically they will have a tremendous impact
on stellar astrophysics and galaxy evolution, solar-system and extra-solar planet science, extra-galactic astrophysics, and fundamental physics
like dark matter and dark energy physics, highly-precise determination of natural constants, testing equivalence principle, determination
of Nordtvedt parameter, possible temporal variation of the gravitational constant, and last but not least testing alternative theories of gravity.
Another aspect of highly-precise astrometric data concerns the essential fact that not only more accurate but also qualitatively new tests
of general relativity become possible \cite{Theory_Experiment,Kopeikin_Efroimsky_Kaplan,Klioner2003b,Book_Clifford_Will}.
 
Preliminary results of the Gaia mission have been published in September 2016 by Gaia Data Release 1 (Gaia DR1), providing 
astrometric data which are more precise than those in any of the former star catalogues \cite{GAIA1,GAIA_DR1_1,GAIA_DR1_2}.  
The five-parameter astrometric solution (positions, proper motions, parallaxes) for about $2$ million stars in common between 
the Tycho-2 Catalogue and Gaia is contained in Gaia DR1.  

The results of Gaia Data Release 2 (Gaia DR2) were published very recently in April 2018 by a series of articles. There are specific articles 
and processing papers which  
concern special scientific issues and which give technical details on the processing and calibration of the raw data. A comprehensive overview  
of Gaia DR2 is expounded in \cite{GAIA_DR2_1}, while the full content of Gaia DR2 is available through the Gaia archive \cite{Gaia_Archive}. 
In particular, Gaia DR2 provides precise positions, proper motions, and parallaxes for more than $1300$ million stars. Furthermore, the 
Gaia DR2 contains positions for more than $550$ thousand quasars which allow for the definition of a new celestial reference frame fully based  
on optical observations of extra-galactic sources (Gaia-CRF2) \cite{Gaia_CRF2}. Based on these results  
the third realization of the International Celestial Reference Frame (ICRF-3) has recently been adopted by the  
XXXth General Assembly of the International Astronomical Union (IAU) in 2018 \cite{ICRF3_A}, which is based on the accurate measurement  
of over $4000$ extragalactic radio sources. The ICRF-3 replaces ICRF-2   
which was adopted at the XXVIIth General Assembly of IAU in 2009. These reference frames are of  
utmost importance for many branches is astronomy, like stellar catalogues, space navigations,  
or determination of the rotational motion of the Earth.  

Of specific importance for our investigations here is the impressive advancement in astrometric accuracy of positional measurements 
arrived within the Gaia DR2. For parallaxes, uncertainties are typically around $30$ \muas $\;$  
for sources brighter than $V \!\! = \!\! 15\;{\rm mag}$, around $100$ \muas$\;$ for sources with a magnitude about  
$V \!\! = \!\! 17\;{\rm mag}$, and around $700$ \muas $\;$ for sources with about $V \!\! = \!\! 20\;{\rm mag}$ \cite{GAIA_DR2_1,GAIA_DR2_2}.  
These results represent a giant advancement in astrometric science and comprise the fact that todays astrometry  
has reached the micro-arcsecond level of accuracy in astrometric measurements.  

Another astrometric space mission aiming at the micro-arcsecond level of accuracy is JASMINE, an approved long-term project 
developed by the National Astronomical Observatory of Japan,  
and which consists of altogether three astrometry satellites, called Nano-JASMINE, Small-JASMINE, and (Medium-Sized) JASMINE \cite{Jasmine1}, 
where the two last satellites shall observe in the infrared. The Nano-JASMINE (nominal mission: $2$ years) is a mission in the optical  
based on CCD (charge-coupled device) and the technical demonstrator of the entire JASMINE project, which represents the  
first space astrometry satellite mission in Japan and the third space-based astrometry mission ever following the ESA missions Hipparcos and Gaia.  
Meanwhile, the technical equipment of the satellite has fully been completed and the launch of Nano-JASMINE is expected within the very 
few next months. The launch of Small-JASMINE is expected around 2024, while there is no concrete plan for the launch of (Medium-Sized) JASMINE.  
Within the series of altogether three JASMINE missions, the target accuracy in the positional measurements 
of stellar objects will be increasing step-by-step, ranging from $3\,{\rm mas}$ by the Nano-JASMINE mission up to $10\,\muas$ within
the (Medium-Sized) JASMINE mission.

\subsection{Future astrometry on the sub-micro-arcsecond level}

It is quite obvious that a long term goal of astrometric science is to arrive at the sub-micro-arcsecond (sub-$\mu{\rm as}$) or even the 
nano-arcsecond (nas) level of accuracy. The scientific objectives for such ultra-highly precise astrometry are overwhelming and it 
is almost impossible to enumerate all advances in science which astrometry on such scales would initiate. For instance,  
astrometry on sub-$\mu{\rm as}$ scale would make it possible to survey hundreds of thousands of stars up to a distance 
of about $100 \,{\rm pc}$ for detecting earth-like planets, would allow for much more stringent tests of General Relativity through light bending,  
would enable the measurement of the energy density of stochastic gravitational wave background, allows for precise mapping of dark matter from  
the areas beyond the Milky Way, enables direct distance measurement of various stellar standard candles up to the closest galaxy clusters, 
would allow for further tests of alternative theories of gravity with much better precision than in the weak-gravitational-field  
regime \cite{article_sub_micro_1,article_sub_micro_2,article_sub_micro_3,article_sub_micro_4,Conference_Cambridge}. 
Especially, the proposed mission Theia \cite{Theia} is primarily designed to study the local dark matter properties, 
the detection of Earth-like exoplanets in our nearest star systems and the physics of highly compact objects like  
white dwarfs, neutron stars, black holes.  
For a more comprehensive list of astronomical and astrophysical problems which can be solved by sub-$\mu{\rm as}$ astrometry we refer to  
the article \cite{Kopeikin_Gwinn}.  
 
Furthermore, as soon as the third (Gaia DR3) and final Gaia Data Release (Gaia Final DR), expected in the fall of 2020 and  
around the end of 2022, respectively, are achieved and analyzed, new questions will emerge, which will require new space-based astrometry  
missions, either in the form of a Gaia-like observer or in the form of satellites aiming at the sub-$\mu{\rm as}$ or even the  
nas level of accuracy. In fact, the impressive progress, made during the realization of the both ESA astrometry  
missions Hipparcos and Gaia, has already encouraged the astrometric science community to further proceed in such directions in foreseeable future.  
Among several astrometry missions suggested to ESA we mention the recent medium-sized (M-5) mission proposals Gaia-NIR \cite{Gaia_NIR}, 
Theia \cite{Theia}, and NEAT \cite{NEAT1,NEAT2,NEAT3}, which in this order are aiming at the $\mu{\rm as}$, sub-$\mu{\rm as}$,  
and even the ${\rm nas}$ level of precision.  

The envisaged advancement from \muas-astrometry to sub-\muas-astrometry implies many subtle effects and new kind of 
challenges in technology and science such as: (a) determination of Solar System ephemerides precisely enough for sub-\muas-astrometry, 
(b) modeling the influence of interstellar medium on light propagation, (c) synchronization of atomic clocks between observer and ground  
stations on the sub-nano-second scale, (d) tracking the spacecraft's worldline and velocity with sufficient accuracy 
for being able to account for aberrational effects, (e) development of new CCD-based technologies in the optical or infrared 
to achieve astrometric data on the sub-\muas-level,  
etc. Each of these and many other challenges have to be clarified before sub-\muas-astrometry becomes feasible.  
But it is clear that astrometric information is mainly carried by light signals of the celestial light sources, hence  
astrometric measurements are intrinsically  
related to the problem about how to trace a light ray detected by the observer back to the celestial light source.  
Therefore, the fundamental assignment in astrometry remains the precise description of the trajectory  
of the light signal as function of coordinate time.   
The foreseen progress in the accuracy of observations and new observational techniques necessitates to account for several  
relativistic effects in the theory of light propagation. A detailed review about the recent progress in the theory of light propagation  
has been given in text books \cite{Kopeikin_Efroimsky_Kaplan,Brumberg1991} as well as in several articles  
\cite{Klioner2003b,Zschocke1,Zschocke2,KK1992,Kopeikin1997,KS1999,KSGE,Klioner2003a,KopeikinMashhoon2002}. 
So in what follows an introduction of the theory of light propagation is just given to the extent that it proves necessary for our investigations.

\subsection{The exact field equations of gravity} 

According to the theory of general relativity \cite{Einstein1,Einstein2} the space-time is not considered as rigidly given once and for all, 
but a differentiable manifold and subject to dynamical laws.   
Therefore, the determination of the (inner) geometry of space-time is the foundation for any  
measurement in relativistic astrometry. The (inner) geometry of the four-dimensional manifold is fully determined by the  
metric tensor $g_{\alpha\beta}$ whose components are identified with
tensorial gravitational potentials generalizing the scalar gravitational potential of Newtonian theory of gravity. 
In compliance with Einstein's field equations \cite{Einstein1,Einstein2}, the metric tensor $g_{\alpha\beta}$ is related to the
stress-energy tensor $T_{\alpha\beta}$ of matter via a set of $10$ coupled non-linear partial differential equations  
given by \cite{Kopeikin_Efroimsky_Kaplan,Einstein1,Einstein2,MTW,Landau_Lifschitz,Fock} (e.g. Sec. 17.1 in \cite{MTW})  
\begin{eqnarray}
R_{\alpha\beta} - \frac{1}{2}\,g_{\alpha\beta}\,R &=& \frac{8\,\pi\,G}{c^4}\,T_{\alpha\beta}  
\label{Field_Equations_5}
\end{eqnarray}

\noindent
where $R_{\alpha \beta} = \Gamma^{\rho}_{\alpha\beta,\rho} - \Gamma^{\rho}_{\alpha\rho,\beta} 
+ \Gamma^{\rho}_{\sigma\rho}\,\Gamma^{\sigma}_{\alpha\beta}
- \Gamma^{\rho}_{\sigma\beta}\,\Gamma^{\sigma}_{\alpha\rho}\,$ is the Ricci tensor (cf. Eq.~(8.47) in \cite{MTW}), 
$R = g^{\alpha\beta} R_{\alpha\beta}$ is the Ricci scalar, and   
\begin{eqnarray}
\Gamma^{\rho}_{\alpha\beta} &=& \frac{1}{2}\,g^{\rho\sigma}
\left(g_{\sigma\alpha , \beta} + g_{\sigma\beta , \alpha} - g_{\alpha\beta , \sigma} \right)  
\label{Christoffel_Symbols}
\end{eqnarray}

\noindent
are the Christoffel symbols which are functions of the metric tensor. 

The field equations of gravity (\ref{Field_Equations_5}) are valid in any coordinate system.  
The final ambition in theoretical astrometry remains of course the determination of observables 
(scalars), which are, by definition, gauge-independent (coordinate-independent) quantities \cite{Observables}.  
There are three possibilities to get such observables \cite{Brumberg2010}:
\begin{enumerate}
\item[1.] performing the calculations solely in terms of coordinate-independent quantities.
\item[2.] using any coordinate system in the calculations.
\item[3.] adopting one coordinate system and determine observables in the final step.
\end{enumerate}

\noindent
The IAU has adopted the third way by recommending the use of harmonic coordinates in 
celestial mechanics and in the astrometric science \cite{IAU_Resolution1}.  
These harmonic coordinates considerably simplify the calculations in celestial mechanics and in the theory of light propagation.  
They are denoted by $x^{\mu} = \left(c t, \ve{x}\right)$, where $t$ is the coordinate time and $\ve{x} =\left(x^1,x^2,x^3\right)$  
is a triplet of spatial coordinates. The harmonic coordinates are curvilinear and they are defined by the harmonic  
gauge condition \cite{Kopeikin_Efroimsky_Kaplan,Brumberg1991,MTW},
\begin{eqnarray}
\frac{\partial \left(\sqrt{- g}\,g^{\alpha \beta}\right)}{\partial x^{\beta}} = 0\,,
\label{harmonic_gauge_condition_1}
\end{eqnarray}

\noindent
where $g = {\rm det}\left(g^{\alpha\beta}\right)$ is the determinant of metric tensor.
The condition (\ref{harmonic_gauge_condition_1}) is called de Donder gauge in honor of its inventor \cite{Donder},
which was also found independently by Lanczos \cite{Lanczos}; we note that (\ref{harmonic_gauge_condition_1}) determines  
(a class of) concrete reference systems, hence it is not surprising that condition (\ref{harmonic_gauge_condition_1}) is not covariant.  
The harmonic coordinates can be treated like Cartesian coordinates besides that they are curvilinear
\cite{Kopeikin_Efroimsky_Kaplan,Brumberg1991,Thorne,Poisson_Lecture_Notes,Poisson_Will};
cf. text below Eq.~(3.1.45) in \cite{Brumberg1991} or the statement above Eq.~(1.1) in \cite{Thorne}, while more detailed 
explanations for this fact are provided in Sections 1.5. and 1.6 in \cite{Poisson_Will}.

In line with these statements, in practical calculations in celestial mechanics and astrometry it is very useful to express the 
exact field equations of gravity (\ref{Field_Equations_5}) in terms of harmonic coordinates. In this so-called Landau-Lifschitz formulation  
of the field equations \cite{Landau_Lifschitz}, the contravariant components of the gothic metric density are decomposed as follows  
\begin{eqnarray}
\sqrt{-g}\,g^{\alpha\beta} &=& \eta^{\alpha \beta} - \overline{h}^{\alpha \beta}\,,
\label{metric_20}
\end{eqnarray}

\noindent
which is especially useful in case of an asymptotically flat space-time.
Here, $\overline{h}^{\alpha \beta}$ is the trace-reversed metric perturbation which describes the deviation of the gothic metric  
tensor density of curved space-time from the metric tensor of Minkowskian space-time.  

The exact field equations (\ref{Field_Equations_5}) in terms of harmonic coordinates can be written as follows
(cf. Eq.~(36.37) in \cite{MTW} or Eq.~(5.2b) in \cite{Thorne}):
\begin{eqnarray}
\opensquare\;
\overline{h}^{\alpha \beta} &=& - \frac{16\,\pi\,G}{c^4}\,\left(\tau^{\alpha \beta} + t^{\alpha \beta}\right),
\label{Field_Equations_10}
\end{eqnarray}

\noindent
where
$\opensquare = \eta^{\mu\nu}\,\partial_{\mu}\,\partial_{\nu}$ is the (flat) d'Alembert operator and
\begin{eqnarray}
\tau^{\alpha \beta} &=& \left( - g\right)\,T^{\alpha \beta}\,,
\label{metric_35}
\\
\nonumber\\
t^{\alpha \beta} &=& \left( - g\right)\,t_{\rm LL}^{\alpha \beta} + \frac{c^4}{16\,\pi\,G}\;
\left(\overline{h}^{\alpha\mu}_{\;\;\;,\;\nu}\;\overline{h}^{\beta \nu}_{\;\;\;,\;\mu}
- \overline{h}^{\alpha\beta}_{\;\;\;,\;\mu\nu}\;\overline{h}^{\mu \nu}\right),
\label{metric_40}
\end{eqnarray}

\noindent
where $t_{\rm LL}^{\alpha \beta}$ is the Landau-Lifschitz pseudotensor of gravitational field \cite{Landau_Lifschitz}, which is symmetric in the   
indices and in explicit form given by Eq.~(20.22) in \cite{MTW} or by Eqs.~(3.503) - (3.505) in \cite{Kopeikin_Efroimsky_Kaplan}. We shall assume  
that the gravitational system is isolated, that means flatness of the metric at spatial infinity and the constraint of no-incoming gravitational  
radiation is imposed at past null infinity ${\cal J}^{-}$ (cf. notation in Section 34 in \cite{MTW} and Figure 34.2. in \cite{MTW}). These  
so-called Fock-Sommerfeld boundary conditions, for instance given by Eqs.~(4.64) and (4.65) in \cite{Kopeikin_Efroimsky_Kaplan}, have been  
adopted from classical electrodynamics \cite{Sommerfeld1,Sommerfeld2} and later formulated for the general theory of gravity \cite{Fock}.  
By imposing the Fock-Sommerfeld boundary conditions, a formal solution of (\ref{Field_Equations_10}) 
is then provided by the implicit integro-differential equation, 
\begin{eqnarray}
\overline{h}^{\alpha \beta} \left(t,\ve{x}\right) &=& \frac{4\,G}{c^4}\,
\int d^3 x^{\prime}\,
\frac{\tau^{\alpha\beta}\left(u, \ve{x}^{\prime}\right) + t^{\alpha\beta}\left(u, \ve{x}^{\prime}\right)}{\left| \ve{x} - \ve{x}^{\prime} \right|}\,,
\label{Introduction_2}
\end{eqnarray}

\noindent
where  
\begin{eqnarray}
u = t - \frac{\displaystyle \left| \ve{x} - \ve{x}^{\prime} \right|}{\displaystyle c} 
\label{Introduction_2a}
\end{eqnarray}

\noindent 
is the retarded time, which is associated with the finite speed of gravitational action and not with the finite speed of light, as one may   
recognize from the fact that electromagnetic fields are not necessarily involved in the stress-energy tensor on the r.h.s.  
of (\ref{Field_Equations_10}) or (\ref{Introduction_2}). In order to deduce the formal solution (\ref{Introduction_2}) from  
the differential equation (\ref{Field_Equations_10}) the Cartesian-like harmonic coordinates $\left(ct,\ve{x}\right)$ have been treated like  
Cartesian coordinates besides that they are curvilinear; cf. text below Eq.~(36.38) in \cite{MTW}.  
The approach about how to solve (\ref{Introduction_2}) iteratively  
is described in some detail in \cite{Kopeikin_Efroimsky_Kaplan}; cf. Eqs.~(3.530a) - (3.530d) in \cite{Kopeikin_Efroimsky_Kaplan}.  
In the first iteration (first post-Minkowskian approximation) the integral runs only over the three-dimensional volume of the matter  
source, while from the second iteration on (second post-Minkowskian approximation and higher) the integral (\ref{Introduction_2}) gets 
also support from the metric perturbation, hence runs over the entire three-dimensional space.  

Four comments are in order about the exact field equations of gravity. 

$\bullet$ First, the retarded time $u$, which is hidden in the exact field equations of gravity (\ref{Field_Equations_5}),  
appears explicitly in the formal solution of the exact field equations (\ref{Introduction_2}),  
which states that a space-time point $\left(u,\ve{x}^{\prime}\right)$ (e.g. located
inside the matter distribution) is in causal contact with a space-time point $\left(t,\ve{x}\right)$ (e.g. located outside the matter source).

$\bullet$ Second, one may consider the propagation of electromagnetic action  
in a curved space-time with background metric $g_{\alpha\beta}$.  
That means, the metric of the curved space-time is determined by some matter distribution $T_{\alpha\beta}$, while 
the impact of the electromagnetic field on the metric of space-time is neglected. The electromagnetic fields are generated 
by some electromagnetic four-current $j^{\mu} = \left(c \rho, \ve{j}\right)$ with $\rho$ and $\ve{j}$ being charge-density and current-density, respectively. 
The covariant field equations of Maxwell's electrodynamics in curved space-time read  
$F^{\mu\nu}_{\quad\, ;\,\nu} = 4\,\pi\,j^{\mu}$ and $F_{\mu\nu\,;\,\rho} + F_{\nu\rho\,;\,\mu} + F_{\rho\mu\,;\,\nu} = 0$ 
(cf. Eqs.~(22.17a) and (22.17b) in \cite{MTW}), where
$F_{\mu\nu} = A_{\nu\, ; \,\mu} - A_{\mu \,;\, \nu}$ is the field-tensor of electromagnetic field (cf. Eq.~(22.19a) in \cite{MTW}), 
the semicolon denotes   
covariant derivative, and $A^{\mu} =\left(\varphi/c, \ve{A}\right)$ is the four-potential, where $\varphi$ is the scalar potential  
and $\ve{A}$ is the vector potential.  
The Characteristics (also called characteristical surface) of the covariant Maxwell equations are governed by the following non-linear partial  
differential equation (non-linear PDE) of first order \cite{Fock,Whittaker,Article_Charaketristiken,Sexl_Urbantke,Iverno},
\begin{eqnarray}
g^{\alpha \beta}\; \frac{\partial \phi}{\partial x^{\alpha}}\; \frac{\partial \phi}{\partial x^{\beta}} = 0 \,,  
\label{Characteristics_ED}
\end{eqnarray}

\noindent
which is valid in the near-zone as well as in the far-zone of the four-current $j^{\alpha}\left(t,\ve{x}\right)$ and is valid in any reference  
system. The Characteristics are three-dimensional curved sub-manifolds, $\phi\left(x^0, x^1, x^2, x^3\right) = {\rm const}$, of the Riemannian  
space-time. In case of flat space-time, i.e. $g^{\alpha \beta} = \eta^{\alpha \beta}$, the characteristical surface 
at the event $\left(x^0_0,x^1_0,x^2_0,x^3_0\right)$ is given by the Minkowskian light-cone,  
\begin{eqnarray}
\fl \hspace{1.0cm} 
\phi\left(x^0,x^1,x^2,x^3\right) = \left(x^0_0-x^0\right)^2 - \left(x^1_0-x^1\right)^2 - \left(x^2_0-x^2\right)^2 - \left(x^3_0-x^3\right)^2\,.   
\label{Light_Cone_ED} 
\end{eqnarray}

\noindent  
That means, an electromagnetic discontinuity (abrupt electromagnetic signal) generated at $\left(x^0_0,x^1_0,x^2_0,x^3_0\right)$  
propagates in the flat space-time along the light-cone (\ref{Light_Cone_ED}).  
The generalization of the light-cone (\ref{Light_Cone_ED}) in flat space-time is the light-conoid in curved space-time as governed by  
Eq.~(\ref{Characteristics_ED}), which for the curved space-time of the Solar system can only be solved approximately, for instance by iteration. 
The PDE of the Characteristics (\ref{Characteristics_ED}) can be derived by means of the following consideration.  
Let $a^{\mu}$ be a continuous (smoothly changing) electromagnetic four-potential generated by some current $j^{\mu}$ somewhere located 
in the Riemannian space-time with metric $g^{\mu\nu}$. Now suppose that the four-current $j^{\mu}$ changes rapidly and generates an abrupt  
Theta-like discontinuity (perturbation) in the electromagnetic field with amplitude $u^{\mu}$, which propagates along some  
hypersurface $\phi$. Then, the entire electromagnetic four-potential $A^{\mu}$ is given by the following expression:  
$A^{\mu}\left(x^0,\ve{x}\right) = a^{\mu}\left(x^0,\ve{x}\right) + u^{\mu}\left(x^0,\ve{x}\right)\,\Theta\left(\phi\left(x^0,\ve{x}\right)\right)$ 
\cite{Sexl_Urbantke}. By inserting this ansatz into the covariant Maxwell equations one just obtains the equation (\ref{Characteristics_ED}) which  
governs the evolution of the hypersurface $\phi$ in the curved space-time on which any discontinuity of the electromagnetic field is located. 
Thus, the three-dimensional sub-manifolds $\phi\left(x^0, x^1, x^2, x^3\right)$ of the Riemannian space-time can be identified with the front  
of electromagnetic action (e.g. abrupt discontinuity in the near-zone of the four-current or wave-front of an electromagnetic wave in the 
far-zone of the four-current) caused by some rapid change in the electromagnetic four-current.  
 
Furthermore, one may introduce a trajectory, $x^{\alpha}\left(\lambda\right)$ where $\lambda$ is an affine curve-parameter,  
which is orthogonal on the surface $\phi$ \cite{Fock,Article_Charaketristiken,Sexl_Urbantke,Iverno},  
\begin{eqnarray}
\frac{d x^{\alpha}\left(\lambda\right)}{d \lambda} = g^{\alpha \beta} \frac{\partial \phi}{\partial x^{\beta}}\,,
\label{Biharacteristics_ED}
\end{eqnarray}

\noindent 
that means is normal to the front of electromagnetic action; we will come back to that issue later,   
cf. text below Eqs.~(\ref{Four_Vector_sigma}) - (\ref{Four_Vector_r_A}). Such trajectories are called Bicharacteristics.  
The Bicharacteristics can be identified with the light rays,  
which propagate with the finite speed of light. Therefore, also the Characteristics, that is the surface of electromagnetic action, propagates  
with the finite speed of light.  
The light-conoid in curved space-time is built by all Bicharacteristics emanating from some (arbitrary) event. As mentioned above, besides  
that $c$ is defined as the fundamental speed of light in vacuum in the flat Minkowski space, it is clear   
that the retardation, that means the natural constant $c$ in the denominator on the r.h.s. in Eq.~(\ref{Introduction_2a}),  
is caused by the finite speed of gravitational action and not due to the finite speed of light. Even in case the stress-energy tensor 
of matter would only consist of electromagnetic fields, 
$4\,\pi\,T^{\alpha\beta} = F^{\alpha \mu}\,F^{\beta}_{\;\;\;\mu} - \frac{1}{4}\,g^{\alpha\beta}\,F_{\mu\nu}\,F^{\mu\nu}$ \cite{MTW},  
then, nevertheless, the retardation would also originate from the finite speed of gravitational fields (in this case with the well-known  
property that the Ricci scalar vanishes but of course not the Ricci tensor) which, in this specific case, 
would entirely be generated by these electrodynamical fields.  

$\bullet$ Third, let us now consider the non-linear PDE for the Characteristics of the exact field equations of gravity (\ref{Field_Equations_5}),  
which is given by \cite{Fock,Article_Charaketristiken,Sexl_Urbantke,Iverno},  
\begin{eqnarray}
g^{\alpha \beta}\; \frac{\partial \omega}{\partial x^{\alpha}}\; \frac{\partial \omega}{\partial x^{\beta}} = 0 \,,   
\label{Characteristics_GR}
\end{eqnarray}
 
\noindent
which is valid in the near-zone as well as in the far-zone of the matter source $T_{\alpha\beta}\left(t,\ve{x}\right)$ and is valid in any  
reference system. The derivation of the PDE (\ref{Characteristics_GR}) for the Characteristics is similar to the above considerations in 
case of the covariant Maxwell equations. Consider a continuous (smoothly changing) background metric $g^{\mu\nu}_0$ which is generated by some  
matter $T^{\alpha\beta}$. Then assume a rapid acceleration of the matter which results in an abrupt Theta-like discontinuity (perturbation) with  
metric $h^{\alpha\beta}$. Hence, the entire metric is given by: $g^{\alpha\beta}\left(x^0,\ve{x}\right) = g^{\alpha\beta}_0\left(x^0,\ve{x}\right)  
+ h^{\alpha\beta}\left(x^0,\ve{x}\right)\,\Theta\left(\omega\left(x^0,\ve{x}\right)\right)$ \cite{Sexl_Urbantke}.  
Now, if one wants to investigate how the gravitational discontinuity (non-analytic gravitational signal) propagates in space and time, one has  
to insert this ansatz into the exact Einstein equations, which yields the PDE (\ref{Characteristics_GR}). 
Thus, the Characteristics $\omega$ can be identified with the front of gravitational action 
(e.g. abrupt discontinuity in the near-zone of matter source or wave-front of a gravitational wave in the far-zone of matter source) 
caused by the matter source. The front of gravitational action $\omega$ is a curved three-dimensional sub-manifold,  
$\omega\left(x^0,x^1,x^2,x^3\right) = {\rm const}$, of the Riemannian space-time, that means a three-dimensional surface on which 
any discontinuities of the gravitational field must lie \cite{Fock,Article_Charaketristiken,Sexl_Urbantke,Iverno}. 
In case of flat background metric, i.e. $g^{\alpha\beta}_0 = \eta^{\alpha\beta}$, the  
solution of the PDE (\ref{Characteristics_GR}) at the event $\left(x^0_0,x^1_0,x^2_0,x^3_0\right)$ is given by the null-cone,  
\begin{eqnarray}
\fl \hspace{1.0cm} 
\omega\left(x^0,x^1,x^2,x^3\right) = \left(x^0_0-x^0\right)^2 - \left(x^1_0-x^1\right)^2 - \left(x^2_0-x^2\right)^2 - \left(x^3_0-x^3\right)^2\,.
\label{Null_Cone_GR}  
\end{eqnarray}
 
\noindent 
That means, a gravitational discontinuity (abrupt gravitational signal) generated at $\left(x^0_0,x^1_0,x^2_0,x^3_0\right)$ propagates in the 
flat space-time along the null-cone (\ref{Null_Cone_GR}).  
The generalization of the null-cone (\ref{Null_Cone_GR}) of gravitational action in flat background metric is the null-conoid in curved space-time  
as governed by (\ref{Characteristics_GR}), which for the curved space-time of the Solar system can only be solved approximately, 
for instance by iteration.  
 
One may also introduce Bicharacteristics for the field equations of gravity,  
$z^{\alpha}\!\left(\rho\right)$ where $\rho$ is an affine curve-parameter, which are trajectories  
orthonormal on the hypersurface $\omega$ \cite{Fock,Article_Charaketristiken,Sexl_Urbantke,Iverno},  
\begin{eqnarray}
\frac{d z^{\alpha}\!\left(\rho\right)}{d \rho} = g^{\alpha \beta} \frac{\partial \omega}{\partial x^{\beta}}\,,  
\label{Biharacteristics_GR}
\end{eqnarray}

\noindent 
that means is normal to the front of gravitational action; we will come back to that issue later,  
cf. text below Eqs.~(\ref{Four_Vector_sigma}) - (\ref{Four_Vector_r_A}).  
These Bicharacteristics can be considered as gravitational rays.  
Such an idealized picture is well justified for a gravitational wave when the wavelength is negligibly small in comparison with  
the spatial region of propagation of the wave. Such condition is satisfied in the far-zone of the Solar System, but not in the near-zone  
of the Solar System where the wavelength of gravitational radiation is larger than the boundary of the near-zone. That is why the  
Bicharacteristics in the near-zone should be considered as a mathematical concept of being normals onto the characteristic hypersurface $\omega$,  
while in the far-zone the Bicharacteristics can physically be interpreted as gravitational rays. But what is important here is the 
fact that the speed of gravity equals the speed of light, because the equations (\ref{Characteristics_ED}) and (\ref{Biharacteristics_ED}) are 
identical with (\ref{Characteristics_GR}) and (\ref{Biharacteristics_GR}), respectively; cf. Section 7.2 in \cite{Kopeikin_Efroimsky_Kaplan}.  
Therefore, as just mentioned above, the natural constant $c$ in the denominator on the r.h.s.  
in Eq.~(\ref{Introduction_2a}) is related to the finite speed of gravity which equals the finite speed of light. The  
null-conoid (at some arbitrary event),
can also be defined as the set of all Bicharacteristics emanating from that (arbitrary) event in the curved space-time. 

$\bullet$ Fourth, as stated above, the equations for the Characteristics, Eq.~(\ref{Characteristics_ED}) and Eq.~(\ref{Characteristics_GR}), 
are fundamental consequences of the  
exact field equations of electrodynamics in curved space-time and the exact field equations of gravity, respectively. They state that there 
is no difference between the speed of light in curved space-time and the speed of gravitational action; cf. \S 53 in \cite{Fock}.  
Nevertheless, the propagation of electromagnetic action and the propagation of gravitational action are two different physical processes,  
and besides that their velocities are numerically equal, it does not mean that they can not be distinguished from each other. 
For instance, if the directions of electromagnetic wave propagation and propagation of gravitational action are different from each other,  
then one may distinguish between the directions of both these velocities; cf. Section 7.2 in \cite{Kopeikin_Efroimsky_Kaplan}.  
Furthermore, the important theoretical prediction of Einstein's theory that both velocities are equal to each other, 
has recently been confirmed by the first detection of gravitational waves generated by the inspiral and merger   
of a binary neutron star and the determination of the location of the source by subsequent observations in the electromagnetic spectrum 
\cite{Ligo1,Ligo2}. This measurement has constrained the difference between the speed of gravity  
and the speed of light to be between $ - 3 \times 10^{- 15}$ and $ + 7 \times 10^{- 16}$ times the speed of light \cite{Ligo3}.  
Needless to say that in this case both physical processes have clearly been separated, besides that the gravitational wave and 
the electromagnetic signal were parallel to each other.  

A detailed description about how the finite speed of gravity in the near-zone of the Solar System could in principle be determined
by means of Very Long Baseline Interferometry (VLBI) has been presented in \cite{Kopeikin_A}. The suggested approach is based on the  
increasing precision of VLBI facilities which allow to determine the impact of the orbital velocity $\ve{v}_A$ of a massive Solar System body 
on the Shapiro time-delay, which states that the total time of the propagation of a light signal from the four-coordinate of a light source  
$\left(c t_0,\ve{x}_0\right)$ to the four-coordinate of an observer $\left(c t_1,\ve{x}_1\right)$ is given by 
(e.g. Eq.~(43) in \cite{KS1999})  
\begin{eqnarray}
c\left(t_1 - t_0\right) = \left|\ve{x}_1 - \ve{x}_0\right| + c\, \Delta \left(t_1,t_0\right)\,, 
\label{Shapiro_0}
\end{eqnarray}

\noindent 
where $\left|\ve{x}_1 - \ve{x}_0\right|$ is the Euclidean distance between source and observer and $\Delta \left(t_1,t_0\right)$ is the 
time-delay of the light signal caused by the gravitational field of the massive body in motion. 

In the first post-Minkowskian (1PM) approximation, which is exact up to terms to order ${\cal O}\left(G^2\right)$ and exact to all orders 
in the speed of the body, the time-delay is given by Eq.~(51) in \cite{KS1999}, which, by neglecting all terms proportional to 
the acceleration of the body (series expansion (\ref{Series_A}) is also employed), reads:  
\begin{eqnarray}
\fl \hspace{0.75cm} \Delta \left(t_1,t_0\right) = - 2\,\frac{G\,M_A}{c^3}\,\left(1 - \frac{\ve{k}\cdot \ve{v}_A\left(s_1\right)}{c}\right)
\ln \frac{r_A\left(s_1\right) - \ve{k} \cdot \ve{r}_A\left(s_1\right)}{r_A\left(s_0\right) - \ve{k} \cdot \ve{r}_A\left(s_0\right)}
+ {\cal O}\left(G^2\right)\,,
\label{Shapiro_1}
\end{eqnarray}

\noindent
which is valid for light propagation in the field of one monopole in arbitrary motion, irrespective of the fact that acceleration terms  
of the body were neglected. The unit-vector $\ve{k}$ points from the light source towards the position of the observer, and the three-vectors  
$\ve{r}_A\left(s_0\right) = \ve{x}\left(t_0\right) - \ve{x}_A\left(s_0\right)$ and 
$\ve{r}_A\left(s_1\right) = \ve{x}\left(t_1\right) - \ve{x}_A\left(s_1\right)$, where $\ve{x}\left(t_0\right)$ and $\ve{x}\left(t_1\right)$ 
are the spatial coordinates of the light signal at source and observer, respectively, while $\ve{x}_A\left(s_0\right)$ 
and $\ve{x}_A\left(s_1\right)$ are the spatial position of the body at the retarded time $s_0$ and $s_1$, as  
defined in the below standing equations (\ref{retarded_time_s_0}) and (\ref{retarded_time_s_1}). 
Let us notice here that (\ref{Shapiro_1}) also agrees with Eqs.~(146) - (148) in \cite{Zschocke2}.  

In the 1.5 post-Newtonian (1.5PN) approximation, which is exact up to terms to order ${\cal O}\left(c^{-4}\right)$ 
that means only exact to the first order in the speed of the body, the time-delay is given by Eq.~(7) in \cite{Will_2003} and reads:  
\begin{eqnarray}
\fl \hspace{0.75cm} \Delta \left(t_1,t_0\right) = - 2\,\frac{G\,M_A}{c^3}\,\left(1 - \frac{\ve{k}\cdot \ve{v}_A}{c}\right) 
\ln \frac{r_A\left(t_1\right) - \ve{K} \cdot \ve{r}_A\left(t_1\right)}{r_A\left(t_0\right) - \ve{K} \cdot \ve{r}_A\left(t_0\right)} 
+ {\cal O}\left(c^{-4}\right)\,,  
\label{Shapiro_2}
\end{eqnarray}

\noindent 
which is valid for light propagation in the field of one monopole in uniform motion, that means all acceleration terms of the body are zero.  
The three-vectors $\ve{r}_A\left(t_0\right) = \ve{x}\left(t_0\right) - \ve{x}_A\left(t_0\right)$ and 
$\ve{r}_A\left(t_1\right) = \ve{x}\left(t_1\right) - \ve{x}_A\left(t_1\right)$, where $\ve{x}_A\left(t_0\right)$ and 
$\ve{x}_A\left(t_1\right)$ are the spatial positions of the body at time of emission $t_0$ and time of reception $t_1$ of the light signal.  
Furthermore, in Eq.~(\ref{Shapiro_2}) the three-vector 
$\displaystyle \ve{K}=\ve{k} - \ve{k} \times \left(\frac{\ve{v}_A}{c} \times \ve{k}\right)$. Let us notice here that  
Eqs.~(137) - (139) in \cite{Zschocke2} are valid for light propagation in the field of one arbitrarily moving body in slow motion, which in  
case of uniform motion coincide with (\ref{Shapiro_2}), as one may show by series expansion.

For grazing light rays or radio waves at massive bodies of the Solar System, the velocity dependent  
terms in (\ref{Shapiro_1}) or (\ref{Shapiro_2}) contribute of the order of a few picoseconds in time-delay; cf. Table II in \cite{Zschocke2} for  
grazing rays at Sun or giant planets. At this order of precision it becomes possible to measure such velocity-dependent terms in  
time-delay (\ref{Shapiro_1}) or (\ref{Shapiro_2}) by means of the most modern VLBI techniques. In fact, such a concrete experiment 
by VLBI facilities has been  
suggested in \cite{Kopeikin_A}, and has finally been performed in 2002 with remarkable effort and precision \cite{Kopeikin_B}.  
In particular, in \cite{Kopeikin_B} the Shapiro time delay of a radio wave, emitted by the quasar ${\rm QSO}\;{\rm J}0842 + 1835$ and passing near   
Jupiter, has been determined with extremely high precision, in order to determine the finite speed of the gravity fields of that moving body.  
This experiment has, at the very first time, succeeded in determining the impact of the orbital velocity effects to order $v_A/c$ of Jupiter on the  
Shapiro time-delay. Subsequently, these results have initiated a controversial debate in the literature about the correct interpretation  
of this experiment 
\cite{Theory_Experiment,Will_2003,Kopeikin_E,Samuel_1,Faber,Asada1,Asada2,Carlip,Pascual,Samuel_2,Kopeikin_I,Kopeikin_CQG,Kopeikin_D,Kopeikin_F,Kopeikin_G,Kopeikin_H}; 
further comments about the Kopeikin-Formalont experiment can be found in \cite{Kopeikin_Efroimsky_Kaplan,Frittelli,Mignard_Crosta,Malkin,Zhu}.  
While there is no doubt at all in the literature about the correctness of the expressions (\ref{Shapiro_1}) and (\ref{Shapiro_2}), 
a central topic of this conversion 
was about the correct physical meaning of the natural constant $c$ in the velocity-dependent terms in (\ref{Shapiro_1}) and (\ref{Shapiro_2}).  

That remarkable debate had arisen just because of the above discussed fundamental prediction of general relativity  
that the speed of gravity and the speed of light are numerically equal.  
That is why it becomes a highly sophisticated assignment of a task to disentangle these both velocities in concrete astrometrical measurements. 
In \cite{KS1999} it was shown that the retarded instant of time $s_0$ and $s_1$ in (\ref{Shapiro_1}) are caused by the retarded time of the 
Li\'enard-Wiechert potential of the metric tensor (Eq.~(10) in \cite{KS1999}), hence they are caused by the finite speed of gravity so that 
the natural constant $c$ is related to the finite speed of gravity.  
And due to the fact that (\ref{Shapiro_2}) can be deduced from (\ref{Shapiro_1}) by series expansion and by assuming a uniform motion of the body, 
one might be inclined to assume that the natural constant $c$ in (\ref{Shapiro_2}) is related to the finite speed of gravity.  
On the other side, in \cite{Will_2003} it was shown that the natural constant $c$ in (\ref{Shapiro_2}) is caused by the finite speed of light 
and is not related to the finite speed of gravity.  
So it might be that a unique interpretation of the experiment is impossible as long as one is restricted to terms of the first order in $v_A/c$.  
But it should be noticed that the controversy was not about the correctness of the theory of general relativity, but
mainly about the question of whether the velocity-dependent term in the Shapiro time-delay is
related to the finite speed of gravity (retardation of gravitational action) or to the finite speed of light (aberration of light). 

In this context it should also be noticed that there is agreement among the participants of this controversy with respect to the 
following minimal set of issues:  
\begin{enumerate}
\item[(a)] the retarded time in (\ref{Introduction_2a}) is caused by the finite speed of gravity.
\item[(b)] the finite speed of gravity has surely an impact on the Shapiro time-delay.
\item[(c)] the impact of orbital velocity of Jupiter on time-delay has been detected in \cite{Kopeikin_B}. 
\end{enumerate}

\noindent 
While in principle the experiment suggested in \cite{Kopeikin_A} is capable to measure the speed of gravity, there is no general consensus about  
the correct interpretation of the results of the concrete experiment in \cite{Kopeikin_B}, as it was also formulated in \cite{Mignard_Crosta}.  
It seems that the fact that the retarded time $u$ in (\ref{Introduction_2a}) as well as the retarded time $s_1$ and $s_0$  
in the below standing equations (\ref{retarded_time_s_0}) and (\ref{retarded_time_s_1}) are due to the finite speed of gravity  
might not necessarily be convincing for a unique and correct interpretation of these astrometrical VLBI measurements \cite{Will_2003}.  
Moreover, in \cite{Will_2003} it was argued that acceleration terms or terms of the order $v_A^2/c^2$ give the first level at which retardation 
effects due to the motion of the massive bodies occur.  
However, in order to determine the next higher order terms, that means terms proportional to the acceleration of the body 
or terms of the order $v_A^2/c^2$ in the 
Shapiro time-delay, one needs to improve the precision in time measurements by VLBI experiments by a factor of about $10^4$, which requires  
an ultra-high precision in the time-resolution of VLBI measurements of about $10^{-3}\,{\rm picoseconds}$, which is far out of reach of  
present-day VLBI facilities. Furthermore, the theoretical interpretation of the experiment might also depend on the generalized theoretical models  
beyond general relativity which allow to distinguish between the speed of light and speed of gravity \cite{Carlip}. Even the semantics in use  
could have an impact on the correct interpretation of these VLBI results \cite{Frittelli}.  
Here, also in view of the exceptional number of articles in the literature related to this subject,  
a detailed and correct interpretation of this famous experiment would be far beyond the intention of our investigation.  
For the moment being, it seems sensible to keep in mind the problem and to realize that further careful investigations  
and higher precisions in VLBI measurements are necessary in order to clarify such involved difficulties  
regarding the distinction between the speed of electromagnetic fields and the speed of gravitational action in the near-zone of the Solar System. 

Finally, having said all that we emphasize again that the retarded instant of time (\ref{Introduction_2a}) originates from the 
finite speed of gravity which equals the speed of light, a fact that is in meanwhile sufficient for our considerations here; 
cf. also the comments in the text below Eqs.~(\ref{retarded_time_s}) and (\ref{retarded_time_s_00}) 
as well as in the text below Eqs.~(\ref{retarded_time_s_0}) and (\ref{retarded_time_s_1}).

\subsection{The exact geodesic equation for light propagation} 

Throughout the investigation the propagation of a light signal in vacuum is considered.  
The most simplest light tracking model presupposes a four-dimensional flat space-time with Minkowskian metric 
$\eta_{\alpha\beta} = {\rm diag}\left(-1,+1,+1,+1\right)$  
which implicitly involves Cartesian coordinates, where the light ray propagates along a straight line. Then, a light signal emitted at some  
spatial point $\ve{x}_0$ at time $t_0$ propagates along it's initial direction $\ve{\sigma}$, so that the light trajectory in the global system  
reads as follows,  
\begin{eqnarray}
\ve{x}_{\rm N}\left(t\right) = \ve{x}_0 + c \left(t-t_0\right) \ve{\sigma}\,,
\label{unperturbed_lightray_1}  
\end{eqnarray}
 
\noindent
where suffix ${\rm N}$ labels Newtonian approximation. Such a simple light propagation model is not sufficient for todays  
precision of astrometric measurements which, as stated above, implicates a corresponding advancement in the theory of light propagation. 
Especially, relativistic astrometry has necessarily to account for the fact that the space-time is not flat but a four-dimensional curved manifold.  
Because the space-time is curved, a light signal propagates along a geodesic which is the generalization of the concept of a straight line  
because a geodesic is a curve that parallel-transports  
its own tangent vector. Consequently, a fundamental assignment in relativistic astrometry concerns the precise modeling of the time track  
of a light signal through the curved space-time of Solar System, that is to say the determination the trajectory of the light signal,  
$\ve{x}\left(t\right)$, in some reference system which covers the entire curved space-time (at least those part of the entire  
space-time which contains the light source and the observer) and, therefore, is called global coordinate system.  

The trajectory of a light signal propagating in curved space-time is determined by the geodesic equation and isotropic condition, which in terms  
of coordinate time read as follows \cite{Kopeikin_Efroimsky_Kaplan,Brumberg1991,MTW}  
(e.g. Eqs.~(3.220) - (3.224) in \cite{Kopeikin_Efroimsky_Kaplan}): 
\begin{eqnarray}
\frac{\ddot{x}^{i}\left(t\right)}{c^2} + \Gamma^{i}_{\alpha\beta} \frac{\dot{x}^{\alpha}\left(t\right)}{c} \frac{\dot{x}^{\beta}\left(t\right)}{c}
- \Gamma^{0}_{\alpha\beta} \frac{\dot{x}^{\alpha}\left(t\right)}{c} \frac{\dot{x}^{\beta}\left(t\right)}{c} \frac{\dot{x}^{i}\left(t\right)}{c} = 0\;,
\label{Geodetic_Equation1}
\\
\nonumber\\
g_{\alpha\beta}\,\frac{\dot{x}^{\alpha}\left(t\right)}{c}\,\frac{\dot{x}^{\beta}\left(t\right)}{c} = 0\,,
\label{Null_Condition1}
\end{eqnarray}

\noindent
where a dot denotes total derivative with respect to coordinate time, hence $\dot{x}^{i}\left(t\right)$ are the three-components of the  
coordinate velocity of the photon. The null condition (\ref{Null_Condition1}) and geodesic equation (\ref{Geodetic_Equation1}) have 
equivalent physical content because (\ref{Null_Condition1}) is a first integral of (\ref{Geodetic_Equation1}). 
As mentioned above, the natural constant $c$ explicitly seen in both these equations (\ref{Geodetic_Equation1}) and (\ref{Null_Condition1}) means  
actually the speed of light, while the natural constant $c$ contained in the Christoffel symbols and metric tensor is related to the finite speed 
of gravity as stated already in the text below Eqs.~(\ref{Introduction_2}) and (\ref{Introduction_2a}). 
Furthermore, it should be mentioned that the coordinate velocity of a light signal in the global system of curved space-time differs  
from the speed of light in flat space-time $\left|\dot{\ve x}\right| \neq c\,$; only in the local system of a free-falling observer both are equal.

\subsubsection{The initial value problem}  

The light signal is assumed to be emitted at the four-position of the light source, $\left(t_0,\ve{x}_0\right)$, as given in 
some global coordinate system $\left(t,\ve{x}\right)$. Then, a  
unique solution of the partial differential equation (\ref{Geodetic_Equation1}) is well-defined by the so-called initial-value problem  
(Cauchy problem), where the spatial position of the light source, $\ve{x}_0$, and the initial unit direction of the light ray,  
$\ve{\mu} = \dot{\ve{x}}\left(t_0\right) / \left|\dot{\ve{x}}\left(t_0\right)\right|$, are given.  
Usually, the initial value problem is often replaced by the so-called initial-boundary conditions  
\cite{Kopeikin_Efroimsky_Kaplan,Brumberg1991,KK1992,Kopeikin1997,KSGE,Zschocke1,Zschocke2}:  
\begin{eqnarray}
\ve{x}_0 = \ve{x}\left(t\right)\bigg|_{t=t_0} \quad {\rm and} \quad \ve{\sigma} = \frac{\dot{\ve{x}}\left(t\right)}{c}\bigg|_{t = - \infty}\,,
\label{Initial_Boundary_Conditions}
\end{eqnarray}

\noindent
with $\ve{\sigma}$ being the unit-direction ($\ve{\sigma} \cdot \ve{\sigma} = 1$) of the light ray at past null infinity ${\cal J}^{-}$  
(cf. notation in Section 34 in \cite{MTW} and Figure 34.2. in \cite{MTW}). The advantage for using initial-boundary conditions 
(\ref{Initial_Boundary_Conditions}) rather than initial-value conditions when integrating the geodesic equation (\ref{Geodetic_Equation1})  
is solely based on the integration constant which becomes simpler at past null infinity.  
One may easily find a unique relation between the tangent vectors $\ve{\sigma}$ and $\ve{\mu}$  
(e.g. Section 3.2.3 in \cite{Brumberg1991}), so one verifies that there is a unique one-to-one correspondence between the initial-boundary  
problem (\ref{Initial_Boundary_Conditions}) and the initial-value problem; 
more precisely, these statements are valid in case of a weak gravitational field and ordinary topology of space-time. 
According to (\ref{Initial_Boundary_Conditions}), the solution for  
the light trajectory is a function of these initial-boundary conditions: $\ve{x}\left(t\right)= \ve{x}\left(t,\ve{x}_0,\ve{\sigma}\right)$.

\subsubsection{The boundary value problem}  

A unique solution of geodesic equation (\ref{Geodetic_Equation1}) can also be defined by the so-called boundary-value problem
rather than the initial-boundary problem (\ref{Initial_Boundary_Conditions}).  
In the boundary-value problem a light signal is supposed to be emitted at some initial space-time point
$\left(t_0,\ve{x}_0\right)$ (source) which is received at another space-time point $\left(t_1,\ve{x}_1\right)$ (observer)
\cite{Kopeikin_Efroimsky_Kaplan,Klioner2003b,Book_Clifford_Will,Brumberg1991,KK1992,Klioner2003a}:  
\begin{eqnarray}
\ve{x}_0 = \ve{x}\left(t\right)\bigg|_{t=t_0} \quad {\rm and} \quad \ve{x}_1 = \ve{x}\left(t\right)\bigg|_{t=t_1}\,.
\label{Boundary_Value_Conditions}
\end{eqnarray}

\noindent
Accordingly, the solution of the light trajectory will be a function of these boundary conditions:  
$\ve{x}\left(t\right)= \ve{x}\left(t,\ve{x}_0,\ve{x}_1\right)$. 
 
Because in reality any light source is located at some finite distance, the solution of the boundary-value problem is of decisive importance
in practical astrometry \cite{Kopeikin_Efroimsky_Kaplan,Brumberg1991,KK1992}. Accordingly, the primary aim of our investigation is  
to determine the solution of the boundary-value problem (\ref{Boundary_Value_Conditions}) when the solution of the initial-boundary problem  
(\ref{Initial_Boundary_Conditions}) is given.

\subsection{The geodesic equation for light propagation in 2PN approximation} 

The metric enters the geodesic equation (\ref{Geodetic_Equation1}) in virtue of the Christoffel symbols (\ref{Christoffel_Symbols}). It
is, however, impossible to determine the Solar System metric without taking recourse to an approximation scheme. Such an approximative  
approach is possible, because in the Solar System the gravitational fields are weak, $m_A/P_A \ll 1$ 
(Schwarzschild radius $m_A = G\,M_A/c^2$ with $M_A$ and $P_A$ being mass and equatorial radius of body $A$) and the motions of matter are slow 
as compared with the speed of light  
$v_A/c \ll 1$ (we have in mind that $v_A$ is just the orbital velocity of the body, but 
in general could also be rotational motion of extended bodies, convection currents  
inside the massive bodies, oscillations of the bodies, etc.).  
Accordingly, a series expansion in inverse powers of the natural constant $c$ is meaningful,   
\begin{eqnarray}
\fl \hspace{1.0cm} g_{\alpha \beta}\left(t,\ve{x}\right) = \eta_{\alpha \beta}
+ h^{(2)}_{\alpha\beta}\left(t,\ve{x}\right) + h^{(3)}_{\alpha\beta}\left(t,\ve{x}\right)
+ h^{(4)}_{\alpha\beta}\left(t,\ve{x}\right) + {\cal O} \left(c^{-5}\right), 
\label{2PN_A}
\end{eqnarray}

\noindent
where $h^{(n)}_{\alpha\beta} \sim {\cal O} \left(c^{-n}\right)$ are tiny perturbations of the flat Minkowskian metric, 
that is $\left|h^{(n)}_{\alpha\beta}\right| \ll 1$ for any $\alpha,\beta$. Here, in line of the comments made above regarding 
the physical meaning of the natural constant $c$, we just notice that the post-Newtonian expansion of the metric tensor (\ref{2PN_A}) 
is of course an expansion with respect to the inverse power of the speed of gravity.  

The series expansion (\ref{2PN_A}) includes all terms  
up to the fifth order and is called post-post-Newtonian (2PN) approximation of the metric tensor.  
The validity of the post-Newtonian expansion (\ref{2PN_A}) is restricted to the near-zone region of the Solar System where
the retardations are small by definition \cite{Kopeikin_Efroimsky_Kaplan,MTW,Poisson_Lecture_Notes,Poisson_Will,Expansion_2PN};  
see also the Fig.~7.7 in \cite{Kopeikin_Efroimsky_Kaplan} or Fig.~36.3 in \cite{MTW}. The near-zone of a gravitating system
is defined as spatial region with the boundary $\left|\ve{x}\right| \ll \lambda_{\rm gr}\,$, where $\lambda_{\rm gr}$ is a  
characteristic wavelength of gravitational waves emitted by the system and the origin of spatial axes is assumed to be located 
at the center-of-mass of the gravitational system or somewhere nearby. 
For the Solar System one obtains about $\lambda_{\rm gr} \sim 10^{17}\,{\rm m}$ which is the lowest wavelength of gravitational
radiation emitted by Jupiter during its revolution around the barycenter of the Solar System \cite{Kopeikin_Efroimsky_Kaplan,Zschocke2,MTW}.
A more accurate statement is achieved by the fact that the term near-zone is intrinsically connected with 
orbital accelerations $a_A$ of the massive bodies $A=1,...,N$ which constitute gravitational system. In mathematical terms it requires  
\begin{eqnarray}
\frac{a_A\left(t\right)\,r_A\left(t\right)}{c^2} \ll \frac{v_A\left(t\right)}{c} \ll 1  
\label{Near_Zone_1}
\end{eqnarray}

\noindent
for each massive body $A$; here $v_A\left(t\right)$ is the orbital velocity and 
$r_A\left(t\right) = \left| \ve{x} - \ve{x}_A\left(t\right)\right|$
is the spatial distance of some field point $\ve{x}$ from the massive body $A$ located at $\ve{x}_A\left(t\right)$. 
The condition (\ref{Near_Zone_1}) has already been stated by by Eq.~(B7) in \cite{Zschocke3} or Eq.~(97) in \cite{Zschocke4} and follows
from $\left|h^{(4)}_{\alpha\beta}\right| \ll \left|h^{(2)}_{\alpha\beta}\right| \ll 1$,
where the metric coefficients for a system of $N$ moving monopoles are given by Eqs.~(24) - (27) in \cite{Zschocke4} 
\footnote{Let us recall that the harmonic gauge condition (\ref{harmonic_gauge_condition_1}) still inherits a residual gauge freedom,
so the harmonic coordinates actually refer to a class of reference systems.
A unique choice of harmonic coordinates is provided by the Barycentric Celestial Reference System (BCRS)
\cite{IAU_Resolution1}, which defines the origin of spatial coordinates at the
barycenter of the Solar System, a stipulation which removes the residual gauge freedom. The metric coefficients for a system 
of $N$ moving monopoles, which have been presented by Eqs.~(24) - (27) in \cite{Zschocke4}, are given in the BCRS, 
so they do not contain any gauge terms.}.  
Using the numerical values
of the most massive Solar System bodies as given in Table~\ref{Table1} we find the spatial radius of the near-zone to be about
\begin{eqnarray}
\left| \ve{x}\right| \le 10^{14}\,{\rm m}\,.
\label{Near_Zone_2}
\end{eqnarray}

\noindent
The results and considerations of our investigation are valid within this spatial region, which corresponds to about $4$ light-days.

By inserting the post-Newtonian expansion of the metric tensor (\ref{2PN_A}) into the geodesic equation (\ref{Geodetic_Equation1})
via the Christoffel symbols (\ref{Christoffel_Symbols}) one obtains the geodesic equation in the so-called post-post-Newtonian (2PN) approximation,  
which is given, for instance, in \cite{Zschocke4,Zschocke3,Bruegmann2005}. 
The formal solution of the geodesic equation in 2PN approximation reads  
\footnote{The notation in Eq.~(\ref{2PN_B}) has been adjusted to the standard notation commonly used in the
literature \cite{Brumberg1991,Zschocke1,Zschocke2,Zschocke3,Zschocke4}. A reconcilable notation for the 
series expansions (\ref{2PN_A}) and (\ref{2PN_B}) can be achieved by noticing that   
$\Delta \ve{x}^{\left(2\right)} \equiv \Delta \ve{x}^{1 {\rm PN}}$ and
$\Delta \ve{x}^{\left(3\right)} \equiv \Delta \ve{x}^{1.5 {\rm PN}}$ and $\Delta \ve{x}^{\left(4\right)} \equiv \Delta \ve{x}^{2 {\rm PN}}$.},
\begin{eqnarray}
\fl \hspace{0.9cm} \ve{x}\left(t\right) = \ve{x}_0 + c \left(t - t_0\right) \ve{\sigma} + \Delta \ve{x}^{1 {\rm PN}}\left(t\right)
+ \Delta \ve{x}^{1.5 {\rm PN}}\left(t\right) + \Delta \ve{x}^{2 {\rm PN}}\left(t\right) + {\cal O} \left(c^{-5}\right).
\label{2PN_B}
\end{eqnarray}

\noindent 
The first two terms on the r.h.s. in (\ref{2PN_B}) represent the unperturbed light ray (\ref{unperturbed_lightray_1}), while 
the subsequent terms represent corrections to the unperturbed light ray. 
The physical meaning of the natural constant $c$ in the unperturbed light ray, $\ve{x}_0 + c \left(t - t_0\right) \ve{\sigma}$, is 
of course the speed of light in flat Minkowskian space-time; cf. comment below Eqs.~(\ref{Geodetic_Equation1}) - (\ref{Null_Condition1}) 
regarding the geodesic equation and isotropic condition for light rays. It should be noticed  
that the post-Newtonian correction terms $\Delta \ve{x}^{n {\rm PN}}\left(t\right)$ originate from the post-Newtonian expansion of 
the metric tensor (\ref{2PN_A}), which is an expansion in inverses powers of $c$, meaning the speed of gravity.  
However, in order to compute these correction terms $\Delta \ve{x}^{n {\rm PN}}\left(t\right)$, the integration of geodesic equation  
proceeded along the unperturbed light ray \cite{Zschocke3,Zschocke4}, where the meaning of $c$ is the speed of light. Therefore, the   
correction terms $\Delta \ve{x}^{n {\rm PN}}\left(t\right)$ in (\ref{2PN_B}) contain the natural constant $c$ in two 
different meanings, namely the speed of light and the speed of gravity. One might believe that this kind of entanglement makes it 
impossible to separate the impact of the finite speed of gravity and the finite speed of light in these correction terms. This is,  
however, not true. The terms related to the characteristics of the gravity field and the terms related to the light characteristics can 
clearly be separated in the solution of the light-ray trajectory (\ref{2PN_B});  
cf. comments below Eqs.~(\ref{Four_Vector_sigma}) - (\ref{Four_Vector_r_A}).  

For an overview of the state-of-the-art in the theory of light propagation we refer to the text books 
\cite{Kopeikin_Efroimsky_Kaplan,Brumberg1991} and the articles
\cite{Klioner2003b,Zschocke1,Zschocke2,KK1992,Kopeikin1997,KS1999,KSGE,Klioner2003a,KopeikinMashhoon2002}. 
According to these references, an impressive progress in the determination of the correction terms $\Delta \ve{x}^{1 {\rm PN}}\left(t\right)$ 
and $\Delta \ve{x}^{1.5 {\rm PN}}\left(t\right)$ has been made during recent decades.  

On the other side, the knowledge of the correction terms $\Delta \ve{x}^{2 {\rm PN}}\left(t\right)$ is pretty much limited thus far.  
In fact, the problem of light propagation in 2PN approximation, that means the determination of the light trajectory (\ref{2PN_B}) 
as function of coordinate time, has only been considered for the following rather restricted situations  
\footnote{Let us notice here that the light deflection in 2PN approximation in the field of one monopole at rest has been determined a long time ago
\cite{EpsteinShapiro,FischbachFreeman,RM1,RM2,RM3,Cowling,BodennerWill2003}. But a unique interpretation of astrometric observations
requires the knowledge of the propagation of the light signal, i.e. the determination of the light trajectory as function of coordinate time 
(\ref{2PN_B}). We also notice the investigation in \cite{Moving_Kerr_Black_Hole2} where the problem of time delay in the field of one  
monopole in uniform motion has been considered, but this investigation was not aiming at astrometric measurements in the Solar System.}:
\begin{enumerate}
\item[$\bullet$] 2PN light trajectory in the field of one monopole at rest \cite{Brumberg1991,Brumberg1987} 
\footnote{The results of \cite{Brumberg1991,Brumberg1987} were later confirmed in several related investigations
\cite{Deng_Xie,Deng_2015,Minazzoli2,Article_Zschocke1,LLT2004,TL2008,Teyssandier,HBL2014b,AshbyBertotti2010,Moving_Kerr_Black_Hole1}.},   
\item[$\bullet$] 2PN light trajectory in the field of two point-like bodies in slow motion \cite{Bruegmann2005},   
\end{enumerate}

\noindent
where \cite{Bruegmann2005} was not intended for light propagation in the Solar System.  

It is, however, clear that for astrometry on the micro-arcsecond and sub-micro-arcsecond level it is indispensable to determine
the light trajectory in the second post-Newtonian approximation for more realistic gravitational systems, especially where the motion of the 
bodies is taken into account  
\cite{Conference_Cambridge,Deng_Xie,Deng_2015,Minazzoli2,Xu_Wu,Xu_Gong_Wu_Soffel_Klioner,Minazzoli1,2PN_Light_PropagationA,Xie_Huang}. 
Already for micro-arcsecond astrometry it is necessary to account for the motion of the Solar System bodies, where it is 
sufficient to determine the light trajectory in the field of one monopole at rest, $\ve{x}_A = {\rm const}$,  
and then simply to insert the retarded position of the body, $\ve{x}_A = \ve{x}_A\left(s_1\right)$, where $s_1$ is the retarded instant of time 
as defined by Eq.~(\ref{retarded_time_s_1}).  
But for the sub-micro-arcsecond astrometry such a simplified access is insufficient, 
because terms which are proportional to the orbital velocity of the body contribute on such level of precision in light deflection.   
In order to account for those terms in the 2PN solution of the light trajectory which are proportional to the orbital velocity of the body, 
one needs to consider the equation of motion for light signals propagating in the gravitational field of moving bodies. On these grounds,  
an analytical solution for the light trajectory in 2PN approximation in the gravitational field of one arbitrarily moving  
pointlike monopole has recently been determined in \cite{Zschocke3,Zschocke4},  
where the so-called initial-value problem (\ref{Initial_Boundary_Conditions}) has been solved:  
\begin{enumerate}
\item[$\bullet$] 2PN light trajectory in field of one arbitrarily moving monopole \cite{Zschocke3,Zschocke4}.
\end{enumerate}
 
\noindent 
Because in reality any light source is located at some finite distance, the consideration of the boundary-value problem  
(\ref{Boundary_Value_Conditions}) is of fundamental importance for the unique interpretation of astrometric  
observations \cite{Kopeikin_Efroimsky_Kaplan,Brumberg1991,KK1992}. Needless to say, that this fact becomes of particular importance 
for astrometry of Solar System objects, say for astrometric measurements in the near-zone of the Solar System, which will be the 
primary topic of this investigation.  
 
The organization of the article is aligned as follows.  
In Section \ref{Section3} the main results of the initial-boundary value problem of 2PN light propagation are summarized,  
which were recently obtained in \cite{Zschocke3,Zschocke4}. Section \ref{Section4} defines the boundary-value problem,    
and series expansions in the near-zone of the Solar System are considered. The three fundamental transformations  
of the boundary-value problem are derived in the Sections \ref{Section5} and \ref{Section6} and \ref{Section7}. An estimation of the numerical magnitude  
of each individual term and the resulting simplified transformations are also given in these Sections. 
The impact of higher order terms beyond 2PN approximation is considered in Section \ref{Section_3PN}.  
The summary and outlook can be found in Section \ref{Section8}.  
The notation, some relations, and details of the calculations are delegated to appendices.

\section{The initial-boundary value problem in 2PN approximation}\label{Section3} 

So as not to have to look up in the literature the main results of our articles \cite{Zschocke3,Zschocke4},  
that is the solution in 2PN approximation for coordinate velocity and trajectory of a light signal propagating in the field of one moving monopole,  
will be summarized for subsequent considerations.  

As formulated in the introductory section, a unique solution of (\ref{Geodetic_Equation1}) is well-defined by initial-boundary conditions,
\begin{eqnarray}
\ve{x}_0 &=& \ve{x}\left(t\right)\bigg|_{t=t_0} \quad {\rm and} \quad \ve{\sigma} = \frac{\dot{\ve{x}}\left(t\right)}{c}\bigg|_{t = - \infty}\,,
\label{Introduction_6}
\end{eqnarray}

\noindent
with $\ve{x}_0$ being the position of the light source at the moment $t_0$ of emission of the light-signal and $\ve{\sigma}$
being the unit-direction ($\ve{\sigma} \cdot \ve{\sigma} = 1$) of the light ray at past null infinity.

\begin{figure}[!ht]
\begin{indented}
\item[]
\includegraphics[scale=0.14]{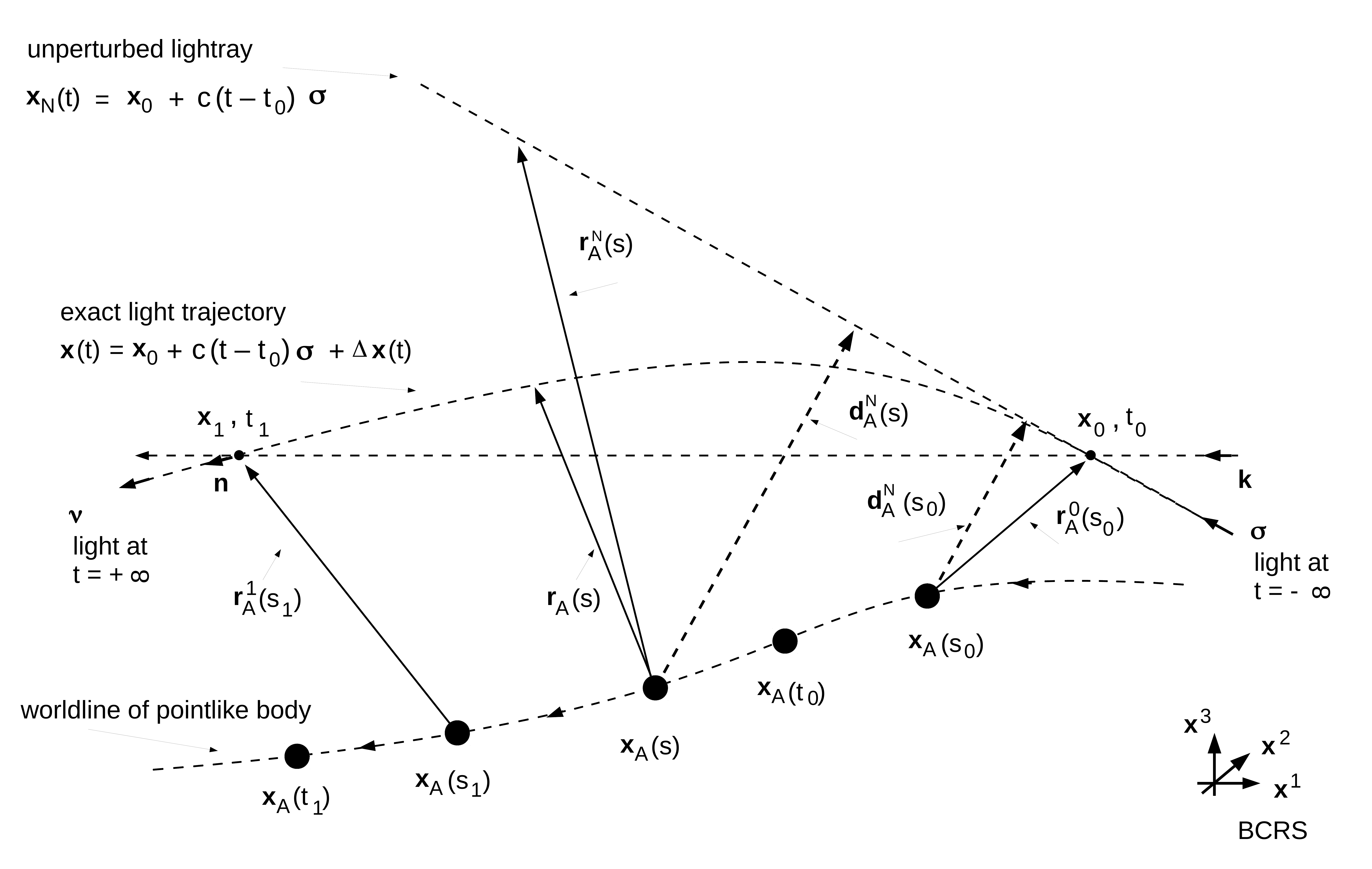}
\end{indented}
\vspace{10pt}
\caption{\label{Diagram1} A geometrical representation of light propagation through the gravitational field
of one pointlike massive body $A$ moving along its worldline $\ve{x}_A\left(t\right)$; the diagram is not Minkowskian but a purely 
spatial picture, i.e. $\left(x^1,x^2,x^3\right)$ denote the three spatial axes of the BCRS.  
The three-vectors $\ve{r}_A\left(s\right)$, $\ve{r}^{\,0}_A\left(s_0\right)$, and $\ve{r}^{\,1}_A\left(s_1\right)$ are defined by 
Eqs.~(\ref{vector_B}), (\ref{vector_rA_0}), and (\ref{vector_rA_1}), respectively; for $\ve{r}^{\rm N}_A\left(s\right)$ see footnote on p.18$\,$. 
The impact vectors $\ve{d}^{\rm N}\left(s\right)$ and $\ve{d}^{\rm N}\left(s_0\right)$ are given by 
Eqs.~(\ref{Impact_Vector_Sigma_s0_Newtonian}).  
The three-vectors $\ve{\sigma}$, $\ve{k}$, and $\ve{n}$ 
are shown, which are defined by the Eqs.~(\ref{Introduction_6}), (\ref{Boundary_3}), and (\ref{Tangent_Vector1}), respectively. 
Their transformations among each other represent the fundamental aspects of the boundary value problem.}  
\end{figure}

\subsection{The coordinate velocity of a light signal in 2PN approximation} 

The first integration of geodesic equation in 2PN approximation yields the coordinate velocity of a light signal 
and is given by (cf. Eq.~(99) in \cite{Zschocke4}):  
\begin{eqnarray}
\fl \hspace{1.0cm} \frac{\dot{\ve{x}}\left(t\right)}{c} = \ve{\sigma}
+ m_A\,\ve{A}_1\left(\ve{r}_A\left(s\right)\right)
+ m_A\,\ve{A}_2\left(\ve{r}_A\left(s\right),\ve{v}_A\left(s\right)\right)
\nonumber\\
\fl \hspace{2.6cm} + \,m_A^2\,\ve{A}_3\left(\ve{r}_A\left(s\right)\right)  
+ m_A\,\ve{\epsilon}_1\left(\ve{r}_A\left(s\right),\ve{v}_A\left(s\right)\right) + {\cal O}\left(c^{-5}\right),
\label{First_Integration}
\end{eqnarray}

\noindent
where the vectorial functions $\ve{A}_1$, $\ve{A}_2$, $\ve{A}_3$, and $\ve{\epsilon}_1$ are given in \ref{Appendix2}  
by Eqs.~(\ref{Vectorial_Function_A1}) - (\ref{Vectorial_Function_A3}) and Eq.~(\ref{epsilon_1}), respectively.  
The argument $\ve{r}_A\left(s\right)$ in the vectorial functions in (\ref{First_Integration}) is 
\footnote{The approximative arguments in the vectorial functions in Eqs.~(99) and (128) in \cite{Zschocke4} can be replaced  
by their exact value $\ve{r}_A\left(s\right)$, because such replacement causes an error of the order ${\cal O}\left(c^{-5}\right)$ 
which is beyond 2PN approximation.}  
\begin{eqnarray}
\ve{r}_A\left(s\right) &=& \ve{x}\left(t\right) - \ve{x}_A\left(s\right),
\label{vector_B}
\end{eqnarray}

\noindent
with $\ve{x}\left(t\right)$ being the exact spatial coordinate of the light signal at global coordinate time $t$, 
while $\ve{x}_A\left(s\right)$ is the spatial position of the body at retarded time $s$,  
which is defined by an implicit relation,  
\begin{eqnarray}
s = t - \frac{r_A\left(s\right)}{c}\,,   
\label{retarded_time_s}
\end{eqnarray}

\noindent 
where $r_A\left(s\right) = \left|\ve{r}_A\left(s\right)\right|$; here it should be noticed that the retardation 
(\ref{retarded_time_s}) is due to the finite speed of propagation of gravity which equals the speed of light.  
The other argument $\ve{v}_A\left(s\right)$ in the  
vectorial functions in (\ref{First_Integration}) is the orbital velocity of the body at the retarded instant of time $s$.  
The retarded time (\ref{retarded_time_s}) is a function of coordinate time and cannot be solved in closed form;  
only for the simple case of linear motion of the 
body a solution is possible as given by Eq.~(3.14) in \cite{Kopeikin_CQG} or Eq.~(9) in \cite{Zschocke_Soffel}.

\subsection{The trajectory of a light signal in 2PN approximation} 

The second integration of geodesic equation in 2PN approximation yields the trajectory of a light signal
and is given by (cf. Eq.~(128) in \cite{Zschocke4}):  
\begin{eqnarray}
\fl \hspace{1.0cm} \ve{x}\left(t\right) = \ve{x}_0 + c \left(t - t_0\right) \ve{\sigma}
+ \,m_A\,\bigg(
\ve{B}_1\left(\ve{r}_A\left(s\right)\right) - \ve{B}_1\left(\ve{r}_A\left(s_0\right)\right)
\bigg)
\nonumber\\
\nonumber\\
\fl \hspace{2.7cm} + \,m_A\,\bigg(\ve{B}^A_2\left(\ve{r}_A\left(s\right),\ve{v}_A\left(s\right)\right)
- \ve{B}^A_2\left(\ve{r}_A\left(s_0\right),\ve{v}_A\left(s\right)\right)\bigg)
\nonumber\\
\nonumber\\
\fl \hspace{2.7cm} + \,m_A\,\bigg(\ve{B}^B_2\left(\ve{r}_A\left(s\right),\ve{v}_A\left(s\right)\right)
- \ve{B}^B_2\left(\ve{r}_A\left(s_0\right),\ve{v}_A\left(s_0\right)\right)\bigg)
\nonumber\\
\nonumber\\
\fl \hspace{2.7cm} + \,m_A^2\,\bigg(\ve{B}_3\left(\ve{r}_A\left(s\right)\right) - \ve{B}_3\left(\ve{r}_A\left(s_0\right)\right) \bigg)
+ m_A\,\ve{\epsilon}_2\left(s,s_0\right) + ¬{\cal O}\left(c^{-5}\right),  
\label{Second_Integration}
\end{eqnarray}

\noindent
where the vectorial functions $\ve{B}_1$, $\ve{B}^A_2$, $\ve{B}^B_2$, $\ve{B}_3$, and $\ve{\epsilon}_2$ are given in \ref{Appendix2}
by Eqs.~(\ref{Vectorial_Function_C1}) - (\ref{Vectorial_Function_C3}) and Eqs.~(\ref{epsilon_3}) - (\ref{epsilon_3b}), respectively.  
The argument $\ve{r}_A\left(s_0\right)$ reads  
\begin{eqnarray}
\ve{r}_A\left(s_0\right) = \ve{x}\left(t_0\right) - \ve{x}_A\left(s_0\right), 
\label{vector_rA_00}
\end{eqnarray}

\noindent
with $\ve{x}\left(t_0\right)$ being the exact spatial coordinate of the light signal at the light source, while $\ve{x}_A\left(s_0\right)$
is the spatial position of the body at retarded time $s_0$, which reads  
\begin{eqnarray}
s_0 &=& t_0 - \frac{r_A\left(s_0\right)}{c}\,,
\label{retarded_time_s_00}
\end{eqnarray}

\noindent 
where $r_A\left(s_0\right) = \left|\ve{r}_A\left(s_0\right)\right|$; let us notice here that the retarded time in (\ref{retarded_time_s_00}) 
is due to the finite speed of propagation of gravity which equals the speed of light.  

The other argument $\ve{v}_A\left(s_0\right)$ in the vectorial functions 
in (\ref{Second_Integration}) is the orbital velocity of the body at the retarded instant of time $s_0$.  
As it has been emphasized in \cite{Zschocke4}, it is important to realize that the velocity in the vectorial functions 
$\ve{B}^A_2$ in (\ref{Second_Integration}) is taken at the very same instant of retarded time $s$,  
which ensures the logarithm in (\ref{Vectorial_Function_C2_A}) in combination with (\ref{Second_Integration}) to be well-defined. 
 
There seems to be a marginal difference between Eq.~(\ref{Second_Integration}) and Eq.~(128) in \cite{Zschocke4},
namely the argument of the velocity term in the second line of both these equations are different.
However, this difference is only apparent, because the relation (cf. Eq.~(121) in \cite{Zschocke4}) 
\begin{eqnarray}
\frac{\ve{v}_A\left(s_0\right)}{c} = \frac{\ve{v}_A\left(s\right)}{c} + \frac{\ve{a}_A\left(s\right)}{c^2}\;c\left(s_0 - s\right) 
+ {\cal O}\left(c^{-3}\right),  
\label{Series_A}
\end{eqnarray}

\noindent
allows to replace $\ve{v}_A\left(s_0\right)$ by $\ve{v}_A\left(s\right)$. But according to this relation,
such a replacement implies the occurrence of a term $\ve{a}_A\left(s\right) \left(s_0 - s\right)$ which is
taken into account in the vectorial function $\ve{\epsilon}_2\left(s,s_0\right)$; cf. last term in (\ref{epsilon_3b})
and text below that equation. Here we also notice the following important relation (cf. Eq.~(127) in \cite{Zschocke4}),  
\begin{eqnarray}
\fl c \left(s_0 - s\right) = r_A\left(s\right) - \ve{\sigma} \cdot \ve{r}_A\left(s\right) - r_A\left(s_0\right) 
+ \ve{\sigma} \cdot \ve{r}_A\left(s_0\right) - \ve{\sigma} \cdot \ve{x}_A\left(s\right) + \ve{\sigma} \cdot \ve{x}_A\left(s_0\right),  
\label{Series_B}
\end{eqnarray}

\noindent 
which is valid up to terms of the order ${\cal O}\left(c^{-2}\right)$ and follows from (\ref{retarded_time_s}) and (\ref{retarded_time_s_00}) 
in virtue of (\ref{Second_Integration}) with (\ref{vector_B}) and (\ref{vector_rA_00}). It should be noticed that the  
solutions of coordinate velocity (\ref{First_Integration}) and trajectory (\ref{Second_Integration}) of a light signal as well as  
relation (\ref{Series_B}) are valid for any kind of configuration between source, body and observer.

\subsection{Impact vectors in the initial value problem} 

In the solution for the coordinate velocity (\ref{First_Integration}) and trajectory (\ref{Second_Integration})  
of a light signal, the following expressions naturally appear,  
\begin{eqnarray}
\ve{d}_A\left(s\right) = \ve{\sigma} \times \left(\ve{r}_A\left(s\right) \times \ve{\sigma}\right),  
\label{Impact_Vector_Sigma_s}
\\
\ve{d}_A\left(s_0\right) = \ve{\sigma} \times \left(\ve{r}_A\left(s_0\right) \times \ve{\sigma}\right),  
\label{Impact_Vector_Sigma_s0}
\end{eqnarray}

\noindent
where the three-vectors $\ve{r}_A\left(s\right)$ and $\ve{r}_A\left(s_0\right)$ are defined by Eqs.~(\ref{vector_B}) and (\ref{vector_rA_00}).   
The three-vectors (\ref{Impact_Vector_Sigma_s}) and (\ref{Impact_Vector_Sigma_s0}) and their absolute values are called  
impact vectors and impact parameters 
\footnote{One may also define impact vectors with respect to the unperturbed light ray,  
\begin{eqnarray}
\ve{d}^{\rm N}_A\left(s\right) = \ve{\sigma} \times \left(\ve{r}^{\rm N}_A\left(s\right) \times \ve{\sigma}\right)  
\quad {\rm and} \quad \ve{d}^{\rm N}_A\left(s_0\right) = \ve{\sigma} \times \left(\ve{r}^{\rm N}_A\left(s_0\right) \times \ve{\sigma}\right),   
\label{Impact_Vector_Sigma_s0_Newtonian}
\end{eqnarray}
 
\noindent
where $\ve{r}^{\rm N}_A\left(s\right) = \ve{x}_0 + c\,\ve{\sigma} \left(t-t_0\right) - \ve{x}_A\left(s\right)$ 
and $\ve{r}^{\rm N}_A\left(s_0\right) = \ve{x}_0 - \ve{x}_A\left(s_0\right) = \ve{r}_A\left(s_0\right)$. 
They are illustrated in Figure~\ref{Diagram1}. Due to  
$\ve{d}_A\left(s\right) = \ve{d}^{\rm N}_A\left(s\right) + {\cal O}\left(c^{-2}\right)$ and  
$\ve{d}_A\left(s_0\right) = \ve{d}^{\rm N}_A\left(s_0\right)$,  
the impact vector $\ve{d}_A\left(s\right)$ differs marginal from $\ve{d}^{\rm N}_A\left(s\right)$, while impact vector  
$\ve{d}_A\left(s_0\right)$ is even identical to $\ve{d}^{\rm N}_A\left(s_0\right)$. The graphical representation 
of $\ve{d}^{\rm N}_A\left(s\right)$ and $\ve{d}^{\rm N}_A\left(s_0\right)$ in Figure~\ref{Diagram1} makes it evident 
why these terms are called impact vectors.}, respectively.

An important condition for the impact parameter $d_A\left(s\right)$ is imposed, which follows from the requirement that the  
light source should not be screened by the finite disk of the body,     
\begin{eqnarray}
d_A\left(s\right) \ge P_A \quad {\rm for} \quad \ve{\sigma} \cdot \ve{r}_A\left(s\right) \ge 0\,, 
\label{Impact_Vector_Sigma_Constraint_1}
\end{eqnarray}
 
\noindent 
cf. Section 4.2. in \cite{Article_Zschocke1} for the case of body at rest. 
If $\ve{\sigma} \cdot \ve{r}_A\left(s\right) < 0$ then there is no constraint imposed for the impact parameter, 
\begin{eqnarray}
d_A\left(s\right) \ge 0 \quad {\rm for} \quad \ve{\sigma} \cdot \ve{r}_A\left(s\right) < 0\,. 
\label{Impact_Vector_Sigma_Constraint_2}
\end{eqnarray}

\noindent
One may show that (\ref{Impact_Vector_Sigma_Constraint_1}) implies $d_A\left(s_0\right) \ge P_A$ 
for $\ve{\sigma} \cdot \ve{r}_A\left(s_0\right) \ge 0$, which is not an additional request but 
has the same meaning as (\ref{Impact_Vector_Sigma_Constraint_1}). But because in the near-zone of the Solar System 
the impact parameter $d_A\left(s_0\right)$  
is related to $d_A\left(s\right)$ via a series expansion, there is no need to impose additional constraints  
on $d_A\left(s_0\right)$.  
This issue will be considered in more detail within the boundary value problem.

\section{The boundary value problem in 2PN approximation}\label{Section4}  

As formulated in the introductory section, a unique solution of (\ref{Geodetic_Equation1}) is also well-defined by boundary conditions,
\begin{eqnarray}
\ve{x}_0 = \ve{x}\left(t\right)\bigg|_{t=t_0} \quad {\rm and} \quad \ve{x}_1 = \ve{x}\left(t\right)\bigg|_{t=t_1}\,, 
\label{Boundary_1}  
\end{eqnarray}

\noindent
where $\ve{x}_0$ is the point of emission of the light signal by the source and $\ve{x}_1$ is the point of reception of the light signal   
by the observer.  
The position of the observer $\ve{x}_1$ in the BCRS is known, while the position of the light source $\ve{x}_0$  
has to be determined by a unique interpretation of astronomical observations, for it is the primary aim of astrometric data reduction.  

In the theory of light propagation the unit-vector $\ve{k}$,  
which points from the light source towards the position of the observer, is of fundamental importance,  
\begin{eqnarray}
\ve{k} &=& \frac{\ve{R}}{R} \quad {\rm with} \quad \ve{R} = \ve{x}_1 - \ve{x}_0 \quad {\rm and} \quad R = \left|\ve{x}_1 - \ve{x}_0\right| \,. 
\label{Boundary_3}
\end{eqnarray}

\noindent
A further important unit-vector is the normalized tangent along the light ray at the observer's position,  
\begin{eqnarray}  
\ve{n} = \frac{\dot{\ve{x}}\left(t_1\right)}{\left|\dot{\ve{x}}\left(t_1\right)\right|}\,.
\label{Tangent_Vector1} 
\end{eqnarray}  

\noindent 
In Figure~\ref{Diagram1} these unit-vectors $\ve{n}$ and $\ve{k}$ are depicted which play the key role in the boundary value problem.  

There are two specific cases for the retarded moment of time (\ref{retarded_time_s}) which are of relevance 
in the boundary value problem:  

(i) The retarded instant of time $s_0$ with respect to the emission of the light signal 
at the four-coordinate of source $\left(c t_0, \ve{x}_0\right)$ (cf.~Eq.~(\ref{retarded_time_s_00})),  
\begin{eqnarray}
s_0 &=& t_0 - \frac{r_A^{\,0}\left(s_0\right)}{c} \quad {\rm with} \quad r_A^{\,0}\left(s_0\right) = \left|\ve{r}^{\,0}_A\left(s_0\right)\right|\,,
\label{retarded_time_s_0}
\end{eqnarray}

\noindent
where
\begin{eqnarray}
\ve{r}^{\,0}_A\left(s_0\right) &=& \ve{x}_0 - \ve{x}_A\left(s_0\right), 
\label{vector_rA_0}
\end{eqnarray}

\noindent  
where the upper index $0$ refers to $\ve{x}_0$ and the argument $s_0$ refers to the body's position $\ve{x}_A\left(s_0\right)$; 
here we notice again that the retarded time in (\ref{retarded_time_s_0}) is caused by the finite speed of propagation of gravity 
which equals the speed of light.  
 
Actually, (\ref{vector_rA_0}) coincides with (\ref{vector_rA_00}) in view of $\ve{x}_0 = \ve{x}\left(t_0\right)$,  
but we will keep the notation (\ref{vector_rA_00}) as is, in order not to change the notation for the initial-value problem  
as used in \cite{Zschocke4}. 

(ii) The retarded instant of time with respect to the reception  
of the light signal at the four-coordinate of observer $\left(c t_1, \ve{x}_1\right)$,  
\begin{eqnarray}
s_1 &=& t_1 - \frac{r^{\,1}_A\left(s_1\right)}{c} \quad {\rm with} \quad r^{\,1}_A\left(s_1\right) = \left|\ve{r}^{\,1}_A\left(s_1\right)\right|\,,  
\label{retarded_time_s_1}
\end{eqnarray}

\noindent
where
\begin{eqnarray}
\ve{r}^{\,1}_A\left(s_1\right) &=& \ve{x}_1 - \ve{x}_A\left(s_1\right),  
\label{vector_rA_1}
\end{eqnarray}

\noindent
where the upper index $1$ refers to $\ve{x}_1$ and the argument $s_1$ refers to the body's position $\ve{x}_A\left(s_1\right)$; 
let us recall that the retarded time in (\ref{retarded_time_s_1}) is due to the finite speed of propagation of gravity which equals 
the speed of light.  

For the difference between these retarded instants of time the following relation holds 
\begin{eqnarray}
\fl c \left(s_0 - s_1\right) = \! \bigg(\!r_A^{\,1}\left(s_1\right) - \ve{k} \cdot \ve{r}^{\,1}_A\left(s_1\right)
- r^{\,0}_A\left(s_0\right) \! + \! \ve{k} \cdot \ve{r}^{\,0}_A\left(s_0\right)\!\!\bigg)
 \bigg(\!1 + \frac{\ve{k} \cdot \ve{v}_A\left(s_1\right)}{c}\!\bigg)
\! + \! {\cal O}\left(c^{-2}\right),
\nonumber\\
\label{relation_k_1}
\end{eqnarray}

\noindent
which follows from (\ref{retarded_time_s_0}) and (\ref{retarded_time_s_1}) as well as (\ref{Second_Integration}) and (\ref{Boundary_3}); 
cf. Eq.~(\ref{Series_B}) in combination with the fact that and $\ve{\sigma} = \ve{k} + {\cal O}\left(c^{-2}\right)$ and taking account 
of the below standing series expansion (\ref{series_expansion_body_1}).   
The relation (\ref{relation_k_1}) is valid for any kind of configuration between source, body and observer.

\subsection{Series expansion of the spatial position of the body} 

In the near-zone of the Solar System a series expansion of the spatial position of the body becomes meaningful.  
It is clear that the determination of $s_0$ requires the knowledge of the four-coordinate of the
light source $\left(c t_0, \ve{x}_0\right)$, which initially is unknown but results from data reduction of astrometric observations.
On the other side, the determination of $s_1$ requires the four-coordinate of the observer
$\left(c t_1, \ve{x}_1\right)$ as well as the worldline of the body $\ve{x}_A\left(t\right)$, both of which 
are fundamental prerequisites for astrometric observations in the near-zone of the Solar System.  
Usually, the four-coordinates of the observer are provided by optical tracking of the spacecraft,  
while $\ve{x}_A\left(t\right)$ is provided by some Solar System ephemeris \cite{JPL}.  
Accordingly, we consider a series expansion of the body's position around $s_1$,  
\begin{eqnarray}
\fl \ve{x}_A\left(s_0\right) = \ve{x}_A\left(s_1\right) + \frac{1}{1!} \,\frac{\ve{v}_A\left(s_1\right)}{c}\,c \left(s_0 - s_1\right) 
+ \frac{1}{2!}\,\frac{\ve{a}_A\left(s_1\right)}{c^2}\,c^2 \left(s_0 - s_1\right)^2 + {\cal O}\left(c^{-3}\right), 
\label{series_expansion_body_1}
\end{eqnarray}

\noindent 
which relates the spatial position of the body at retarded time $s_1$ and at retarded time $s_0$, and where  
the expression for $c \left(s_0 - s_1\right)$ is given by Eq.~(\ref{relation_k_1}). 
The r.h.s. of (\ref{series_expansion_body_1}) still depends on $s_0$. So it turns out to be meaningful to introduce a  
further three-vector which is defined as follows, 
\begin{eqnarray}
\ve{r}^{\,0}_A\left(s_1\right) &=& \ve{x}_0 - \ve{x}_A\left(s_1\right) \quad {\rm and} \quad
r^{\,0}_A\left(s_1\right) = \left| \ve{r}^{\,0}_A\left(s_1\right) \right|,
\label{vector_rA_0_s1}
\end{eqnarray}

\noindent
where the upper index $0$ refers to $\ve{x}_0$ and the argument $s_1$ refers to the body's position $\ve{x}_A\left(s_1\right)$.
Using this three-vector one may show by iterative use of relation (\ref{series_expansion_body_1}) that 
the expression for $c \left(s_0 - s_1\right)$ as given by Eq.~(\ref{relation_k_1}) can also be expressed solely in terms of 
$s_1$ as follows, 
\begin{eqnarray}
\fl c \left(s_0 - s_1\right) = \! \bigg(\!r_A^{\,1}\left(s_1\right) - \ve{k} \cdot \ve{r}^{\,1}_A\left(s_1\right)
- r^{\,0}_A\left(s_1\right) + \ve{k} \cdot \ve{r}^{\,0}_A\left(s_1\right)\!\!\bigg)
 \bigg(\!1 + \frac{\ve{k} \cdot \ve{v}_A\left(s_1\right)}{c}\!\bigg)
\! + \! {\cal O}\left(c^{-2}\right). 
\nonumber\\
\label{relation_k_2}
\end{eqnarray}
 
\noindent 
The series expansion (\ref{series_expansion_body_1}) is absolutely convergent in the near-zone of the Solar System where the time of  
light propagation is certainly less than the orbital period of any massive body orbiting around the barycenter of the Solar System.  
That means, according to the convergence criterion \cite{Mathematical_Methods},
the following limit exists
\footnote{For instance, the worldline of a body in a two-dimensional circular orbit of radius $r$ is
$\displaystyle \ve{x}_A\left(t\right) = \left(r\,\cos \omega t\;,\;r\,\sin \omega t \right)^{\rm T}$ where $\omega = 2\,\pi/T$ is the
angular frequency with $T$ being the orbital period. One gets $\left|\ve{x}_A^{(n)}\right| = r\,\omega^n$, hence the limit
$\displaystyle L = \lim_{n \rightarrow \infty} \frac{2\,\pi}{T}\,\frac{\left|s_0 - s_1\right|}{n + 1} = 0$.}
\begin{eqnarray}
\fl L = \lim_{n \rightarrow \infty}
\frac{\displaystyle \left|\ve{x}_A^{\left(n+1\right)}\left(s_1\right)\right|\,\frac{\left|s_0 - s_1\right|^{n+1}}{\left(n + 1\right)!}}
{\displaystyle \left|\ve{x}_A^{\left(n\right)}\left(s_1\right)\right|\,\frac{\left|s_0 - s_1\right|^{n}}{n!}}
 < 1 \quad {\rm where} \quad
\ve{x}_A^{\left(n\right)}\left(s_1\right) = \frac{d^n \ve{x}_A\left(s\right)}{d\,s^n}\bigg|_{s=s_1}\,.
\label{Convergence_Criterion}
\end{eqnarray}

\noindent
Even though that terms proportional to the velocity of the body, $\ve{v}_A$, can be of the same magnitude or even much larger
than the first term on the r.h.s. of the series expansion (\ref{series_expansion_body_1}), 
the series expansion converges
so rapidly that just the first few terms up to order ${\cal O}\left(c^{-3}\right)$ were represented,
while higher derivatives of the body's position (jerk-term, snap-term, jounce-term, etc.) are not given explicitly.
This fact can be seen by inserting the numerical parameters in Table~\ref{Table1} into the series expansion (\ref{series_expansion_body_1}).  
Finally, we notice that the expansion (\ref{series_expansion_body_1}) implies a series expansion of the spatial velocity of the body,
\begin{eqnarray}
\frac{\ve{v}_A\left(s_0\right)}{c} = \frac{\ve{v}_A\left(s_1\right)}{c}
+ \frac{\ve{a}_A\left(s_1\right)}{c^2}\,c \left(s_0 - s_1\right) + {\cal O}\left(c^{-3}\right), 
\label{series_expansion_vA_2}
\end{eqnarray}

\noindent
where for $c \left(s_0 - s_1\right)$ one has to use relation (\ref{relation_k_2}).  

\subsection{Impact vectors in the boundary value problem} 

For the boundary value problem the relevant impact vectors are defined with respect to the unit vector $\ve{k}$ in Eq.~(\ref{Boundary_3}). As  
we will see, the impact vector $\ve{d}^k_A$ at retarded time $s_0$ and $s_1$ will naturally appear in the solution of the boundary value problem,  
\begin{eqnarray}
\ve{d}^k_A\left(s_0\right) = \ve{k} \times \left(\ve{r}^{\,0}_A\left(s_0\right) \times \ve{k}\right) 
= \ve{k} \times \left(\ve{r}^{\,1}_A\left(s_0\right) \times \ve{k}\right)\,,
\label{Impact_Vector_k0}
\\
\ve{d}^k_A\left(s_1\right) = \ve{k} \times \left(\ve{r}^{\,1}_A\left(s_1\right) \times \ve{k}\right) 
= \ve{k} \times \left(\ve{r}^{\,0}_A\left(s_1\right) \times \ve{k}\right)\,,  
\label{Impact_Vector_k1}
\end{eqnarray}

\noindent
where in the second expression on the r.h.s. in (\ref{Impact_Vector_k0}) the three-vector
\begin{eqnarray}
\ve{r}^{\,1}_A\left(s_0\right) = \ve{x}_1 - \ve{x}_A\left(s_0\right)  
\label{vector_rA_1_s0}
\end{eqnarray}

\noindent
has been introduced. The first expression on the r.h.s. in (\ref{Impact_Vector_k0}) and (\ref{Impact_Vector_k1}) is regarded as the  
actual definition of the impact vector, while the second expression on the r.h.s. in (\ref{Impact_Vector_k0}) and (\ref{Impact_Vector_k1})  
just establishes an equality.  
The notation impact vector for the three-vectors (\ref{Impact_Vector_k0}) and (\ref{Impact_Vector_k1}) becomes evident by their  
graphical representations as given by the Figures~\ref{Diagram2}, \ref{Diagram3} and \ref{Diagram4}. For the same reason their absolute values,   
\begin{eqnarray}
d^k_A\left(s_0\right) = \left|\ve{k} \times \ve{r}^{\,0}_A\left(s_0\right) \right|
= \left| \ve{k} \times \ve{r}^{\,1}_A\left(s_0\right) \right|\,,
\label{Impact_Parameter_k0}
\\
d^k_A\left(s_1\right) = \left| \ve{k} \times \ve{r}^{\,1}_A\left(s_1\right) \right|
= \left| \ve{k} \times \ve{r}^{\,0}_A\left(s_1\right)\right|\,, 
\label{Impact_Parameter_k1}
\end{eqnarray}

\noindent  
are called impact parameter.  
Like in Eq.~(\ref{Impact_Vector_Sigma_Constraint_1}), for the impact parameter at retarded time $s_1$ the following constraint is imposed,  
\begin{eqnarray}
d^k_A\left(s_1\right) \ge P_A \quad {\rm for} \quad \ve{k} \cdot \ve{r}^{\,1}_A\left(s_1\right) \ge 0\,, 
\label{Impact_Vector_k_Constraint_1}
\end{eqnarray}

\noindent
which generalizes the constraint $d^k_A \ge P_A$ for light propagation in the field of bodies at rest 
(cf. Section 4.2 in \cite{Article_Zschocke1}) and just represents the  
fact that configurations where the light source can be seen by the observer in front of the finite sized body are excluded from the 
light propagation model. If $\ve{k} \cdot \ve{r}^{\,1}_A\left(s_1\right) < 0$ then there is no constraint for the impact vector, 
\begin{eqnarray}
d^k_A\left(s_1\right) \ge 0 \quad {\rm for} \quad \ve{k} \cdot \ve{r}^{\,1}_A\left(s_1\right) < 0\,. 
\label{Impact_Vector_k_Constraint_2}
\end{eqnarray}
 
\noindent 
Actually, one may show that (\ref{Impact_Vector_k_Constraint_1}) implies $d^k_A\left(s_0\right) \ge P_A$ if  
$\ve{k} \cdot \ve{r}^{\,0}_A\left(s_1\right) \ge 0$; such a configuration has been represented in Figure~\ref{Diagram3}.  
But there is no need for any constraint on the  
impact parameter $d^k_A\left(s_0\right)$, because this impact parameter is not independent of $d^k_A\left(s_1\right)$.  
This important issue will be considered in more detail in what follows.  

As stated, the impact vectors (\ref{Impact_Vector_k0}) and (\ref{Impact_Vector_k1}) are not independent of each other but related 
via a series expansion. Such a relation is obtained by inserting (\ref{series_expansion_body_1}) into (\ref{vector_rA_1_s0})  
and subsequently into the second term on the r.h.s. of (\ref{Impact_Vector_k0}), which yields  
\begin{eqnarray}
\fl \hspace{1.75cm} \ve{d}^k_A\left(s_0\right) = \ve{d}^k_A\left(s_1\right)
- \frac{1}{1!}\,\ve{k} \times \left(\frac{\ve{v}_A\left(s_1\right)}{c} \times \ve{k}\right)\,c \left(s_0 - s_1\right)
\nonumber\\
\nonumber\\
\fl \hspace{4.5cm} - \frac{1}{2!}\,\ve{k} \times \left(\frac{\ve{a}_A\left(s_1\right)}{c^2} \times \ve{k} \right)\,c^2 \left(s_0 - s_1\right)^2
+ {\cal O}\left(c^{-3}\right), 
\label{Impact_Vector_Relation1}
\end{eqnarray}

\noindent 
where $c \left(s_0 - s_1\right)$ is given by Eq.~(\ref{relation_k_2}). For the absolute value we obtain from (\ref{Impact_Vector_Relation1})  
\begin{eqnarray}
\fl \left(d^k_A\left(s_0\right)\right)^2 \! = \left(d^k_A\left(s_1\right)\right)^2
- 2\,\ve{d}^k_A\left(s_1\right) \cdot \frac{\ve{v}_A\left(s_1\right)}{c} c \left(s_0 - s_1\right)
- \ve{d}^k_A\left(s_1\right) \cdot \frac{\ve{a}_A\left(s_1\right)}{c^2} c^2 \left(s_0 - s_1\right)^2 
\nonumber\\ 
\nonumber\\ 
\fl \hspace{1.75cm} + \left|\ve{k} \times \frac{\ve{v}_A\left(s_1\right)}{c}\right|^2\,c^2 \left(s_0 - s_1\right)^2  
+ \, {\cal O}\left(c^{-3}\right),  
\label{series_expansion_dA_2}
\end{eqnarray}

\noindent 
where $\displaystyle \ve{k} \cdot \ve{d}_A^k\left(s_1\right) = 0$ has been used. 
Whatever we need is a relation between the inverse of $d^k_A\left(s_0\right)$ and the inverse of $d^k_A\left(s_1\right)$.
As mentioned above, the terms proportional to the velocity and acceleration of the body might become larger than the
first term, hence a series expansion of the inverse of (\ref{series_expansion_dA_2}) is not necessarily possible in general. 
So we will have to use the exact identity,  
\begin{eqnarray}
\frac{1}{d^k_A\left(s_0\right)} = \frac{1}{d^k_A\left(s_1\right)}
+ \frac{\left(d^k_A\left(s_1\right)\right)^2 - \left(d^k_A\left(s_0\right)\right)^2}
{d^k_A\left(s_0\right)\,d^k_A\left(s_1\right)\,\left(d^k_A\left(s_0\right) + d^k_A\left(s_1\right)\right)} \,.  
\label{Relation_d0_1}
\end{eqnarray}

\noindent
The latter is used in evaluating the following expansion of the inverse impact parameter,  
\begin{eqnarray}
\fl \frac{1}{d^k_A\left(s_0\right)} = \frac{1}{d^k_A\left(s_1\right)}
+ \frac{\displaystyle 2 \,\ve{d}^k_A\left(s_1\right) \cdot \frac{\displaystyle \ve{v}_A \left(s_1\right)}{c} c \left(s_0 - s_1\right)} 
{d^k_A\left(s_0\right)\,d^k_A\left(s_1\right)\,\left(d^k_A\left(s_0\right) + d^k_A\left(s_1\right)\right)}
\nonumber\\ 
\nonumber\\ 
\fl \hspace{1.68cm}  
+ \, \frac{\displaystyle \ve{d}^k_A\left(s_1\right) \cdot \frac{\displaystyle \ve{a}_A \left(s_1\right)}{c^2} c^2 \left(s_0 - s_1\right)^2}
{d^k_A\left(s_0\right)\,d^k_A\left(s_1\right)\,\left(d^k_A\left(s_0\right) + d^k_A\left(s_1\right)\right)}
- \frac{\displaystyle \left|\ve{k} \times \frac{\ve{v}_A\left(s_1\right)}{c}\right|^2 c^2 \left(s_0 - s_1\right)^2}
{d^k_A\left(s_0\right)\,d^k_A\left(s_1\right)\,\left(d^k_A\left(s_0\right) + d^k_A\left(s_1\right)\right)}
\nonumber\\ 
\nonumber\\ 
\fl \hspace{1.68cm} + \, {\cal O}\left(c^{-3}\right), 
\label{Relation_d0_4}
\end{eqnarray}

\noindent
which is an incomplete series expansion because the r.h.s. still depends on $d^k_A\left(s_0\right)$.  

A comment should be in order about these relations in (\ref{Relation_d0_1}) and (\ref{Relation_d0_4}).  
In contrast to $d^k_A\left(s_1\right)$, which must be larger than the equatorial radius $P_A$ of the massive body as long as 
$\ve{k} \cdot \ve{r}^{\,1}_A\left(s_1\right) > 0$, there is no such kind of constraint for the impact parameter $d^k_A\left(s_0\right)$. 
In other words, the impact parameter $d^k_A\left(s_0\right)$ can become arbitrarily small and might even vanish, so that the  
limit $d^k_A\left(s_0\right) \rightarrow 0$ is quite possible; cf. the related comment below Eq.~(B.12) in \cite{Zschocke4}.  
For such cases the relations (\ref{Relation_d0_1}) and (\ref{Relation_d0_4}) remain strictly valid, but  
the expressions on the l.h.s. and r.h.s. of these relations would become arbitrarily large. 
One has, however, to keep in mind that the inverse of the impact parameter $d^k_A\left(s_0\right)$ is only one piece  
of a more complex expression which, up to terms of the order ${\cal O}\left(c^{-5}\right)$, remains finite when 
inserting the r.h.s. of (\ref{Relation_d0_4}), even in the limit $d^k_A\left(s_0\right) \rightarrow 0$.  
It is a remarkable feature of the 2PN solution that the constraint (\ref{Impact_Vector_k_Constraint_1}) turns out to be sufficient  
to keep each term finite in each of the transformations of boundary value problem, regardless of how small 
the impact parameter $d^k_A\left(s_0\right)$ can be. But one has to bear in mind the fact that the impact vectors,  
the impact parameters, and the inverse of the impact parameters are not independent of each other, but related 
via Eqs.~(\ref{Impact_Vector_Relation1}), (\ref{series_expansion_dA_2}) and (\ref{Relation_d0_4}).

\subsection{Notation of four-vectors} 
 
In what follows we will determine three fundamental transformations which comprise the boundary value problem,  
that means the transformations between $\ve{\sigma}$ in Eq.~(\ref{Introduction_6}),
$\ve{k}$ in Eq.~(\ref{Boundary_3}), $\ve{n}$ in Eq.~(\ref{Tangent_Vector1}), in their chain of reasoning.
But before we proceed further, the following simplifying notation of four-dimensional vectors is introduced, as adopted from 
\cite{Kopeikin_Efroimsky_Kaplan,KS1999,KopeikinMashhoon2002},  
\begin{eqnarray}
\sigma^{\mu} = \left(1,\ve{\sigma}\right), \hspace{3.9cm} \eta_{\mu\nu}\,\sigma^{\mu} \sigma^{\nu} = 0 \,,
\label{Four_Vector_sigma}
\\
k^{\mu} = \left(1,\ve{k}\right), \hspace{4.0cm} \eta_{\mu\nu}\,k^{\mu} k^{\nu} = 0 \,,
\label{Four_Vector_k}
\\
\hspace{-0.5cm} r_A^{\mu}\left(s\right) = \left(r_A\left(s\right),\ve{r}_A\left(s\right)\right),
\hspace{1.47cm} \eta_{\mu\nu}\,r_A^{\mu}\left(s\right) r_A^{\nu}\left(s\right) = 0\,. 
\label{Four_Vector_r_A}
\end{eqnarray}

\noindent
Each of these four-dimensional quantities, (\ref{Four_Vector_sigma}) and (\ref{Four_Vector_k}) and (\ref{Four_Vector_r_A}), is a null-vector with  
respect to the metric tensor $\eta_{\mu\nu}$ of the flat Minkowskian space-time. But one has to take care about their different meaning:  
the four-vectors (\ref{Four_Vector_sigma}) and (\ref{Four_Vector_k}) are, up to terms of the order ${\cal O}\left(c^{-2}\right)$,  
directed along the light ray which is 
a Bicharacteristic (\ref{Biharacteristics_ED}) of the covariant Maxwell equations in the curved space-time of the Solar System, while 
the four-vector (\ref{Four_Vector_r_A}) is directed along the Bicharacteristic (\ref{Biharacteristics_GR}) of the field equations 
of gravity; cf. comments made below Eq.~(7.82) in \cite{Kopeikin_Efroimsky_Kaplan}. 
These facts allow formally to clearly separate the terms related to the characteristics of the gravity field from those terms 
related to the light characteristics; cf. text below Eq.~(\ref{2PN_B}). But these remarks do not mean, that in concrete experiments the  
effects related to the speed of gravity can easily and clearly be separated from the effects related to the speed of light; cf. comments  
below Eqs.~(\ref{Shapiro_2}). Furthermore, one should keep in mind that only (\ref{Four_Vector_sigma}) is actually a physical four-vector,  
because it is defined in the asymptotic region of the Solar System which is Minkowskian, hence can be interpreted as a four-dimensional  
arrow pointing from one event to another. On the other side, the four-quantities (\ref{Four_Vector_k}) and (\ref{Four_Vector_r_A}) are  
introduced as difference of two events in Riemannian space-time, hence they cannot be considered as physical four-vectors in the common sense,  
because in Riemannian space-time a physical four-vector is a (class of) directional derivative acting on some (arbitrary) scalar function; 
cf Sec. 9.2. in \cite{MTW}. Here, we consider four-quantities like (\ref{Four_Vector_sigma}) - (\ref{Four_Vector_r_A}) as purely 
mathematical objects with whom it is allowed to apply usual vectorial operations; cf. text below Eq.~(\ref{Relation_alpha}). In the solution of  
the light trajectory one encounters terms which, in the sense just described, are called four-scalars between the four-vectors 
$\sigma_{\mu} = \left(-1,\ve{\sigma}\right)$, $k_{\mu} = \left(-1,\ve{k}\right)$ and $r_A^{\mu}\left(s\right)$, given by  
\footnote{The  notation $k^{\mu}$ is employed in \cite{Kopeikin_Efroimsky_Kaplan} (cf. text below Eq.~(7.82) in \cite{Kopeikin_Efroimsky_Kaplan})  
for what we call $\sigma^{\mu}$ (cf. Eq.~(\ref{Four_Vector_sigma})).  
Furthermore, our three-vector $\ve{k}$ in Eq.~(\ref{Boundary_3}) coincides, up to a minus sign,  
with the three-vector $\ve{K}$ used in \cite{Kopeikin_Efroimsky_Kaplan,KS1999,KopeikinMashhoon2002}  
(cf. Eq.~(7.66) in \cite{Kopeikin_Efroimsky_Kaplan} or Eq.~(36) in \cite{KS1999} or Eq.~(44) in \cite{KopeikinMashhoon2002}).
It will certainly not cause any kind of confusion that $r_A\left(s\right) \equiv r_A^{\mu}\left(s\right)$ on the l.h.s.  
in (\ref{scalar_product}) denotes the four-vector, while $r_A\left(s\right) \equiv \left|\ve{r}_A\left(s\right)\right|$ on  
the r.h.s. in (\ref{scalar_product}) denotes the absolute value of the three-vector.  
Throughout the manuscript a single four-vector carries always a Lorentz-index. Only in four-scalar products the four-vectors do not carry
a Lorentz index, but then there will always be a dot among these four-vectors. In three-scalar products there is also a dot among the
three-vectors, but the three-vectors are always in bold. For the details of notation in use we refer to \ref{Appendix0}.},
\begin{eqnarray}
\sigma \cdot r_A\left(s\right) \equiv \sigma_{\mu}\,r_A^{\mu}\left(s\right)
= - \left(r_A\left(s\right) - \ve{\sigma} \cdot \ve{r}_A\left(s\right)\right),
\label{scalar_product_sigma}
\\
k \cdot r_A\left(s\right) \equiv k_{\mu}\,r_A^{\mu}\left(s\right) = - \left(r_A\left(s\right) - \ve{k} \cdot \ve{r}_A\left(s\right)\right).
\label{scalar_product}
\end{eqnarray}

\noindent 
These four-vectors in (\ref{Four_Vector_sigma}) and (\ref{Four_Vector_r_A}) and their scalar-product (\ref{scalar_product_sigma}) 
do naturally appear as arguments of vectorial functions in the solution of the initial-boundary value problem for the 
light trajectory, while the four-vectors in (\ref{Four_Vector_k}) and (\ref{Four_Vector_r_A}) and their scalar-product (\ref{scalar_product})  
do naturally appear as arguments of vectorial functions in the solution of the boundary value problem for the light trajectory. 
 
Here, we just have introduced the above standing notation in order to simplify the mathematical expressions in the 
boundary value problem of the theory of light propagation.  
In particular, for the two specific four-vectors,   
\begin{eqnarray}
r^{\,0\;\mu}_A \left(s_0\right) = \left(r^{\,0}_A \left(s_0\right), \ve{r}^{\,0}_A \left(s_0\right)\right) \quad {\rm where} \quad  
r^{\,0}_A \left(s_0\right) = \left|\ve{r}^{\,0}_A \left(s_0\right)\right|,  
\label{Four_Vector_1}
\\
r^{\,1\;\mu}_A \left(s_1\right) = \left(r^{\,1}_A\left(s_1\right), \ve{r}^{\,1}_A\left(s_1\right)\right) \quad {\rm where} \quad  
r^{\,1}_A \left(s_1\right) = \left|\ve{r}^{\,1}_A \left(s_1\right)\right|,  
\label{Four_Vector_2}
\end{eqnarray}

\noindent
we obtain the following specific cases of four-scalar products
\begin{eqnarray}
k \cdot r_A^{\,0}\left(s_0\right) = - \left(r_A^{\,0}\left(s_0\right) - \ve{k} \cdot \ve{r}^{\,0}_A\left(s_0\right)\right),
\label{scalar_product_1}
\\
k \cdot r^{\,1}_A\left(s_1\right) = - \left(r^{\,1}_A\left(s_1\right) - \ve{k} \cdot \ve{r}^{\,1}_A\left(s_1\right)\right), 
\label{scalar_product_2}
\end{eqnarray}

\noindent
where the upper indices $0$ and $1$ refer to $\ve{x}_0$ and $\ve{x}_1$, respectively, as introduced in 
Eqs.~(\ref{vector_rA_0}) and (\ref{vector_rA_1}), so they are of course not Lorentz indices.  
In line with this notation we also need to introduce the four-vector  
\begin{eqnarray}
r^{\,0\;\mu}_A \left(s_1\right) = \left(r^{\,0}_A \left(s_1\right), \ve{r}^{\,0}_A \left(s_1\right)\right) \quad {\rm where} \quad
r^{\,0}_A \left(s_1\right) = \left|\ve{r}^{\,0}_A \left(s_1\right)\right|,
\label{Four_Vector_3}
\end{eqnarray}

\noindent 
and the four-scalar product
\begin{eqnarray}
k \cdot r_A^{\,0}\left(s_1\right) = - \left(r_A^{\,0}\left(s_1\right) - \ve{k} \cdot \ve{r}^{\,0}_A\left(s_1\right)\right),
\label{scalar_product_3}
\end{eqnarray}

\noindent
where the three-vector $\ve{r}^{\,0}_A\left(s_1\right)$ and its absolute value 
$r^{\,0}_A\left(s_1\right) = \left|\ve{r}^{\,0}_A\left(s_1\right)\right|$ were defined by (\ref{vector_rA_0_s1}).

\section{Transformation from $\ve{k}$ to $\ve{\sigma}$}\label{Section5}  

The most important relation in the formulation of the boundary value problem concerns the transformation from $\ve{k}$ to $\ve{\sigma}$,  
where the unit tangent vector $\ve{\sigma}$ of the light ray at past null infinity is defined by Eq.~(\ref{Introduction_6}), while the 
unit vector $\ve{k}$ is defined by Eq.~(\ref{Boundary_3}) and determines the unit direction from the light source towards the observer.  

\subsection{The implicit expression for the transformation from $\ve{k}$ to $\ve{\sigma}$}  

From (\ref{Second_Integration}) one finds the following formal expression, 
\begin{eqnarray}
\fl _{\rm N} & \hspace{-1.7cm} \biggr|&  \hspace{-1.5cm} {\ve{\sigma}} = {\ve{k}}
\nonumber\\
\nonumber\\
\fl _{\rm 1PN}& \hspace{-1.7cm}\biggr|& \hspace{-1.5cm}  
+ \frac{m_A}{R} \,
\Bigg(\ve{k} \times \bigg[ \ve{k} \times
\left(\ve{B}_1 \left(\ve{r}^{\,1}_A\left(s_1\right)\right) - \ve{B}_1 \left(\ve{r}^{\,0}_A\left(s_0\right)\right)\right) \bigg] \Bigg)
\nonumber\\
\nonumber\\
\fl _{\rm 1.5PN}& \hspace{-1.7cm}\biggr|& \hspace{-1.5cm}   
+ \frac{m_A}{R} \,
\Bigg(\ve{k} \times \bigg[ \ve{k} \times
\left(\ve{B}^A_2 \left(\ve{r}^{\,1}_A\left(s_1\right),\ve{v}_A\left(s_1\right)\right) - \ve{B}^A_2\left(\ve{r}^{\,0}_A\left(s_0\right),\ve{v}_A\left(s_1\right)\right)\right) 
\bigg] \Bigg)
\nonumber\\
\nonumber\\
\fl _{\rm 1.5PN}& \hspace{-1.7cm}\biggr|& \hspace{-1.5cm}   
+ \frac{m_A}{R} \,
\Bigg(\ve{k} \times \bigg[ \ve{k} \times
\left(\ve{B}^B_2 \left(\ve{r}^{\,1}_A\left(s_1\right),\ve{v}_A\left(s_1\right)\right) - \ve{B}^B_2\left(\ve{r}^{\,0}_A\left(s_0\right),\ve{v}_A\left(s_0\right)\right)\right)
\bigg] \Bigg)
\nonumber\\
\nonumber\\
\fl _{\rm 2PN}& \hspace{-1.7cm}\biggr|& \hspace{-1.5cm}   
+ \frac{m_A^2}{R} \,
\Bigg(\ve{k} \times \bigg[ \ve{k} \times
\left(\ve{B}_3\left(\ve{r}^{\,1}_A\left(s_1\right)\right) - \ve{B}_3\left(\ve{r}^{\,0}_A\left(s_0\right)\right)\right) \bigg] \Bigg) 
\nonumber\\
\nonumber\\
\fl _{\rm 2PN}& \hspace{-1.7cm}\biggr|& \hspace{-1.5cm}   
+ \frac{m_A^2}{R^2}\,\bigg[\ve{B}_1\left(\ve{r}^{\,1}_A\left(s_1\right)\right) - \ve{B}_1\left(\ve{r}^{\,0}_A\left(s_0\right)\right)\bigg] 
\times \bigg[\ve{k} \times \left(\ve{B}_1\left(\ve{r}^{\,1}_A\left(s_1\right)\right) - \ve{B}_1\left(\ve{r}^{\,0}_A\left(s_0\right)\right)\right)\bigg]  
\nonumber\\
\nonumber\\
\fl _{\rm 2PN}& \hspace{-1.7cm}\biggr|& \hspace{-1.5cm}  
- \frac{3}{2}\,\frac{m_A^2}{R^2}\;\ve{k}\; 
\bigg|\ve{k} \times \left(\ve{B}_1 \left(\ve{r}^{\,1}_A\left(s_1\right)\right) - \ve{B}_1 \left(\ve{r}^{\,0}_A\left(s_0\right)\right)\right)\bigg|^2  
+ \hat{\ve \epsilon}_2\left(s_1,s_0\right)  
\nonumber\\
\nonumber\\
\fl _{\rm 2.5PN}& \hspace{-1.7cm}\biggr|& \hspace{-1.5cm}  
 + {\cal O}\left(c^{-5}\right),  
\label{Transformation_k_to_sigma_5}
\end{eqnarray}
 
\noindent
where relation (\ref{appendix_E_5}) has been used in order to deduce (\ref{Transformation_k_to_sigma_5}).  
From (\ref{Transformation_k_to_sigma_5}) follows that $\ve{\sigma} \cdot \ve{\sigma} = 1 + {\cal O}\left(c^{-5}\right)$ so
that $\ve{\sigma}$ is still a unit vector up to terms beyond 2PN approximation. 
In the limit of body at rest the transformation (\ref{Transformation_k_to_sigma_5}) agrees with
Eq.~(3.2.50) in \cite{Brumberg1991} and with Eq.~(68) in \cite{Article_Zschocke1}.

The meaning of the notation in the transformation (\ref{Transformation_k_to_sigma_5}) and in each of the subsequent transformations  
is the following: 1PN terms are proportional to $m_A$, 1.5PN terms are proportional to $m_A\,v_A/c$,  
2PN terms are proportional to either $m^2_A$ or $m_A\,v_A^2/c^2$.  

The vectorial functions  
$\ve{B}_1\,,\dots\,,\ve{B}_3$ are given by Eqs.~(\ref{Vectorial_Function_C1}) - (\ref{Vectorial_Function_C3}) in the \ref{Appendix2},  
while $\hat{\ve \epsilon}_2$ is given by Eq.~(\ref{epsilon2}) in the \ref{Appendix_epsilon}. The expression for $R$ is given by  
Eq.~(\ref{Boundary_3}). Furthermore, the three-vector $\ve{r}^{\,0}_A\left(s_0\right)$ is given by (\ref{vector_rA_0}), while the  
three-vector $\ve{r}^{\,1}_A\left(s_1\right)$ is given by (\ref{vector_rA_1}).  

The vectorial functions $\ve{B}_1\,,\dots\,,\ve{B}_3$ as well as $\hat{\ve \epsilon}_2$ depend on $\ve{\sigma}$ rather than $\ve{k}$.  
Therefore, the expression (\ref{Transformation_k_to_sigma_5}) represents, as it stands, an implicit form of the transformation $\ve{k}$ 
to $\ve{\sigma}$. The explicit transformation $\ve{k}$ to $\ve{\sigma}$ is arrived within the next section.

\subsection{The explicit expression for the transformation from $\ve{k}$ to $\ve{\sigma}$}  

In the given approximation one may immediately replace $\ve{\sigma}$ by $\ve{k}$ in the 1.5PN and 2PN terms, because it would cause  
an error of the order ${\cal O}\left(c^{-5}\right)$ which is beyond 2PN approximation. That means, in the vectorial functions of the third   
until the seventh line in (\ref{Transformation_k_to_sigma_5}) one may substitute $\ve{\sigma}$ by $\ve{k}$, while in the vectorial function 
in the second line in (\ref{Transformation_k_to_sigma_5}) one needs to have the relation between $\ve{\sigma}$ and $\ve{k}$ in 1PN approximation  
as given by (\ref{appendix_E_10}), which subsequently yields Eqs.~(\ref{Relation_Impact_Vectors_1}) and (\ref{appendix_E_15}).  
Using these relations one finally arrives at the following explicit expression for the transformation from $\ve{k}$ to $\ve{\sigma}$:  
\begin{eqnarray}
\fl _{\rm N} & \hspace{-0.15cm} \biggr|& {\ve{\sigma}} = {\ve{k}}
\nonumber\\
\nonumber\\
\fl _{\rm 1PN}& \hspace{-0.5cm} _{\ve{\rho}_1}\biggr|& 
- 2\,\frac{m_A}{R}
\left(\frac{\ve{d}^k_A\left(s_1\right)}{k \cdot r^{\,1}_A\left(s_1\right)}
- \frac{\ve{d}^k_A\left(s_0\right)}{k \cdot r_A^{\,0}\left(s_0\right)}\right)
\nonumber\\
\nonumber\\
\fl _{\rm 1.5PN}& \hspace{-0.5cm} _{\ve{\rho}_2}\biggr|& 
+ 2\,\frac{m_A}{R}\,\ve{k} \times \left(\frac{\ve{v}_A\left(s_1\right)}{c} \times \ve{k}\right) 
\ln \frac{k \cdot r^{\,1}_A\left(s_1\right)}{k \cdot r_A^{\,0}\left(s_0\right)}  
\nonumber\\
\nonumber\\
\fl _{\rm 1.5PN}& \hspace{-0.5cm} _{\ve{\rho}_3}\biggr|& 
- 2\,\frac{m_A}{R}\,\ve{k} \times \left(\frac{\ve{v}_A\left(s_1\right)}{c} \times \ve{k}\right) 
+ 2\,\frac{m_A}{R}\,\ve{k} \times \left(\frac{\ve{v}_A\left(s_0\right)}{c} \times \ve{k}\right)  
\nonumber\\
\nonumber\\
\fl _{\rm 1.5PN}& \hspace{-0.5cm} _{\ve{\rho}_4}\biggr|& 
+ 2\,\frac{m_A}{R}\,\frac{\ve{k} \cdot \ve{v}_A\left(s_1\right)}{c}\,
\frac{\ve{d}^k_A\left(s_1\right)}{k \cdot r^{\,1}_A\left(s_1\right)}  
- 2\,\frac{m_A}{R}\,\frac{\ve{k} \cdot \ve{v}_A\left(s_0\right)}{c}\,
\frac{\ve{d}^k_A\left(s_0\right)}{k \cdot r_A^{\,0}\left(s_0\right)}  
\nonumber\\
\nonumber\\
\fl _{{\rm scaling}\;{\rm 2PN}}& \hspace{-0.5cm} _{\ve{\rho}_5} \biggr|& 
- 2\,\frac{m_A^2}{R^2}\ve{k}
\left|\frac{\ve{d}^k_A\left(s_1\right)}{k \cdot r^{\,1}_A\left(s_1\right)}
- \frac{\ve{d}^k_A\left(s_0\right)}{k \cdot r_A^{\,0}\left(s_0\right)}\right|^2
\nonumber\\
\nonumber\\
\fl _{{\rm enhanced}\;{\rm 2PN}}& \hspace{-0.5cm} _{\ve{\rho}_6}\biggr|& 
- 2 \frac{m_A^2}{R^2}
\left(\frac{\ve{d}^k_A\left(s_1\right)}{k \cdot r^{\,1}_A\left(s_1\right)} + \frac{\ve{d}^k_A\left(s_0\right)}{k \cdot r_A^{\,0}\left(s_0\right)}\right)
\left|\frac{\ve{d}^k_A\left(s_1\right)}{k \cdot r^{\,1}_A\left(s_1\right)}
- \frac{\ve{d}^k_A\left(s_0\right)}{k \cdot r_A^{\,0}\left(s_0\right)}\right|^2
\nonumber\\
\nonumber\\
\fl _{{\rm enhanced}\;{\rm 2PN}}& \hspace{-0.5cm} _{\ve{\rho}_7}\biggr|& 
- 4\,\frac{m_A^2}{R}\,\left(\frac{\ve{d}^k_A\left(s_1\right)}{\left(k \cdot r^{\,1}_A\left(s_1\right)\right)^2}
- \frac{\ve{d}^k_A\left(s_0\right)}{\left(k \cdot r_A^{\,0}\left(s_0\right)\right)^2}\right)
\nonumber\\
\nonumber\\
\fl _{\rm 2PN}& \hspace{-0.55cm} _{\ve{\rho}^A_8}\biggr|& 
+ \frac{15}{4} \frac{m_A^2}{R}
\frac{\ve{d}^k_A\left(s_1\right)}{\left|\ve{k} \times \ve{r}^{\,1}_A\left(s_1\right)\right|^3} \,
\left(\ve{k} \cdot \ve{r}^{\,1}_A\left(s_1\right)\right)
\left(\arctan \frac{\ve{k} \cdot \ve{r}^{\,1}_A\left(s_1\right)}{\left|\ve{k} \times \ve{r}^{\,1}_A\left(s_1\right)\right|} + \frac{\pi}{2} \right)
\nonumber\\
\fl _{\rm 2PN}& \hspace{-0.55cm} _{\ve{\rho}^B_8}\biggr|& 
- \frac{15}{4} \frac{m_A^2}{R} 
\frac{\ve{d}^k_A\left(s_0\right)}{\left|\ve{k} \times \ve{r}^{\,0}_A\left(s_0\right)\right|^3} \,
\left(\ve{k} \cdot \ve{r}^{\,0}_A\left(s_0\right)\right)
\left( \arctan \frac{\ve{k} \cdot \ve{r}^{\,0}_A\left(s_0\right)}{\left|\ve{k} \times \ve{r}^{\,0}_A\left(s_0\right)\right|} + \frac{\pi}{2}\right) 
\nonumber\\
\nonumber\\
\fl _{\rm 2PN}& \hspace{-0.5cm} _{\ve{\rho}_9}\biggr|& 
- \frac{1}{4}\frac{m_A^2}{R}
\left(\frac{\ve{d}^k_A\left(s_1\right)}{\left(r^{\,1}_A\left(s_1\right)\right)^2} 
- \frac{\ve{d}^k_A\left(s_0\right)}{\left(r^{\,0}_A\left(s_0\right)\right)^2}\right)
\nonumber\\
\nonumber\\
\fl _{\rm 2PN}& \hspace{-0.15cm} \biggr|& + \hat{\ve \epsilon}_2\left(s_1,s_0\right) 
\nonumber\\
\nonumber\\
\fl _{\rm 2.5PN}& \hspace{-0.15cm} \biggr|& 
+ {\cal O}\left(c^{-5}\right),   
\label{Transformation_k_to_sigma}
\end{eqnarray}

\noindent 
where $\ve{\rho}_i = \ve{\rho}_i\left(s_1,s_0\right)$ with $i=1\,, \cdots \,,9$ that appear before the vertical lines  
are by definition equal to the expressions on the right of the vertical bars in each line, and   
the term $\hat{\ve \epsilon}_2$ is given by Eq.~(\ref{Transformation_k_to_sigma_epsilon}) in the \ref{Appendix_epsilon}.  

The transformation (\ref{Transformation_k_to_sigma}) 
allows the determination of $\ve{\sigma}$ for the given boundary conditions $\ve{x}_0$ and $\ve{x}_1$.  
In the limit of bodies at rest the relation (\ref{Transformation_k_to_sigma}) is in agreement with the expression as given by 
Eq.~(3.2.52) in \cite{Brumberg1991} and Eq.~(74) in \cite{Article_Zschocke1}.  

The term $\ve{\rho}_5$ in (\ref{Transformation_k_to_sigma}) is proportional to vector $\ve{k}$ and originates from the
terms in the last two lines of (\ref{Transformation_k_to_sigma_5}), where the vectorial relation  
$\ve{a} \times \left(\ve{b} \times \ve{c}\right) = \ve{b} \left(\ve{a} \cdot \ve{c} \right) - \ve{c} \left(\ve{a} \cdot \ve{b} \right)$
has been used. In the transformation (\ref{Transformation_k_to_sigma}) a term proportional to vector $\ve{k}$ does not influence 
the angle $\delta\left(\ve{\sigma},\ve{k}\right)$, which can be computed from the vector product $\ve{\sigma} \times \ve{k}$.  
The only impact of that term proportional to vector $\ve{k}$ is to keep the vector $\ve{\sigma}$ to have unit length.  
Therefore, all terms proportional to vector $\ve{k}$ will be called {\it scaling terms}.  

In anticipation of subsequent considerations, the notation {\it enhanced 2PN terms} in (\ref{Transformation_k_to_sigma}) 
for the 2PN terms $\ve{\rho}_6$ and $\ve{\rho}_7$ has been introduced whose meaning is as follows. The estimation of the upper limit  
of the sum of these terms is given by Eq.~(\ref{angle_2PN_1}), which recovers that their upper limit is proportional to the 
large factor $r^{\,1}_A\left(s_1\right)/P_A$ and are, therefore, called {\it enhanced 2PN terms} in (\ref{Transformation_k_to_sigma}) 
in order to distinguish them from standard 2PN terms in (\ref{Transformation_k_to_sigma}) which do not contain such a large factor.  
Originally, {\it enhanced terms} have been recovered for the case of 2PN light propagation in the field of one monopole  
at rest \cite{Article_Zschocke1,Teyssandier,AshbyBertotti2010}. In our detailed investigation in \cite{Article_Zschocke1} 
for light propagation in the gravitational field of one monopole at rest we have demonstrated that the mathematical origin  
of {\it enhanced 2PN terms} is solely caused by iterative procedure of the integration of the geodesic equation. 
The same conclusion is valid for the case of light propagation in the gravitational field of one monopole in motion. 
That means, solving iteratively the geodesic equation in 1PN approximation 
(i.e. the first four terms on the r.h.s. in the first line of Eq.~(31) in \cite{Zschocke4} or Eq.~(45) in \cite{Zschocke1} 
where the metric in 1PN approximation is given by the first two terms in Eq.~(24) in \cite{Zschocke4}) 
then the first iteration contains terms proportional 
to $m_A$, the second iteration contains terms proportional to $m^2_A$, and so on. Using this iterative approach it is inevitable to encounter  
these so-called {\it enhanced terms} all of which contain that typical enhancing factor $r^{\,1}_A\left(s_1\right)/P_A$. 
It should also be noticed that the {\it enhanced terms} impose no limit on the distance between observer and light source,  
but impose only a constraint on the distance between observer and light-ray deflecting body. Here, we consider light deflection caused 
by Solar System bodies, where the distance between observer and massive body is limited by the near-zone of the Solar System  
as given by Eq.~(\ref{Near_Zone_2}). That fact elucidates the limitation of the post-Newtonian approach which is not 
applicable for the far-zone of the Solar System.

\subsection{The simplified expression for the transformation from $\ve{k}$ to $\ve{\sigma}$}

Two comments are in order about the transformation $\ve{k}$ to $\ve{\sigma}$ as given by Eq.~(\ref{Transformation_k_to_sigma}):

1. The transformation (\ref{Transformation_k_to_sigma}) is of rather involved structure.
In order to simplify the transformation one has to neglect all those terms whose magnitude it smaller than the envisaged accuracy of
$1\,{\rm nas}$ in light deflection.

2. The transformation (\ref{Transformation_k_to_sigma}) depends on the variables $m_A$, $\ve{x}_0$, $\ve{x}_1$,
$\ve{x}_A\left(s_0\right)$ and $\ve{x}_A\left(s_1\right)$. As mentioned, while the four-coordinates of the observer
$\left(c t_1,\ve{x}_1\right)$ are precisely known and the fundamental prerequisite of any astrometric measurement,
the four-coordinates of the light source $\left(c t_0,\ve{x}_0\right)$ are not directly accessible but follow
from data reduction of the astronomical observations. Stated differently, while the retarded instant of time $s_1$
as defined by (\ref{retarded_time_s_1}) is precisely known from the very beginning,
the retarded instant of time $s_0$ as defined by Eq.~(\ref{retarded_time_s_0}) is, first of all,
an unknown parameter in the theory of light propagation. 

In conclusion of these comments it becomes clear that practical astrometry necessitates a transformation (\ref{Transformation_k_to_sigma})  
solely in terms of $s_1$ and where all those terms are neglected which contribute less than the given goal accuracy of $1\,{\rm nas}$ in 
light deflection. Such a transformation is obtained by means of a series expansion of each individual term in 
(\ref{Transformation_k_to_sigma}) around $s_1$, which reads 
\begin{eqnarray}
\fl \hspace{1.5cm} \ve{\rho}_i\left(s_1,s_0\right) = \ve{\rho}_i\left(s_1,s_1\right) + \Delta \ve{\rho}_i\left(s_1,s_1\right) 
+ {\cal O}\left(c^{-5}\right) \quad {\rm for} \quad i = 1\,, \cdots \,, 4\,.  
\label{series_expansion_rho} 
\\
\fl \hspace{1.5cm} \ve{\rho}_i\left(s_1,s_0\right) = \ve{\rho}_i\left(s_1,s_1\right) + {\cal O}\left(c^{-5}\right)  
\quad {\rm for} \quad i = 5\,, \cdots\,, 9\,.     
\label{series_expansion_rho_2PN}
\\
\fl \hspace{1.5cm} \hat{\ve \epsilon}_2\left(s_1,s_0\right) = \hat{\ve \epsilon}_2\left(s_1,s_1\right) + {\cal O}\left(c^{-5}\right).
\label{series_expansion_epsilon_2} 
\end{eqnarray}

\noindent
In \ref{Appendix_EstimationA} the results for the upper limits are presented, while the approach is described in \ref{Appendix_Estimation1} 
and a detailed example is given in \ref{Appendix_Example}.   
The results for the upper limits are given by Eqs.~(\ref{Estimation_rho1}) and (\ref{Example_rho1_Delta}),  
Eqs.~(\ref{Term_rho_2_2}) and (\ref{Term_rho_2_1_E}), Eqs.~(\ref{Term_rho_3_3_A}) and (\ref{Term_rho_3_4}),  
Eqs.~(\ref{Estimation_rho_4_1}) and (\ref{Estimation_rho_4_2}),  
as well as Eqs.~(\ref{Term_rho_5_2}), (\ref{Term_rho_6_2}), (\ref{Term_rho_7_2}), (\ref{Term_rho_8_2}), (\ref{Term_rho_9_2}), and  
(\ref{Appendix_estimation_epsilon_3}). Numerical values for the upper limits are given in Table~\ref{Table2}.  
These results can be summarized as follows: 
\begin{eqnarray}
\fl \hspace{1.5cm} \left|\ve{\rho}_i\left(s_1,s_1\right)\right| \;\;\le 1\,{\rm nas}\;, \quad i = 3,5,8,9\,.  
\label{series_expansion_rho_result_1}
\\
\nonumber\\
\fl \hspace{1.5cm} \left|\Delta \ve{\rho}_i\left(s_1,s_1\right)\right| \le 1\,{\rm nas}\;, \quad i = 1\,,2\,,3\,,4\,.       
\label{series_expansion_rho_result_2}
\\
\nonumber\\
\fl \hspace{1.5cm} \left|\hat{\ve \epsilon}_2\left(s_1,s_1\right)\right| \;\;\; \le 1\,{\rm nas}\,.
\label{series_expansion_epsilon_2_result}
\end{eqnarray}

\noindent
Besides the fact that the absolute value of the {\it scaling term} $\ve{\rho}_5$ is less than $1\,{\rm nas}$, that term can be  
omitted anyway, because, as stated above already, it has no impact on the angle $\delta\left(\ve{\sigma},\ve{k}\right)$ between  
$\ve{\sigma}$ and $\ve{k}$.  
For the absolute value of the total sum  
of all those neglected terms (\ref{series_expansion_rho_result_1}) - (\ref{series_expansion_epsilon_2_result}) 
which are not proportional to the three-vector $\ve{k}$, we get 
\begin{eqnarray}
\fl I_1 = \left|\sum\limits_{i=3,8,9} \ve{\rho}_i\left(s_1,s_1\right) 
+ \sum\limits_{i=1,2,3,4} \Delta \ve{\rho}_i\left(s_1,s_1\right) + \hat{\ve \epsilon}_2\left(s_1,s_1\right)\right|  
\nonumber\\ 
\nonumber\\ 
\fl \hspace{0.35cm} \le\, \frac{6\,m_A}{r^{\,1}_A\left(s_1\right)}\,\frac{v_A\left(s_1\right)}{c} + \frac{15}{4}\,\pi\,\frac{m_A^2}{P_A^2} 
+ \frac{6\,m_A}{P_A}\,\frac{v_A^2\left(s_1\right)}{c^2} + \frac{6\,m_A}{r^{\,1}_A\left(s_1\right)}\,\frac{v_A^2\left(s_1\right)}{c^2}  
+ 18\,m_A\,\frac{a_A\left(s_1\right)}{c^2}\,. 
\nonumber\\ 
\label{Sum_1}
\end{eqnarray}

\noindent
For the upper limits of the terms in (\ref{Sum_1}) we have used that
\begin{eqnarray}
\fl \hspace{1.7cm} \left| \ve{\rho}_8 + \ve{\rho}_9 \right| \le \frac{15}{4}\,\pi\,\frac{m_A^2}{P_A^2}\,,
\label{Inequality_Sum_1_A}
\\
\nonumber\\
\fl \hspace{2.2cm} \left|\Delta \ve{\rho}_1 \right|
\le 6\,\frac{m_A}{r^{\,1}_A\left(s_1\right)}\,\frac{v_A\left(s_1\right)}{c}\,,
\label{Inequality_Sum_1_B}
\\
\nonumber\\
\fl \left|\Delta \ve{\rho}_2 + \Delta \ve{\rho}_3 + \Delta \ve{\rho}_4 \right|
\le 6\,\frac{m_A}{r_A^{\,1}\left(s_1\right)}\,\frac{v_A^2\left(s_1\right)}{c^2}
+ 4\,\frac{m_A}{P_A}\,\frac{v_A^2\left(s_1\right)}{c^2}
+ 8\,m_A\,\frac{a_A\left(s_1\right)}{c^2}\,,
\label{Inequality_Sum_1_C}
\end{eqnarray}

\noindent
while $\left|\ve{\rho}_3\left(s_1,s_1\right)\right| = 0$ according to Eq.~(\ref{Term_rho_3_3_A}). 
The inequality (\ref{Inequality_Sum_1_A}) is not shown explicitly, but follows by using the approach as 
described in \ref{Appendix_Estimation1}. The inequality (\ref{Inequality_Sum_1_B}) has been shown in \ref{Appendix_Example}  
and is given by Eq.~(\ref{Term_rho_10_2}).  
The inequality (\ref{Inequality_Sum_1_C}) follows from (\ref{Term_rho_2_1_E}), (\ref{Term_rho_3_4}),
and (\ref{Estimation_rho_4_2}), while the upper limit of $\left|\hat{\ve \epsilon}_2\right|$ is given by
Eq.~(\ref{Appendix_estimation_epsilon_3}).  
Using the numerical parameters as given by Table~\ref{Table1} one obtains  
\begin{eqnarray}
\fl \hspace{0.75cm} I_1 \le \; 1.0\,{\rm nas} \quad {\rm for}\;{\rm Sun}\;{\rm at}\,45^{\circ}\,\left({\rm solar}\;{\rm aspect}\;{\rm angle}\right),  
\nonumber\\
\fl \hspace{1.15cm} \le \; 1.1\,{\rm nas} \quad {\rm for}\;{\rm Jupiter}\,,
\label{Total_Sum_1}
\end{eqnarray}

\noindent
and less than $0.38\,{\rm nas}$ for any other Solar System body. 
Accordingly, the terms (\ref{series_expansion_rho_result_1}) - (\ref{series_expansion_epsilon_2_result})  can be  
neglected for sub-\muas$\,$ astrometry and even for astrometry on the level of $1.1\;{\rm nas}$ in light deflection.
In this way one obtains the simplified transformation $\ve{k}$ to $\ve{\sigma}$ fully in terms of $s_1$:  
\begin{eqnarray}
\fl _{\rm N} & \hspace{-0.15cm} \biggr|& {\ve{\sigma}} = {\ve{k}}
\nonumber\\
\nonumber\\
\fl _{\rm 1PN}& \hspace{-0.5cm} _{\ve{\rho}_1}\biggr|&
- 2\,\frac{m_A}{R}
\left(\frac{\ve{d}^k_A\left(s_1\right)}{k \cdot r^{\,1}_A\left(s_1\right)}
- \frac{\ve{d}^k_A\left(s_1\right)}{k \cdot r_A^{\,0}\left(s_1\right)}\right)
\nonumber\\
\nonumber\\
\fl _{\rm 1.5PN}& \hspace{-0.5cm} _{\ve{\rho}_2}\biggr|&
+ 2\,\frac{m_A}{R}\,\ve{k} \times \left(\frac{\ve{v}_A\left(s_1\right)}{c} \times \ve{k}\right)
\ln \frac{k \cdot r^{\,1}_A\left(s_1\right)}{k \cdot r_A^{\,0}\left(s_1\right)}
\nonumber\\
\nonumber\\
\fl _{\rm 1.5PN}& \hspace{-0.5cm} _{\ve{\rho}_4}\biggr|&
+ 2\,\frac{m_A}{R}\,\frac{\ve{k} \cdot \ve{v}_A\left(s_1\right)}{c}\,
\left(\frac{\ve{d}^k_A\left(s_1\right)}{k \cdot r^{\,1}_A\left(s_1\right)}
- \frac{\ve{d}^k_A\left(s_1\right)}{k \cdot r_A^{\,0}\left(s_1\right)}\right) 
\nonumber\\
\nonumber\\
\fl _{{\rm enhanced}\;{\rm 2PN}}& \hspace{-0.5cm} _{\ve{\rho}_6}\biggr|&
- 2 \frac{m_A^2}{R^2}
\left(\frac{\ve{d}^k_A\left(s_1\right)}{k \cdot r^{\,1}_A\left(s_1\right)} + \frac{\ve{d}^k_A\left(s_1\right)}{k \cdot r_A^{\,0}\left(s_1\right)}\right)
\left|\frac{\ve{d}^k_A\left(s_1\right)}{k \cdot r^{\,1}_A\left(s_1\right)}
- \frac{\ve{d}^k_A\left(s_1\right)}{k \cdot r_A^{\,0}\left(s_1\right)}\right|^2
\nonumber\\
\nonumber\\
\fl _{{\rm enhanced}\;{\rm 2PN}}& \hspace{-0.5cm} _{\ve{\rho}_7}\biggr|&
- 4\,\frac{m_A^2}{R}\,\left(\frac{\ve{d}^k_A\left(s_1\right)}{\left(k \cdot r^{\,1}_A\left(s_1\right)\right)^2}
- \frac{\ve{d}^k_A\left(s_1\right)}{\left(k \cdot r_A^{\,0}\left(s_1\right)\right)^2}\right)
\nonumber\\
\nonumber\\
\fl _{\rm 2.5PN}& \hspace{-0.15cm} \biggr|&
+ {\cal O}\left(c^{-5}\right),   
\label{Simplified_Transformation_k_to_sigma}
\end{eqnarray}

\noindent
where $\ve{\rho}_i = \ve{\rho}_i\left(s_1,s_1\right)$ with $i=1,2,4,6,7$ that appear before the vertical lines
are by definition equal to the expressions on the right of the vertical bars in each line.  
In the limit of monopole at rest this expression coincides with Eqs.~(79) - (80) in \cite{Article_Zschocke1}. 
Using the approach and the results in the appendix one obtains for the upper limits of the 1PN, 1.5PN, and 2PN terms in 
the simplified transformation (\ref{Simplified_Transformation_k_to_sigma}):  
\begin{eqnarray}
\fl {\rm 1PN} \hspace{2.75cm} 
\left|\ve{\rho}_1 \right| \le 4\,\frac{m_A}{P_A}\,,  
\label{angle_1PN_1}
\\
\nonumber\\
\fl {\rm 1.5PN} \hspace{1.75cm} 
\left|\ve{\rho}_2 + \ve{\rho}_4\right| \le 6\,\frac{m_A}{P_A}\,\frac{v_A\left(s_1\right)}{c}\,, 
\label{angle_15PN_1}
\\
\nonumber\\
\fl {\rm enhanced}\;{\rm 2PN} \hspace{0.5cm}  \left|\ve{\rho}_6 + \ve{\rho}_7 \right| \le 16\,  
\frac{m^2_A}{P^2_A}\,\frac{r^{\,1}_A\left(s_1\right)}{P_A}\,. 
\label{angle_2PN_1}
\end{eqnarray} 

\noindent 
The reason of why there is a factor $6$ in (\ref{angle_15PN_1}) rather than a factor $4$ is discussed in the text below Eq.~(\ref{angle_2PN_3}).  
In the limit of body at rest, the results (\ref{angle_1PN_1}) and (\ref{angle_2PN_1}) would coincide with Eqs.~(76) and (77) 
in \cite{Article_Zschocke1}, respectively. 

The simplified transformation (\ref{Simplified_Transformation_k_to_sigma}) depends on the variables
$m_A$, $\ve{x}_0$, $\ve{x}_1$, and $\ve{x}_A\left(s_1\right)$ and does not any longer depend on the retarded time $s_0$.
The only unknown in (\ref{Simplified_Transformation_k_to_sigma}) is the three-coordinate of the light source, $\ve{x}_0$,  
whose determination is the primary aim of data reduction of astronomical observations. 
For the neglected terms of the order ${\cal O}\left(c^{-5}\right)$ (2.5PN approximation) and of the order ${\cal O}\left(c^{-6}\right)$ 
(3PN approximation) we refer to Section \ref{Section_3PN}, where some statements about their impact on light deflection are given.

\section{Transformation from $\ve{\sigma}$ to $\ve{n}$}\label{Section6} 

Now we consider the transformation from $\ve{\sigma}$ to $\ve{n}$, where $\ve{\sigma}$ is the unit tangent vector 
along the light trajectory at past null infinity as defined by (\ref{Introduction_6}), while $\ve{n}$ is the unit tangent vector along the 
light trajectory at the observer's position as defined by Eq.~(\ref{Tangent_Vector1}).  

\subsection{The implicit expression for the transformation from $\ve{\sigma}$ to $\ve{n}$}  

By inserting (\ref{First_Integration}) into (\ref{Tangent_Vector1}) one obtains    
\begin{eqnarray}
\fl _{\rm N} & \hspace{-1.7cm} \biggr|&  \hspace{-1.5cm} {\ve{n}} = {\ve{\sigma}}
\nonumber\\
\nonumber\\
\fl _{\rm 1PN} & \hspace{-1.7cm} \biggr|&  \hspace{-1.5cm} 
+ m_A\,\ve{\sigma} \times \left(\ve{A}_1\left(\ve{r}^{\,1}_A\left(s_1\right)\right) \times \ve{\sigma}\right)  
\nonumber\\ 
\nonumber\\ 
\fl _{\rm 1.5PN} & \hspace{-1.7cm} \biggr|&  \hspace{-1.5cm} 
+ m_A\,
\ve{\sigma} \times \left(\ve{A}_2\left(\ve{r}^{\,1}_A\left(s_1\right), \ve{v}_A\left(s_1\right)\right) \times \ve{\sigma}\right) 
\nonumber\\ 
\nonumber\\ 
\fl _{\rm 2PN} & \hspace{-1.7cm} \biggr|&  \hspace{-1.5cm} 
+ m_A^2\,\ve{\sigma} \times \left(\ve{A}_3\left(\ve{r}^{\,1}_A\left(s_1\right)\right) \times \ve{\sigma}\right)
- m_A^2\, \ve{A}_1\left(\ve{r}^{\,1}_A\left(s_1\right)\right) \left(\ve{\sigma} \cdot \ve{A}_1\left(\ve{r}^{\,1}_A\left(s_1\right)\right)\right)
\nonumber\\ 
\nonumber\\ 
\fl _{\rm 2PN} & \hspace{-1.7cm} \biggr|&  \hspace{-1.5cm} 
- \frac{1}{2}\,m_A^2\,\ve{\sigma} 
\left(\ve{A}_1\left(\ve{r}^{\,1}_A\left(s_1\right)\right) \cdot \ve{A}_1\left(\ve{r}^{\,1}_A\left(s_1\right)\right)\right)
+ \frac{3}{2}\,m_A^2\,\ve{\sigma} \left(\ve{\sigma} \cdot \ve{A}_1\left(\ve{r}^{\,1}_A\left(s_1\right)\right)\right)^2 
+ \hat{\ve \epsilon}_1\left(s_1\right)  
\nonumber\\ 
\nonumber\\ 
\fl _{\rm 2.5PN} & \hspace{-1.7cm} \biggr|&  \hspace{-1.5cm}
+ {\cal O}\left(c^{-5}\right),   
\label{Transformation_sigma_to_n_5}  
\end{eqnarray}

\noindent 
where the vectorial functions $\ve{A}_1$, $\ve{A}_2$, $\ve{A}_3$ are given by Eqs.~(\ref{Vectorial_Function_A1}) - (\ref{Vectorial_Function_A3}) 
in the \ref{Appendix2}, while expression $\hat{\ve \epsilon}_1$ has been given by Eq.~(\ref{epsilon1}) in the \ref{Appendix_epsilon}.  
From (\ref{Transformation_sigma_to_n_5}) follows that $\ve{n} \cdot \ve{n} = 1 + {\cal O}\left(c^{-5}\right)$ so that $\ve{n}$ is still a  
unit vector up to terms beyond 2PN approximation. The vectorial functions $\ve{A}_1\,,\,\dots\,,\ve{A}_3$ as well as $\hat{\ve \epsilon}_1$  
depend on vector $\ve{\sigma}$. However, the aim is to achieve the transformation from $\ve{\sigma}$ to $\ve{n}$ in  
terms of the boundary values (\ref{Boundary_1}), which implies to express these vectorial functions in terms of  
$\ve{k}$ rather than $\ve{\sigma}$. This will be done in the next section.

\subsection{The explicit expression for the transformation from $\ve{\sigma}$ to $\ve{n}$}  

In the 1.5PN and 2PN terms one may immediately replace $\ve{\sigma}$ by $\ve{k}$, because such a replacement would cause 
an error of the order ${\cal O}\left(c^{-5}\right)$ which is beyond 2PN approximation. 
That means in the vectorial functions of the third until the fifth line in (\ref{Transformation_sigma_to_n_5}) one may 
substitute $\ve{\sigma}$ by $\ve{k}$, while in the vectorial function in the second line in (\ref{Transformation_sigma_to_n_5}) 
one has to use relation (\ref{appendix_E_10}) and Eqs.~(\ref{Relation_Impact_Vectors_1}) and (\ref{appendix_E_15}).  
Using these relations one finally arrives at the following explicit expression for the transformation from $\ve{\sigma}$ to $\ve{n}$:
\begin{eqnarray}
\fl _{\rm N} & \hspace{-0.1cm} \biggr|& \hspace{+0.05cm} {\ve{n}} = {\ve{\sigma}}
\nonumber\\
\nonumber\\
\fl _{\rm 1PN}& \hspace{-0.5cm} _{\ve{\varphi}_1}\biggr|& 
+ 2\,m_A\,\frac{\ve{d}_A^k\left(s_1\right)}{r^{\,1}_A\left(s_1\right)}\,\frac{1}{k \cdot r^{\,1}_A\left(s_1\right)}\,   
\nonumber\\
\nonumber\\
\fl _{{\rm scaling}\;{\rm 1.5PN}}& \hspace{-0.5cm} _{\ve{\varphi}_2} \biggr|& 
- 4\,\frac{m_A}{r^{\,1}_A\left(s_1\right)}\,\ve{k}\,\frac{\ve{k} \cdot \ve{v}_A\left(s_1\right)}{c}
\nonumber\\
\nonumber\\
\fl _{\rm 1.5PN} & \hspace{-0.5cm} _{\ve{\varphi}_3} \biggr|& 
- 2\,\frac{m_A}{r^{\,1}_A\left(s_1\right)}\,\frac{\ve{d}^k_A\left(s_1\right)}{k \cdot r^{\,1}_A\left(s_1\right)}\,
\frac{\ve{k}\cdot \ve{v}_A\left(s_1\right)}{c}  
\nonumber\\
\nonumber\\
\fl _{\rm 1.5PN} & \hspace{-0.5cm} _{\ve{\varphi}_4} \biggr|& 
+ 4\,\frac{m_A}{r^{\,1}_A\left(s_1\right)}\,\frac{\ve{v}_A\left(s_1\right)}{c} 
+ \frac{2\,m_A}{\left(r^{\,1}_A\left(s_1\right)\right)^2}\,\ve{d}^k_A\left(s_1\right)\,\frac{\ve{k} \cdot \ve{v}_A\left(s_1\right)}{c} 
\nonumber\\
\nonumber\\
\fl _{\rm 1.5PN} & \hspace{-0.5cm} _{\ve{\varphi}_5} \biggr|&   
+ \frac{2\,m_A}{\left(r^{\,1}_A\left(s_1\right)\right)^2}\,\frac{\ve{d}^k_A\left(s_1\right)}{k \cdot r^{\,1}_A\left(s_1\right)}\,
\frac{\ve{d}^k_A\left(s_1\right) \cdot \ve{v}_A\left(s_1\right)}{c}
\nonumber\\
\nonumber\\
\fl _{{\rm scaling}\;{\rm 2PN}} & \hspace{-0.5cm} _{\ve{\varphi}_6} \biggr|& - 2\,\ve{k}\,\frac{m_A^2}{\left(r^{\,1}_A\left(s_1\right)\right)^2}\,
\frac{\ve{d}^k_A\left(s_1\right) \cdot \ve{d}^k_A\left(s_1\right)}{\left(k \cdot r^{\,1}_A\left(s_1\right)\right)^2}  
\nonumber\\
\nonumber\\
\fl _{{\rm scaling}\;{\rm 2PN}} & \hspace{-0.5cm} _{\ve{\varphi}_7} \biggr|& 
+ 4\,\frac{\ve{k}}{R}\,\frac{m_A^2}{r^{\,1}_A\left(s_1\right)}\,\frac{1}{k \cdot r^{\,1}_A\left(s_1\right)}
\left(\frac{\ve{d}^k_A\left(s_1\right) \cdot \ve{d}^k_A\left(s_1\right)}{k \cdot r^{\,1}_A\left(s_1\right)}
- \frac{\ve{d}^k_A\left(s_0\right) \cdot \ve{d}^k_A\left(s_1\right)}{k \cdot r_A^{\,0}\left(s_0\right)}\right)
\nonumber\\
\nonumber\\
\fl _{{\rm enhanced}\;{\rm 2PN}} & \hspace{-0.5cm} _{\ve{\varphi}_8} \biggr|& 
+ 4\,\frac{m_A^2}{r^{\,1}_A\left(s_1\right)}\,\frac{\ve{d}_A^k\left(s_1\right)}{\left(k \cdot r^{\,1}_A\left(s_1\right)\right)^2} 
\nonumber\\
\nonumber\\
\fl _{{\rm enhanced}\;{\rm 2PN}} & \hspace{-0.5cm} _{\ve{\varphi}_9} \biggr|& 
+ 4\,\frac{m_A^2}{r^{\,1}_A\left(s_1\right)}\,\frac{1}{R}\,
\frac{\ve{d}_A^k\left(s_1\right)}{\left(k \cdot r^{\,1}_A\left(s_1\right)\right)^2}
\left(\frac{\ve{d}^k_A\left(s_1\right) \cdot \ve{d}^k_A\left(s_1\right)}{k \cdot r^{\,1}_A\left(s_1\right)}
- \frac{\ve{d}^k_A\left(s_0\right) \cdot \ve{d}^k_A\left(s_1\right)}{k \cdot r_A^{\,0}\left(s_0\right)} \right)
\nonumber\\
\nonumber\\
\fl _{{\rm enhanced}\;{\rm 2PN}} & \hspace{-0.55cm} _{\ve{\varphi}_{10}} \biggr|&  
+ 4\,\frac{m_A^2}{r^{\,1}_A\left(s_1\right)}\,\frac{1}{R}\,
\frac{\ve{k} \cdot \ve{r}^{\,1}_A\left(s_1\right)}{k \cdot r^{\,1}_A\left(s_1\right)}
\left(\frac{\ve{d}^k_A\left(s_1\right)}{k \cdot r^{\,1}_A\left(s_1\right)}
- \frac{\ve{d}^k_A\left(s_0\right)}{k \cdot r_A^{\,0}\left(s_0\right)} \right)
\nonumber\\
\nonumber\\
\fl _{\rm 2PN} & \hspace{-0.55cm} _{\ve{\varphi}_{11}} \biggr|& 
- 4\,\frac{m_A^2}{\left(r_A^{\,1}\left(s_1\right)\right)^2}\,\frac{\ve{d}_A^k\left(s_1\right)}{k \cdot r^{\,1}_A\left(s_1\right)} 
\nonumber\\
\nonumber\\
\fl _{\rm 2PN} & \hspace{-0.55cm} _{\ve{\varphi}_{12}} \biggr|& 
- \frac{m_A^2}{2}\,\ve{d}_A^k\left(s_1\right)\frac{\ve{k} \cdot \ve{r}^{\,1}_A\left(s_1\right)}{\left(r^{\,1}_A\left(s_1\right)\right)^4}
- \frac{15}{4}\,\frac{m_A^2}{\left(r_A^{\,1}\left(s_1\right)\right)^2}\,\ve{d}^k_A\left(s_1\right)
\frac{\ve{k} \cdot \ve{r}^{\,1}_A\left(s_1\right)}{\left|\ve{k} \times \ve{r}^{\,1}_A\left(s_1\right)\right|^2}
\nonumber\\
\nonumber\\
\fl _{\rm 2PN} & \hspace{-0.55cm} _{\ve{\varphi}_{13}} \biggr|& - \frac{15}{4}\,m_A^2\,
\frac{\ve{d}^k_A\left(s_1\right)}{\left|\ve{k} \times \ve{r}^{\,1}_A\left(s_1\right)\right|^3}
\left(\arctan \frac{\ve{k} \cdot \ve{r}^{\,1}_A\left(s_1\right)}{\left|\ve{k} \times \ve{r}^{\,1}_A\left(s_1\right)\right|} + \frac{\pi}{2}\right)  
\nonumber\\
\nonumber\\
\fl _{\rm 2PN} & \hspace{-0.025cm} \biggr|& + \hat{\ve \epsilon}_1\left(s_1\right) 
\nonumber\\
\nonumber\\
\fl _{\rm 2.5PN} & \hspace{-0.025cm} \biggr|& + {\cal O}\left(c^{-5}\right),   
\label{Transformation_sigma_to_n}
\end{eqnarray}

\noindent
where $\ve{\varphi}_i = \ve{\varphi}_i\left(s_1\right)$ for $i=1\,, \cdots \,,6,8,11,12,13$ and 
$\ve{\varphi}_i = \ve{\varphi}_i\left(s_1,s_0\right)$ for $i = 7,9,10$ that appear before the vertical lines    
are by definition equal to the expressions on the right of the vertical bars in each line,  
while the term $\hat{\ve \epsilon}_1$ is given by Eq.~(\ref{Transformation_sigma_to_n_epsilon}) in the \ref{Appendix_epsilon}.    
With the aid of the transformation (\ref{Transformation_sigma_to_n}) one may determine the difference between the vectors $\ve{n}$ and 
$\ve{\sigma}$ from the given boundary conditions $\ve{x}_0$ and $\ve{x}_1$.  
In the limit of bodies at rest the relation (\ref{Transformation_sigma_to_n}) is in agreement with the expression as given by
Eq.~(81) in \cite{Article_Zschocke1}.  

The 1.5PN {\it scaling term} $\ve{\varphi}_2$ in the third line of (\ref{Transformation_sigma_to_n}) originates from the
term in the third line of (\ref{Transformation_sigma_to_n_5}), where the vectorial relation
$\ve{a} \times \left(\ve{b} \times \ve{c}\right) = \ve{b} \left(\ve{a} \cdot \ve{c} \right) - \ve{c} \left(\ve{a} \cdot \ve{b} \right)$
has been used. The 2PN {\it scaling term} $\ve{\varphi}_6$ in the seventh line of (\ref{Transformation_sigma_to_n}) originates from the 
first term of the fifth line of (\ref{Transformation_sigma_to_n_5}). 
The 2PN {\it scaling term} $\ve{\varphi}_7$ in the eighth line of (\ref{Transformation_sigma_to_n}) originates from the 
term in the second line of (\ref{Transformation_sigma_to_n_5}), where relation (\ref{Relation_Impact_Vectors_1}) has to be used.  
In the transformation (\ref{Transformation_sigma_to_n}) a term proportional to vector $\ve{k}$ influences the 
angle $\delta\left(\ve{\sigma},\ve{n}\right)$ between $\ve{\sigma}$ and $\ve{n}$ only beyond 2PN approximation,  
due to $\ve{\sigma} \times \ve{k} = {\cal O}\left(c^{-2}\right)$.  
Hence, the only impact of these {\it scaling terms}, $\ve{\varphi}_2$, $\ve{\varphi}_6$, $\ve{\varphi}_7$, 
is to keep the vector $\ve{n}$ to have unit length.

\subsection{Simplified expression for the transformation from $\ve{\sigma}$ to $\ve{n}$}

The transformation $\ve{\sigma}$ to $\ve{n}$ as given by Eq.~(\ref{Transformation_sigma_to_n}) contains many terms which 
contribute less than $1\,{\rm nas}$ in light deflection. Furthermore,  
as discussed above in the text before Eq.~(\ref{series_expansion_body_1}) and in the second comment before Eq.~(\ref{series_expansion_rho}), 
while the four-coordinates of the observer $\left(c t_1,\ve{x}_1\right)$ are precisely known and the fundamental basis for any accurate astrometric 
measurement, the four-coordinates of the light source $\left(c t_0,\ve{x}_0\right)$ are, first of all, not available but the  
result of astrometric data reduction. That means, the retarded instant of time $s_1$ defined by (\ref{retarded_time_s_1})
is precisely known, while the retarded time $s_0$ defined by (\ref{retarded_time_s_0}) is, first of all, an unknown parameter
in the theory of light propagation.  
Therefore, practical astrometry necessitates a transformation $\ve{\sigma}$ to $\ve{n}$ as function of $s_1$ and which contains only 
those terms which are above the goal accuracy of $1\,{\rm nas}$.
In (\ref{Transformation_sigma_to_n}) the terms depend only on $s_1$, except of $\ve{\varphi}_7$, $\ve{\varphi}_9$, and $\ve{\varphi}_{10}$.  
Hence, we consider only the series expansion of these three terms, which reads  
\begin{eqnarray}
\fl \hspace{1.5cm} \ve{\varphi}_i\left(s_1,s_0\right) = \ve{\varphi}_i\left(s_1,s_1\right)  
+ {\cal O}\left(c^{-5}\right) \quad {\rm for} \quad i = 7,9,10\,.
\label{series_expansion_varphi}
\end{eqnarray}

\noindent
The upper limit of each individual term in the transformation (\ref{Transformation_sigma_to_n}) has been
determined, given by (\ref{Appendix_Estimation_phi_1}), (\ref{Appendix_Estimation_phi_2}),
(\ref{Appendix_Estimation_phi_3}), (\ref{Appendix_Estimation_phi_4}), (\ref{Appendix_Estimation_phi_5}), (\ref{Appendix_Estimation_phi_6}),
(\ref{Appendix_Estimation_phi_7_2}), (\ref{Appendix_Estimation_phi_8_2}), (\ref{Appendix_Estimation_phi_9_2}), (\ref{Appendix_Estimation_phi_10_2}),
(\ref{Appendix_Estimation_phi_11_2}), (\ref{Appendix_Estimation_phi_12_2}), (\ref{Appendix_Estimation_phi_13_2}),
and (\ref{Appendix_estimation_epsilon_1}). Numerical values are given in Table~\ref{Table3}.  
These results can be summarized as follows:
\begin{eqnarray}
\fl \hspace{1.5cm} \left|\ve{\varphi}_i\left(s_1\right)\right| \;\;\le 1\,{\rm nas}\;, \quad i = 2,4,5,6,11,12,13\,.
\label{series_expansion_varphi_result_1}
\\
\nonumber\\
\fl \hspace{1.0cm} \left|\ve{\varphi}_i\left(s_1,s_1\right)\right| \;\;\le 1.3\,{\rm nas}\;, \quad i = 7\,.
\label{series_expansion_varphi_result_2}
\\
\nonumber\\
\fl \hspace{1.5cm} \left|\hat{\ve \epsilon}_1\left(s_1\right)\right| \;\;\; \le 1\,{\rm nas}\,.
\label{series_expansion_epsilon_1_result}
\end{eqnarray}

\noindent
Besides the fact that the absolute value of the {\it scaling terms} $\ve{\varphi}_2$, $\ve{\varphi}_6$, $\ve{\varphi}_7$ 
is less than $1.3\,{\rm nas}$, these terms are irrelevant for the angle $\delta\left(\ve{\sigma}, \ve{n}\right)$ between 
the vectors $\ve{\sigma}$ and \ve{n}, as stated above. Therefore, these scaling terms will be  
omitted in the simplified transformation. For the absolute value of the total sum of all those neglected terms  
(\ref{series_expansion_varphi_result_1}) - (\ref{series_expansion_epsilon_1_result}) which are not proportional to the three-vector $\ve{k}$, 
we get 
\begin{eqnarray}
\fl \hspace{0.75cm} I_2 = \left|\sum\limits_{i=4,5} \ve{\varphi}_i\left(s_1\right) 
+ \sum\limits_{i=11,12,13} \ve{\varphi}_i\left(s_1,s_1\right) + \hat{\ve \epsilon}_1\left(s_1\right)\right|  
\nonumber\\
\nonumber\\
\fl \hspace{1.15cm} \le\; 8\,\frac{m_A}{r^{\,1}_A\left(s_1\right)}\,\frac{v_A\left(s_1\right)}{c} + \frac{15}{4}\,\pi\,\frac{m_A^2}{P_A^2}
+ 18\,\frac{m_A}{P_A}\,\frac{v_A^2\left(s_1\right)}{c^2} + 8\,\frac{m_A}{r^{\,1}_A}\,\frac{v_A^2\left(s_1\right)}{c^2}\,.  
\label{Sum_2}
\end{eqnarray}

\noindent 
For the upper limits of the terms in (\ref{Sum_2}) we have used that
\begin{eqnarray}
\fl \hspace{3.0cm} \left| \ve{\varphi}_4 + \ve{\varphi}_5 \right| \le 8\,\frac{m_A}{r_A^{\,1}\left(s_1\right)}\,\frac{v_A\left(s_1\right)}{c}\,,  
\label{Inequality_Sum_2_A}
\\
\nonumber\\
\fl \hspace{1.75cm} \left|\ve{\varphi}_{11} + \ve{\varphi}_{12} + \ve{\varphi}_{13} \right| \le \frac{15}{4}\,\pi\,\frac{m_A^2}{P_A^2}\,,  
\label{Inequality_Sum_2_B}
\end{eqnarray}

\noindent
while the upper limit of $\left|\hat{\ve \epsilon}_1\right|$ is given by Eq.~(\ref{Appendix_estimation_epsilon_1}). 
These inequalities, Eqs.~(\ref{Inequality_Sum_2_A}) and (\ref{Inequality_Sum_2_B}), are not shown explicitly, but 
can be demonstrated with the aid of the approach as described in \ref{Appendix_Estimation1}.  
Using the numerical parameters as given by Table~\ref{Table1} one obtains for the absolute value of the total sum  
\begin{eqnarray}
\fl \hspace{0.75cm} I_2 \le \; 1.1\,{\rm nas} \quad {\rm for}\;{\rm Sun}\;{\rm at}\,45^{\circ}\,\left({\rm solar}\;{\rm aspect}\;{\rm angle}\right),
\nonumber\\
\fl \hspace{1.15cm} \le \; 1.2\,{\rm nas} \quad {\rm for}\;{\rm Jupiter}\,,
\label{Total_Sum_2}
\end{eqnarray}

\noindent
and less than $0.5\,{\rm nas}$ for any other Solar System body.
Accordingly, the terms (\ref{series_expansion_varphi_result_1}) - (\ref{series_expansion_epsilon_1_result}) can be  
neglected for sub-\muas$\,$ astrometry and even for astrometry on the level of $1.2\;{\rm nas}$ in light deflection.
In this way one obtains the simplified transformation $\ve{\sigma}$ to $\ve{n}$ fully in terms of $s_1$, which reads  
\begin{eqnarray}
\fl _{\rm N} & \hspace{-0.1cm} \biggr|& \hspace{+0.05cm} {\ve{n}} = {\ve{\sigma}}
\nonumber\\
\nonumber\\
\fl _{\rm 1PN}& \hspace{-0.5cm} _{\ve{\varphi}_1}\biggr|&
+ 2\,m_A\,\frac{\ve{d}_A^k\left(s_1\right)}{r^{\,1}_A\left(s_1\right)}\,\frac{1}{k \cdot r^{\,1}_A\left(s_1\right)}\,
\nonumber\\
\nonumber\\
\fl _{\rm 1.5PN} & \hspace{-0.5cm} _{\ve{\varphi}_3} \biggr|&
- 2\,\frac{m_A}{r^{\,1}_A\left(s_1\right)}\,\frac{\ve{d}^k_A\left(s_1\right)}{k \cdot r^{\,1}_A\left(s_1\right)}\,
\frac{\ve{k}\cdot \ve{v}_A\left(s_1\right)}{c}
\nonumber\\
\nonumber\\
\fl _{{\rm enhanced}\;{\rm 2PN}} & \hspace{-0.5cm} _{\ve{\varphi}_8} \biggr|&
+ 4\,\frac{m_A^2}{r^{\,1}_A\left(s_1\right)}\,\frac{\ve{d}_A^k\left(s_1\right)}{\left(k \cdot r^{\,1}_A\left(s_1\right)\right)^2}
\nonumber\\
\nonumber\\
\fl _{{\rm enhanced}\;{\rm 2PN}} & \hspace{-0.5cm} _{\ve{\varphi}_9} \biggr|&
+ 4\,\frac{m_A^2}{r^{\,1}_A\left(s_1\right)}\,\frac{1}{R}\,
\frac{\ve{d}_A^k\left(s_1\right)}{\left(k \cdot r^{\,1}_A\left(s_1\right)\right)^2}
\left(\frac{\ve{d}^k_A\left(s_1\right) \cdot \ve{d}^k_A\left(s_1\right)}{k \cdot r^{\,1}_A\left(s_1\right)}
- \frac{\ve{d}^k_A\left(s_1\right) \cdot \ve{d}^k_A\left(s_1\right)}{k \cdot r_A^{\,0}\left(s_1\right)} \right)
\nonumber\\
\nonumber\\
\fl _{{\rm enhanced}\;{\rm 2PN}} & \hspace{-0.55cm} _{\ve{\varphi}_{10}} \biggr|&
+ 4\,\frac{m_A^2}{r^{\,1}_A\left(s_1\right)}\,\frac{1}{R}\,
\frac{\ve{k} \cdot \ve{r}^{\,1}_A\left(s_1\right)}{k \cdot r^{\,1}_A\left(s_1\right)}
\left(\frac{\ve{d}^k_A\left(s_1\right)}{k \cdot r^{\,1}_A\left(s_1\right)}
- \frac{\ve{d}^k_A\left(s_1\right)}{k \cdot r_A^{\,0}\left(s_1\right)} \right)
\nonumber\\
\nonumber\\
\fl _{\rm 2.5PN} & \hspace{-0.025cm} \biggr|& + {\cal O}\left(c^{-5}\right),  
\label{Simplified_Transformation_sigma_to_n}
\end{eqnarray}

\noindent
where $\ve{\varphi}_i = \ve{\varphi}_i\left(s_1\right)$ for $i=1,3,8$ and $\ve{\varphi}_i = \ve{\varphi}_i\left(s_1,s_1\right)$ for $i=9,10$  
which appear before the vertical lines are by definition equal to the expressions on the right of the vertical bars in each line.  
In the limit of monopole at rest this expression coincides with Eqs.~(85) - (86) in \cite{Article_Zschocke1}.
By means of the approach and using the results in the appendix one obtains for the upper limits of the 1PN, 1.5PN, and 2PN terms 
in the simplified transformation (\ref{Simplified_Transformation_sigma_to_n}): 
\begin{eqnarray}
\fl {\rm 1PN} \hspace{3.9cm}
\left|\ve{\varphi}_1 \right| \le 4\,\frac{m_A}{P_A}\,, 
\label{angle_1PN_2}
\\
\nonumber\\
\fl {\rm 1.5PN} \hspace{3.65cm}
\left|\ve{\varphi}_3 \right| \le 4\,\frac{m_A}{P_A}\,\frac{v_A\left(s_1\right)}{c}\,,  
\label{angle_15PN_2}
\\
\nonumber\\
\fl {\rm enhanced}\;{\rm 2PN} \hspace{0.5cm}  \left|\ve{\varphi}_8 + \ve{\varphi}_9 + \ve{\varphi}_{10}\right| \le 16\,
\frac{m^2_A}{P^2_A}\,\frac{r^{\,1}_A\left(s_1\right)}{P_A}\,. 
\label{angle_2PN_2}
\end{eqnarray}

\noindent
In the limit of body at rest, the results (\ref{angle_1PN_2}) and (\ref{angle_2PN_2}) would coincide with Eqs.~(82) and (83)  
in \cite{Article_Zschocke1}, respectively. 
The simplified transformation (\ref{Simplified_Transformation_sigma_to_n}) depends on the variables
$m_A$, $\ve{x}_0$, $\ve{x}_1$, and $\ve{x}_A\left(s_1\right)$, but not anymore on the retarded time $s_0$.  
Thus, the only unknown in (\ref{Simplified_Transformation_sigma_to_n}) is the three-coordinate of the light source, $\ve{x}_0$,
whose determination is the fundamental aim of astrometric data reduction.  
Some statement about the neglected terms of the order ${\cal O}\left(c^{-5}\right)$ (2.5PN approximation) and of the 
order ${\cal O}\left(c^{-6}\right)$ (3PN approximation) are given in Section \ref{Section_3PN}.

\section{Transformation from $\ve{k}$ to $\ve{n}$}\label{Section7} 

\subsection{The explicit expression for the transformation from $\ve{k}$ to $\ve{n}$} 

The actual aim of the boundary value problem is to establish a relation between the unit-vectors $\ve{k}$  
and $\ve{n}$. From the transformations (\ref{Transformation_k_to_sigma}) and (\ref{Transformation_sigma_to_n}) 
we immediately obtain the transformation $\ve{k}$ to $\ve{n}$:   
\begin{eqnarray}
\fl _{\rm N} & \hspace{0.1cm} \biggr|& \hspace{0.15cm} {\ve{n}} = {\ve{k}}
\nonumber\\
\nonumber\\
\fl _{\rm 1PN}& \hspace{-0.95cm} _{\ve{\rho}_1} + _{\ve{\varphi}_1} \biggr|& \hspace{0.0cm}
- 2\,\frac{m_A}{R}
\left(\frac{\ve{d}^k_A\left(s_1\right)}{k \cdot r^{\,1}_A\left(s_1\right)}
- \frac{\ve{d}^k_A\left(s_0\right)}{k \cdot r_A^{\,0}\left(s_0\right)}\right)
+ 2\,m_A\,\frac{\ve{d}_A^k\left(s_1\right)}{r^{\,1}_A\left(s_1\right)}\,\frac{1}{k \cdot r^{\,1}_A\left(s_1\right)}\,
\nonumber\\
\nonumber\\
\fl _{\rm 1.5PN} & \hspace{-0.3cm} _{\ve{\rho}_2} \biggr|& \hspace{0.0cm}
+ 2\,\frac{m_A}{R}\,\ve{k} \times \left(\frac{\ve{v}_A\left(s_1\right)}{c} \times \ve{k}\right)
\ln \frac{k \cdot r^{\,1}_A\left(s_1\right)}{k \cdot r_A^{\,0}\left(s_0\right)}
\nonumber\\
\nonumber\\
\fl _{\rm 1.5PN} & \hspace{-0.3cm} _{\ve{\rho}_3} \biggr|& \hspace{0.0cm}
- 2\,\frac{m_A}{R}\,\ve{k} \times \left(\frac{\ve{v}_A\left(s_1\right)}{c} \times \ve{k}\right)
+ 2\,\frac{m_A}{R}\,\ve{k} \times \left(\frac{\ve{v}_A\left(s_0\right)}{c} \times \ve{k}\right)
\nonumber\\
\nonumber\\
\fl _{\rm 1.5PN} & \hspace{-0.3cm} _{\ve{\rho}_4} \biggr|& \hspace{0.0cm}
+ 2\,\frac{m_A}{R}\,\frac{\ve{k} \cdot \ve{v}_A\left(s_1\right)}{c}\,
\frac{\ve{d}^k_A\left(s_1\right)}{k \cdot r^{\,1}_A\left(s_1\right)}
- 2\,\frac{m_A}{R}\,\frac{\ve{k} \cdot \ve{v}_A\left(s_0\right)}{c}\,
\frac{\ve{d}^k_A\left(s_0\right)}{k \cdot r_A^{\,0}\left(s_0\right)}
\nonumber\\
\nonumber\\
\fl _{\rm 1.5PN} & \hspace{-0.3cm} _{\ve{\varphi}_3} \biggr|& \hspace{0.0cm}
- 2\,\frac{m_A}{r^{\,1}_A\left(s_1\right)}\,\frac{\ve{d}^k_A\left(s_1\right)}{k \cdot r^{\,1}_A\left(s_1\right)}\,
\frac{\ve{k}\cdot \ve{v}_A\left(s_1\right)}{c}
\nonumber\\
\nonumber\\
\fl _{\rm 1.5PN} & \hspace{-0.3cm} _{\ve{\varphi}_4} \biggr|& \hspace{0.0cm}
+ 4\,\frac{m_A}{r^{\,1}_A\left(s_1\right)}\,\frac{\ve{v}_A\left(s_1\right)}{c}
+ \frac{2\,m_A}{\left(r^{\,1}_A\left(s_1\right)\right)^2}\,\ve{d}^k_A\left(s_1\right)\,\frac{\ve{k} \cdot \ve{v}_A\left(s_1\right)}{c}
\nonumber\\
\nonumber\\
\fl _{\rm 1.5PN} & \hspace{-0.3cm} _{\ve{\varphi}_5} \biggr|& \hspace{0.0cm}
+ \frac{2\,m_A}{\left(r^{\,1}_A\left(s_1\right)\right)^2}\,\frac{\ve{d}^k_A\left(s_1\right)}{k \cdot r^{\,1}_A\left(s_1\right)}\,
\frac{\ve{d}^k_A\left(s_1\right) \cdot \ve{v}_A\left(s_1\right)}{c}
\nonumber\\
\nonumber\\
\fl _{{\rm scaling}\;{\rm 1.5PN}}& \hspace{-0.3cm} _{\ve{\varphi}_2} \biggr|& \hspace{0.0cm}
- 4\,\frac{m_A}{r^{\,1}_A\left(s_1\right)}\,\ve{k}\,\frac{\ve{k} \cdot \ve{v}_A\left(s_1\right)}{c}
\nonumber\\
\nonumber\\
\fl _{{\rm scaling}\;{\rm 2PN}} & \hspace{-1.00cm} _{\ve{\rho}_5} + _{\ve{\varphi}_6} \biggr|& \hspace{0.0cm} 
- \frac{2\,m_A^2}{\left(r^{\,1}_A\left(s_1\right)\right)^2}\,\ve{k}\, 
\frac{\ve{d}^k_A\left(s_1\right) \cdot \ve{d}^k_A\left(s_1\right)}{\left(k \cdot r^{\,1}_A\left(s_1\right)\right)^2}
- 2\,\frac{m_A^2}{R^2}\ve{k}
\left|\frac{\ve{d}^k_A\left(s_1\right)}{k \cdot r^{\,1}_A\left(s_1\right)}
- \frac{\ve{d}^k_A\left(s_0\right)}{k \cdot r_A^{\,0}\left(s_0\right)}\right|^2
\nonumber\\
\nonumber\\
\fl _{{\rm scaling}\;{\rm 2PN}}& \hspace{-0.35cm} _{\ve{\varphi}_7} \biggr|& \hspace{0.0cm}
+ 4\,\frac{\ve{k}}{R}\,\frac{m_A^2}{r^{\,1}_A\left(s_1\right)}\,\frac{1}{k \cdot r^{\,1}_A\left(s_1\right)}
\left(\frac{\ve{d}^k_A\left(s_1\right) \cdot \ve{d}^k_A\left(s_1\right)}{k \cdot r^{\,1}_A\left(s_1\right)}
- \frac{\ve{d}^k_A\left(s_0\right) \cdot \ve{d}^k_A\left(s_1\right)}{k \cdot r_A^{\,0}\left(s_0\right)}\right)
\nonumber\\
\nonumber\\
\fl _{{\rm enhanced}\;{\rm 2PN}} & \hspace{-0.25cm} _{\ve{\rho}_6} \biggr|& \hspace{0.0cm}
- 2 \frac{m_A^2}{R^2}
\left(\frac{\ve{d}^k_A\left(s_1\right)}{k \cdot r^{\,1}_A\left(s_1\right)} 
+ \frac{\ve{d}^k_A\left(s_0\right)}{k \cdot r_A^{\,0}\left(s_0\right)}\right)
\left|\frac{\ve{d}^k_A\left(s_1\right)}{k \cdot r^{\,1}_A\left(s_1\right)}
- \frac{\ve{d}^k_A\left(s_0\right)}{k \cdot r_A^{\,0}\left(s_0\right)}\right|^2
\nonumber\\
\nonumber\\
\fl _{{\rm enhanced}\;{\rm 2PN}} & \hspace{-0.25cm} _{\ve{\rho}_7} \biggr|& \hspace{0.0cm}
- 4\,\frac{m_A^2}{R}\,\left(\frac{\ve{d}^k_A\left(s_1\right)}{\left(k \cdot r^{\,1}_A\left(s_1\right)\right)^2}
- \frac{\ve{d}^k_A\left(s_0\right)}{\left(k \cdot r_A^{\,0}\left(s_0\right)\right)^2}\right)
\nonumber\\
\nonumber\\
\fl _{{\rm enhanced}\;{\rm 2PN}} & \hspace{-0.25cm} _{\ve{\varphi}_8} \biggr|& \hspace{0.0cm}
+ 4\,\frac{m_A^2}{r^{\,1}_A\left(s_1\right)}\,\frac{\ve{d}_A^k\left(s_1\right)}{\left(k \cdot r^{\,1}_A\left(s_1\right)\right)^2}
\nonumber\\
\nonumber\\
\fl _{{\rm enhanced}\;{\rm 2PN}} & \hspace{-0.25cm} _{\ve{\varphi}_9} \biggr|& \hspace{0.0cm}
+ 4\,\frac{m_A^2}{r^{\,1}_A\left(s_1\right)}\,\frac{1}{R}\,
\frac{\ve{d}_A^k\left(s_1\right)}{\left(k \cdot r^{\,1}_A\left(s_1\right)\right)^2}
\left(\frac{\ve{d}^k_A\left(s_1\right) \cdot \ve{d}^k_A\left(s_1\right)}{k \cdot r^{\,1}_A\left(s_1\right)}
- \frac{\ve{d}^k_A\left(s_0\right) \cdot \ve{d}^k_A\left(s_1\right)}{k \cdot r_A^{\,0}\left(s_0\right)} \right)
\nonumber\\
\nonumber\\
\fl _{{\rm enhanced}\;{\rm 2PN}} & \hspace{-0.35cm} _{\ve{\varphi}_{10}} \biggr|& \hspace{0.05cm}
+ 4\,\frac{m_A^2}{r^{\,1}_A\left(s_1\right)}\,\frac{1}{R}\,
\frac{\ve{k} \cdot \ve{r}^{\,1}_A\left(s_1\right)}{k \cdot r^{\,1}_A\left(s_1\right)}
\left(\frac{\ve{d}^k_A\left(s_1\right)}{k \cdot r^{\,1}_A\left(s_1\right)}
- \frac{\ve{d}^k_A\left(s_0\right)}{k \cdot r_A^{\,0}\left(s_0\right)} \right)
\nonumber\\
\nonumber\\
\fl _{\rm 2PN} & \hspace{-0.2cm} _{\ve{\rho}^A_{8}} \biggr|& \hspace{0.0cm}
+ \frac{15}{4} \frac{m_A^2}{R}
\frac{\ve{d}^k_A\left(s_1\right)}{\left|\ve{k} \times \ve{r}^{\,1}_A\left(s_1\right)\right|^3} \,
\left(\ve{k} \cdot \ve{r}^{\,1}_A\left(s_1\right)\right)
\left(\arctan \frac{\ve{k} \cdot \ve{r}^{\,1}_A\left(s_1\right)}{\left|\ve{k} \times \ve{r}^{\,1}_A\left(s_1\right)\right|} + \frac{\pi}{2} \right)
\nonumber\\
\fl _{\rm 2PN} & \hspace{-0.2cm} _{\ve{\rho}^B_{8}} \biggr|& \hspace{0.0cm}
- \frac{15}{4} \frac{m_A^2}{R}
\frac{\ve{d}^k_A\left(s_0\right)}{\left|\ve{k} \times \ve{r}^{\,0}_A\left(s_0\right)\right|^3} \,
\left(\ve{k} \cdot \ve{r}^{\,0}_A\left(s_0\right)\right)
\left( \arctan \frac{\ve{k} \cdot \ve{r}^{\,0}_A\left(s_0\right)}{\left|\ve{k} \times \ve{r}^{\,0}_A\left(s_0\right)\right|} + \frac{\pi}{2} \right)
\nonumber\\
\nonumber\\
\fl _{\rm 2PN}& \hspace{-0.95cm} _{\ve{\rho}_9} + _{\ve{\varphi}_{11}} \biggr|& \hspace{0.0cm}
- \frac{1}{4}\frac{m_A^2}{R}\left(\!\frac{\ve{d}^k_A\left(s_1\right)}{\left(r_A^{\,1}\left(s_1\right)\right)^2} 
- \frac{\ve{d}^k_A\left(s_0\right)}{\left(r_A^{\,0}\left(s_0\right)\right)^2}\!\right)
- 4\,\frac{m_A^2}{\left(r_A^{\,1}\left(s_1\right)\right)^2}\,\frac{\ve{d}_A^k\left(s_1\right)}{k \cdot r^{\,1}_A\left(s_1\right)}
\nonumber\\
\nonumber\\
\fl _{\rm 2PN} & \hspace{-0.3cm} _{\ve{\varphi}_{12}} \biggr|& \hspace{0.0cm}
- \frac{m_A^2}{2}\,\ve{d}_A^k\left(s_1\right)\frac{\ve{k} \cdot \ve{r}^{\,1}_A\left(s_1\right)}{\left(r^{\,1}_A\left(s_1\right)\right)^4}
- \frac{15}{4}\,\frac{m_A^2}{\left(r_A^{\,1}\left(s_1\right)\right)^2}\,\ve{d}^k_A\left(s_1\right)
\frac{\ve{k} \cdot \ve{r}^{\,1}_A\left(s_1\right)}{\left|\ve{k} \times \ve{r}^{\,1}_A\left(s_1\right)\right|^2}
\nonumber\\
\nonumber\\
\fl _{\rm 2PN} & \hspace{-0.3cm} _{\ve{\varphi}_{13}} \biggr|& \hspace{0.0cm}
- \frac{15}{4}\,m_A^2\,\frac{\ve{d}^k_A\left(s_1\right)}{\left|\ve{k} \times \ve{r}^{\,1}_A\left(s_1\right)\right|^3}  
\left(\arctan \frac{\ve{k} \cdot \ve{r}^{\,1}_A\left(s_1\right)}{\left|\ve{k} \times \ve{r}^{\,1}_A\left(s_1\right)\right|} + \frac{\pi}{2} \right)
\nonumber\\
\nonumber\\
\fl _{\rm 2PN}& \hspace{0.25cm} \biggr|& \hspace{0.0cm}
+ \hat{\ve \epsilon}_1\left(s_1\right) + \hat{\ve \epsilon}_2\left(s_1,s_0\right)  
\nonumber\\
\nonumber\\
\fl _{\rm 2.5PN} & \hspace{0.25cm} \biggr|& \hspace{0.0cm} + {\cal O}\left(c^{-5}\right), 
\label{Transformation_k_to_n}
\end{eqnarray}

\noindent 
where $\ve{\rho}_i = \ve{\rho}_i\left(s_1,s_0\right)$ with $i=1\,, \cdots \,,9$, and 
$\ve{\varphi}_i = \ve{\varphi}_i\left(s_1\right)$ for $i=1\,, \cdots \,,6,8,11,12,13$, and 
$\ve{\varphi}_i = \ve{\varphi}_i\left(s_1,s_0\right)$ for $i = 7,9,10$ which appear before the vertical lines
are by definition equal to the expressions on the right of the vertical bars in each line.  

The transformation (\ref{Transformation_k_to_n}) allows one to determine the unit coordinate direction $\ve{n}$ at 
the observers position from the boundary values $\ve{x}_0$ and $\ve{x}_1$.  
In the limit of body at rest this expression coincides with Eq.~(87) in \cite{Article_Zschocke1}. 
The terms $\hat{\ve \epsilon}_1$ and $\hat{\ve \epsilon}_2$ are given in the \ref{Appendix_epsilon} by 
Eqs.~(\ref{Transformation_sigma_to_n_epsilon}) and (\ref{Transformation_k_to_sigma_epsilon}), respectively. 
In view of the remarkable algebraic structure in (\ref{Transformation_k_to_n}) it is evident how important the estimation  
of the upper limit of the individual terms is. Such an estimation allows for a considerably simpler structure of this transformation, 
which will be the topic of the next section.

\subsection{Simplified expression for the transformation from $\ve{k}$ to $\ve{n}$}

The simplified transformation from $\ve{k}$ to $\ve{n}$ follows from Eqs.~(\ref{Simplified_Transformation_k_to_sigma})
and (\ref{Simplified_Transformation_sigma_to_n}), that means where all {\it scaling terms} are omitted and all terms 
are neglected whose individual contribution is less than $1\,{\rm nas}$ in light deflection  
for Sun at $45^{\circ}$ and all the other Solar System bodies.
For the total sum of all those neglected terms which are not proportional to three-vector $\ve{k}$ one obtains 
\begin{eqnarray}
\fl I_3 = \bigg|\sum\limits_{i=3,8,9} \ve{\rho}_i\left(s_1,s_1\right)
+ \sum\limits_{i=1,2,3,4} \Delta \ve{\rho}_i\left(s_1,s_1\right) 
+ \sum\limits_{i=4,5} \ve{\varphi}_i\left(s_1\right) + \sum\limits_{i=11,12,13} \ve{\varphi}_i\left(s_1,s_1\right)  
\nonumber\\
\nonumber\\
\fl \hspace{1.0cm} + \hat{\ve \epsilon}_1\left(s_1\right) + \hat{\ve \epsilon}_2\left(s_1,s_1\right) \bigg|  
\nonumber\\
\nonumber\\
\fl \hspace{0.5cm} \le \frac{10\,m_A}{r^{\,1}_A\left(s_1\right)}\,\frac{v_A\left(s_1\right)}{c} + \frac{15}{4} \pi \frac{m_A^2}{P_A^2}
+ \frac{14\,m_A}{r_A^{\,1}\left(s_1\right)}\,\frac{v_A^2\left(s_1\right)}{c^2} 
+ 24\,\frac{m_A}{P_A} \frac{v_A^2\left(s_1\right)}{c^2} + 18 \,m_A \,\frac{a_A\left(s_1\right)}{c^2}\,.
\nonumber\\
\label{Sum_3}
\end{eqnarray}

\noindent
For the upper limits of the terms in (\ref{Sum_3}) we have used that  
\begin{eqnarray} 
\fl \left| \ve{\rho}_8 + \ve{\rho}_9 + \ve{\varphi}_{11} + \ve{\varphi}_{12} + \ve{\varphi}_{13} \right| \le \frac{15}{4}\,\pi\,\frac{m_A^2}{P_A^2}\,,
\label{Inequality_B}
\\
\nonumber\\
\fl \hspace{1.75cm} \left|\Delta \ve{\rho}_1 + \ve{\varphi}_4 + \ve{\varphi}_5\right| 
\le 10\,\frac{m_A}{r^{\,1}_A\left(s_1\right)}\,\frac{v_A\left(s_1\right)}{c}\,, 
\label{Inequality_A}
\\
\nonumber\\
\fl \hspace{1.25cm} \left|\Delta \ve{\rho}_2 + \Delta \ve{\rho}_3 + \Delta \ve{\rho}_4 \right| 
\le \frac{6\,m_A}{r_A^{\,1}\left(s_1\right)}\,\frac{v_A^2\left(s_1\right)}{c^2} 
+ \frac{4\,m_A}{P_A}\,\frac{v_A^2\left(s_1\right)}{c^2} 
+ 8\,m_A\,\frac{a_A\left(s_1\right)}{c^2}\,,  
\label{Inequality_C}
\end{eqnarray}

\noindent 
while $\left|\ve{\rho}_3\left(s_1,s_1\right)\right| = 0$ according to Eq.~(\ref{Term_rho_3_3_A}). 
The inequality (\ref{Inequality_B}) is not shown explicitly, but its validity can be demonstrated by means of the approach    
described in \ref{Appendix_Estimation1}. The inequality (\ref{Inequality_A}) is  
shown in \ref{Appendix_Inequality_A}, while the inequality (\ref{Inequality_C}) follows from (\ref{Term_rho_2_1_E}), (\ref{Term_rho_3_4}),  
and (\ref{Estimation_rho_4_2}). The upper limit of $\left|\hat{\ve \epsilon}_1\right|$ and $\left|\hat{\ve \epsilon}_2\right|$ are given by  
Eqs.~(\ref{Appendix_estimation_epsilon_1}) and (\ref{Appendix_estimation_epsilon_3}), respectively.  
Using the numerical parameters as given by Table~\ref{Table1} one obtains  
\begin{eqnarray}
\fl \hspace{0.75cm} I_3 \le \; 1.3\,{\rm nas} \quad {\rm for}\;{\rm Sun}\;{\rm at}\,45^{\circ}\,\left({\rm solar}\;{\rm aspect}\;{\rm angle}\right),
\nonumber\\
\fl \hspace{1.15cm} \le \; 1.3\,{\rm nas} \quad {\rm for}\;{\rm Jupiter}\,,
\label{Total_Sum_3}
\end{eqnarray}

\noindent
and less than $0.64\,{\rm nas}$ for all the other Solar System bodies. 
Accordingly, by neglecting these terms in (\ref{Sum_3}) one obtains the simplified transformation $\ve{k}$ to $\ve{n}$ fully in terms of $s_1$,   
which reads as follows: 
\begin{eqnarray}
\fl _{\rm N} & \hspace{0.1cm} \biggr|& \hspace{0.15cm} {\ve{n}} = {\ve{k}}
\nonumber\\
\nonumber\\
\fl _{\rm 1PN}& \hspace{-0.95cm} _{\ve{\rho}_1} + _{\ve{\varphi}_1} \biggr|& \hspace{0.0cm}
- 2\,\frac{m_A}{R}
\left(\frac{\ve{d}^k_A\left(s_1\right)}{k \cdot r^{\,1}_A\left(s_1\right)}
- \frac{\ve{d}^k_A\left(s_1\right)}{k \cdot r_A^{\,0}\left(s_1\right)}\right)
+ 2\,m_A\,\frac{\ve{d}_A^k\left(s_1\right)}{r^{\,1}_A\left(s_1\right)}\,\frac{1}{k \cdot r^{\,1}_A\left(s_1\right)}\,
\nonumber\\
\nonumber\\
\fl _{\rm 1.5PN} & \hspace{-0.3cm} _{\ve{\rho}_2} \biggr|& \hspace{0.0cm}
+ 2\,\frac{m_A}{R}\,\ve{k} \times \left(\frac{\ve{v}_A\left(s_1\right)}{c} \times \ve{k}\right)
\ln \frac{k \cdot r^{\,1}_A\left(s_1\right)}{k \cdot r_A^{\,0}\left(s_1\right)}
\nonumber\\
\nonumber\\
\fl _{\rm 1.5PN} & \hspace{-0.3cm} _{\ve{\rho}_4} \biggr|& \hspace{0.0cm}
+ 2\,\frac{m_A}{R}\,\frac{\ve{k} \cdot \ve{v}_A\left(s_1\right)}{c}\,
\left(\frac{\ve{d}^k_A\left(s_1\right)}{k \cdot r^{\,1}_A\left(s_1\right)}
- \frac{\ve{d}^k_A\left(s_1\right)}{k \cdot r_A^{\,0}\left(s_1\right)}\right) 
\nonumber\\
\nonumber\\
\fl _{\rm 1.5PN} & \hspace{-0.3cm} _{\ve{\varphi}_3} \biggr|& \hspace{0.0cm}
- 2\,\frac{m_A}{r^{\,1}_A\left(s_1\right)}\,\frac{\ve{d}^k_A\left(s_1\right)}{k \cdot r^{\,1}_A\left(s_1\right)}\,
\frac{\ve{k}\cdot \ve{v}_A\left(s_1\right)}{c}
\nonumber\\
\nonumber\\
\fl _{{\rm enhanced}\;{\rm 2PN}} & \hspace{-0.25cm} _{\ve{\rho}_6} \biggr|& \hspace{0.0cm}
- 2 \frac{m_A^2}{R^2}
\left(\frac{\ve{d}^k_A\left(s_1\right)}{k \cdot r^{\,1}_A\left(s_1\right)}
+ \frac{\ve{d}^k_A\left(s_1\right)}{k \cdot r_A^{\,0}\left(s_1\right)}\right)
\left|\frac{\ve{d}^k_A\left(s_1\right)}{k \cdot r^{\,1}_A\left(s_1\right)}
- \frac{\ve{d}^k_A\left(s_1\right)}{k \cdot r_A^{\,0}\left(s_1\right)}\right|^2
\nonumber\\
\nonumber\\
\fl _{{\rm enhanced}\;{\rm 2PN}} & \hspace{-0.25cm} _{\ve{\rho}_7} \biggr|& \hspace{0.0cm}
- 4\,\frac{m_A^2}{R}\,\left(\frac{\ve{d}^k_A\left(s_1\right)}{\left(k \cdot r^{\,1}_A\left(s_1\right)\right)^2}
- \frac{\ve{d}^k_A\left(s_1\right)}{\left(k \cdot r_A^{\,0}\left(s_1\right)\right)^2}\right)
\nonumber\\
\nonumber\\
\fl _{{\rm enhanced}\;{\rm 2PN}} & \hspace{-0.25cm} _{\ve{\varphi}_8} \biggr|& \hspace{0.0cm}
+ 4\,\frac{m_A^2}{r^{\,1}_A\left(s_1\right)}\,\frac{\ve{d}_A^k\left(s_1\right)}{\left(k \cdot r^{\,1}_A\left(s_1\right)\right)^2}
\nonumber\\
\nonumber\\
\fl _{{\rm enhanced}\;{\rm 2PN}} & \hspace{-0.25cm} _{\ve{\varphi}_9} \biggr|& \hspace{0.0cm}
+ 4\,\frac{m_A^2}{r^{\,1}_A\left(s_1\right)}\,\frac{1}{R}\,
\frac{\ve{d}_A^k\left(s_1\right)}{\left(k \cdot r^{\,1}_A\left(s_1\right)\right)^2}
\left(\!\frac{\ve{d}^k_A\left(s_1\right) \cdot \ve{d}^k_A\left(s_1\right)}{k \cdot r^{\,1}_A\left(s_1\right)}
- \frac{\ve{d}^k_A\left(s_1\right) \cdot \ve{d}^k_A\left(s_1\right)}{k \cdot r_A^{\,0}\left(s_1\right)} \!\right)
\nonumber\\
\nonumber\\
\fl _{{\rm enhanced}\;{\rm 2PN}} & \hspace{-0.35cm} _{\ve{\varphi}_{10}} \biggr|& \hspace{0.05cm}
+ 4\,\frac{m_A^2}{r^{\,1}_A\left(s_1\right)}\,\frac{1}{R}\,
\frac{\ve{k} \cdot \ve{r}^{\,1}_A\left(s_1\right)}{k \cdot r^{\,1}_A\left(s_1\right)}
\left(\frac{\ve{d}^k_A\left(s_1\right)}{k \cdot r^{\,1}_A\left(s_1\right)}
- \frac{\ve{d}^k_A\left(s_1\right)}{k \cdot r_A^{\,0}\left(s_1\right)} \right)
\nonumber\\
\nonumber\\
\fl _{\rm 2.5PN} & \hspace{0.25cm} \biggr|& \hspace{0.0cm} + {\cal O}\left(c^{-5}\right), 
\label{Simplified_Transformation_k_to_n}
\end{eqnarray}

\noindent
where $\ve{\rho}_i = \ve{\rho}_i\left(s_1,s_1\right)$ for $i=1,2,4,6,7$, and $\ve{\varphi}_i = \ve{\varphi}_i\left(s_1\right)$ for $i = 1,3,8$,  
and $\ve{\varphi}_i = \ve{\varphi}_i\left(s_1,s_1\right)$ for $i = 9,10$ which appear before the vertical lines   
are by definition equal to the expressions on the right of the vertical bars in each line.  
In the limit of body at rest this expression coincides with Eqs.~(92) - (93) in \cite{Article_Zschocke1}. 
For the distance $R = \left|\ve{R}\right|$ one should implement the exact expression (\ref{Boundary_3}),
because the approximative expression (\ref{vector_R_series_6}) is slightly more complicated and only in use for the estimations
but not for astrometric data reduction. By means of the approach and results of the appendix,  
one obtains for the upper limits of the 1PN, 1.5PN, and 2PN terms of the simplified transformation (\ref{Simplified_Transformation_k_to_n}):  
\begin{eqnarray}
\fl {\rm 1PN} \hspace{4.6cm} 
\left|\ve{\rho}_1 + \ve{\varphi}_1\right| \le 4\,\frac{m_A}{P_A}\,,  
\label{angle_1PN_3}
\\
\nonumber\\
\fl {\rm 1.5PN} \hspace{3.55cm}
\left|\ve{\rho}_2 + \ve{\rho}_4 + \ve{\varphi}_3 \right| \le 6\,\frac{m_A}{P_A}\,\frac{v_A\left(s_1\right)}{c}\,,
\label{angle_15PN_3}
\\
\nonumber\\
\fl {\rm enhanced}\;{\rm 2PN} \hspace{0.5cm} \left|\ve{\rho}_6 + \ve{\rho}_7 + \ve{\varphi}_8 + \ve{\varphi}_9 + \ve{\varphi}_{10} \right|
\le 16\,\frac{m^2_A}{P^2_A}\,\frac{r^{\,1}_A\left(s_1\right)}{P_A}\,.  
\label{angle_2PN_3}
\end{eqnarray}

\noindent 
As outlined above, the 2PN term (\ref{angle_2PN_3}) is a so-called {\it enhanced term} 
because of the factor $r^{\,1}_A\left(s_1\right)/P_A$. 

A further comment is in order about the upper limit of the 1.5PN terms as given by (\ref{angle_15PN_3}). 
In Eq.~(179) in \cite{Zschocke2} the upper limit of the 1.5PN terms in light deflection was given by 
$\displaystyle \varphi_{\rm 1.5PN} \le 4\,\frac{m_A}{P_A}\,\frac{v_A\left(s_1\right)}{c}$ in  
agreement with the results in \cite{Klioner2003b,KopeikinMakarov2007}. The marginal difference between 
the factor $6$ in Eq.~(\ref{angle_15PN_3}) and the factor $4$ in Eq.~(179) in \cite{Zschocke2}  
originates from the logarithmic term $\ve{\rho}_2$ in the simplified transformation (\ref{Simplified_Transformation_k_to_n}).  
That term has been estimated by Eq.~(\ref{Term_rho_2_2}), according to which the term $\ve{\rho}_2$ would vanish 
in the limit of light sources at infinity. So the term $\ve{\rho}_2$ originates from the boundary value problem,  
which has not been on the scope of the investigations \cite{Klioner2003b,Zschocke2,KopeikinMakarov2007}.  
In particular, without the term $\ve{\rho}_2$ we would get the result as given by Eq.~(179) in \cite{Zschocke2}.  

Let us summarize the variables on which the simplified transformations (\ref{Simplified_Transformation_k_to_sigma}), 
(\ref{Simplified_Transformation_sigma_to_n}), and (\ref{Simplified_Transformation_k_to_n}) depend on, as there are:  
$m_A, \ve{x}_0, \ve{x}_1, \ve{x}_A\left(s_1\right)$. The values $m_A, \ve{x}_1, \ve{x}_A\left(s_1\right)$  
are provided by some ephemerides and tracking of the orbit of the satellite (observer).  
Thus, the only unknown in these transformations remains the spatial position  
of the light source $\ve{x}_0$, which is the primary aim of astrometric data reduction.

\section{Impact of higher order terms}\label{Section_3PN}  

The transformations (\ref{Transformation_k_to_sigma}), (\ref{Transformation_sigma_to_n}), (\ref{Transformation_k_to_n}) 
and their simplified versions (\ref{Simplified_Transformation_k_to_sigma}), (\ref{Simplified_Transformation_sigma_to_n}), 
(\ref{Simplified_Transformation_k_to_n}) are valid up to terms of the order ${\cal O}\left(c^{-5}\right)$. So the 
question arises about the impact of these higher order terms. Are they relevant for nas-astrometry?  

In order to address the problem we consider the light deflection angle $\varphi = \angle \left(\ve{k}, \ve{n}\right)$. 
By including terms up to the order ${\cal O}\left(c^{-7}\right)$ the post-Newtonian expansion 
of the light deflection angle is   
\begin{eqnarray}
\fl \varphi = \arcsin \left| \ve{k} \times \ve{n} \right|  =  \left| \ve{k} \times \ve{n} \right| 
+ \frac{1}{6}\,\left| \ve{k} \times \ve{n} \right|^3 + {\cal O}\left(\left| \ve{k} \times \ve{n} \right|^5\right)  
\label{series_arcsin}
\\
\fl \hspace{0.3cm} = \varphi_{\rm 1PN} + \varphi_{\rm 1.5PN} + \varphi_{\rm 2PN} 
+ \varphi_{\rm 2.5PN} + \varphi_{\rm 3PN} + \varphi_{\rm 3.5PN} + \varphi_{\rm 4PN} + \varphi_{\rm 4.5PN} + {\cal O}\left(c^{-10}\right),   
\nonumber\\ 
\label{light_deflection_angle}
\end{eqnarray}

\noindent
where $\varphi_{\rm n PN} = {\cal O}\left(c^{- 2 n}\right)$. The first term on the r.h.s. of (\ref{series_arcsin}) contributes 
to any order, while the second term on the r.h.s. of (\ref{series_arcsin}) contributes to the order ${\cal O}\left(c^{-6}\right)$ and beyond.  
The 1PN, 1.5PN, and 2PN terms in (\ref{light_deflection_angle}) can be obtained   
from the simplified transformation (\ref{Simplified_Transformation_k_to_n}). One obtains  
\begin{eqnarray}
\varphi_{\rm 1PN} = \left| \ve{k} \times \left(\ve{\rho}_1 + \ve{\varphi}_1\right)\right| 
\le  4\,\frac{m_A}{P_A}\,, 
\label{higher_order_terms_1PN} 
\\
\nonumber\\
\hspace{-0.25cm} \varphi_{\rm 1.5PN} = \left| \ve{k} \times \left(\ve{\rho}_2 + \ve{\rho}_4 + \ve{\varphi}_3\right)\right|  
\le  6\,\frac{m_A}{P_A}\,\frac{v_A\left(s_1\right)}{c}\,, 
\label{higher_order_terms_15PN} 
\\
\nonumber\\
\varphi_{\rm 2PN} = \left| \ve{k} \times \left(\ve{\rho}_6 + \ve{\rho}_7 + \ve{\varphi}_8 + \ve{\varphi}_9 + \ve{\varphi}_{10} \right)\right|  
\le 16\,\frac{m_A^2}{P_A^2}\,\frac{r^{\,1}_A\left(s_1\right)}{P_A} \,,   
\label{higher_order_terms_2PN}
\end{eqnarray}

\noindent
all of which are relevant on the nas-scale of accuracy. 
The next order beyond 2PN approximation would be 2.5PN terms. While they are out of the scope of the present investigation, 
a few comments can be stated already right now. Basically, there are three kind of 2.5PN terms, as there are  
\begin{eqnarray}
\varphi^A_{\rm 2.5PN} \sim \frac{m_A}{P_A}\,\frac{v^3_A\left(s_1\right)}{c^3} \ll 1\,{\rm nas}\;,
\label{higher_order_terms_25PN_A}
\\ 
\varphi^B_{\rm 2.5PN} \sim m_A\,\frac{v_A\left(s_1\right)}{c}\frac{a_A\left(s_1\right)}{c^2} \ll 1\,{\rm nas}\;,
\label{higher_order_terms_25PN_B}
\\
\varphi^C_{\rm 2.5PN} \sim \frac{m_A^2}{P_A^2}\,\frac{v_A\left(s_1\right)}{c}\,\frac{r^{\,1}_A\left(s_1\right)}{P_A}\,.
\label{higher_order_terms_25PN_C}
\end{eqnarray}

\noindent
The structure of the 2.5PN terms in (\ref{higher_order_terms_25PN_A}) and (\ref{higher_order_terms_25PN_B}) 
follows from a series expansion of the first post-Minkowskian (1PM) solution of a light signal propagating in the field of one 
arbitrarily moving monopole as found in \cite{KS1999}; for more explicit expressions
of the coordinate velocity and light trajectory we refer to Eqs.~(C.1) - (C.8) in \cite{Klioner2003a} or
Eqs.~(E.4) - (E.6) and (E.16) in \cite{Zschocke4}. So these terms in (\ref{higher_order_terms_25PN_A}) and (\ref{higher_order_terms_25PN_B}) 
have no enhancement factor and they are negligible even for highly precise measurements on the nas-scale of accuracy, also in case of 
some large numerical factor in front of these terms.   
But what about the 2.5PN terms in (\ref{higher_order_terms_25PN_C})? They are connected with  
an enhancement factor $r^{\,1}_A\left(s_1\right)/P_A$ and might become large enough to be of relevance for nas-astrometry. 
As it stands, the term (\ref{higher_order_terms_25PN_C}) is less than $1\,{\rm nas}$ for any Solar System body (even for grazing rays at 
the Sun), but certainly there will be some large numerical factor in front of this term. Then, the 2.5PN term 
(\ref{higher_order_terms_25PN_C}) would be above the threshold of $1\,{\rm nas}$ for grazing rays at Jupiter. In order to determine more  
precisely the relevance of 2.5PN terms (\ref{higher_order_terms_25PN_C}) for nas-astrometry,  
one should consider the 2PN light trajectory in the field of one monopole at rest, $\ve{x}_A = {\rm const}$,  
and then perform a Lorentz transformation in order to account for the  
translational motion of the body, which would yield all terms proportional to $m_A^2\,\left(v_A/c\right)^n$ with $n=1, 2, 3, ...$.  
Such an approach has already been developed in the first post-Minkowskian approximation \cite{Klioner2003a} and might be generalized  
for the case of 2.5PN light propagation in the field of one body in translational motion.  

Let us now consider terms of the 3PN approximation, that means terms of the order ${\cal O}\left(c^{-6}\right)$ 
in (\ref{light_deflection_angle}). Are they of relevance for nas-astrometry?  
A reliable answer can be found in the following manner. In \cite{Zschocke_Lense_Equation} a lens equation has been derived for the  
light deflection angle $\varphi$ in the field of one spherically symmetric body at rest, $\ve{x}_A = {\rm const}$, given by 
(cf. Eq.~(15) in \cite{Zschocke_Lense_Equation} and shift of the origin of spatial axes  
by three-vector $\ve{x}_A$)  
\begin{eqnarray}
\fl \hspace{2.0cm} \varphi = \frac{1}{2} \left(\sqrt{\left(\frac{d^k_A}{r^{\,1}_A}\right)^2  
+ 8\,\frac{m_A}{r^{\,1}_A}\, \frac{r^{\,0}_A\,r^{\,1}_A - \ve{r}^{\,0}_A \cdot \ve{r}^{\,1}_A}{R\,r^{\,1}_A} } - \frac{d^k_A}{r^{\,1}_A} \right) 
+ {\cal O}\left(\frac{m_A^2}{P_A^2}\right), 
\label{lens_equation}
\end{eqnarray}

\noindent
where $\ve{r}^{\,0}_A = \ve{x}_0 - \ve{x}_A$ and $\ve{r}^{\,1}_A = \ve{x}_1 - \ve{x}_A$ and the impact parameter   
$d^k_A = \left|\ve{k} \times \ve{r}^{\,0}_A\right| = \left|\ve{k} \times \ve{r}^{\,1}_A\right|$ is independent of time. 
The neglected terms of order ${\cal O}\left(m_A^2/P_A^2\right)$ has been shown to be less than 
$\displaystyle \frac{15}{4}\,\pi\,\frac{m_A^2}{P_A^2}$ which 
is less than $1\,{\rm nas}$ for Sun at $45^{\circ}$ (solar aspect angle adopted from the Gaia mission) and all the other Solar System bodies.  
The lens equation (\ref{lens_equation}) represents the total sum of 
all {\it enhanced terms}. Of course, for an observer located in the Solar System the lens effect (i.e. second image of the source 
caused by Solar System bodies) cannot be detected, and that is why the second solution with the lower sign  
in Eq.~(15) in \cite{Zschocke_Lense_Equation} is omitted here for our considerations.  

In the near-zone of the Solar System we have $m_A/d^k_A \ll 1$, which allows for a series expansion of the lens equation (\ref{lens_equation}) 
in terms of this small parameter. This possibility is utilized to get  
\begin{eqnarray}
\varphi = \varphi_{\rm 1PN} + \varphi_{\rm 2PN} + \varphi_{\rm 3PN} + {\cal O}\left(c^{-8}\right) + {\cal O}\left(\frac{m_A^2}{P_A^2}\right),  
\label{series_expansion_lens_equation}
\end{eqnarray}

\noindent 
which has already been given by Eq.~(26) in \cite{Zschocke_Lense_Equation}; because the body is assumed to be at rest 
in (\ref{lens_equation}) there are no 1.5PN terms, 2.5PN terms and so on in the series expansion (\ref{series_expansion_lens_equation}). 
For the upper limits one obtains (cf. Eqs.~(17) and (18) in \cite{Zschocke_Lense_Equation}),   
\begin{eqnarray}
\varphi_{\rm 1PN} \le 4\,\frac{m_A}{P_A}\,,  
\label{series_expansion_lens_equation_1PN}
\\
\varphi_{\rm 2PN} \le 16\,\frac{m_A^2}{P_A^2}\,\frac{r^{\,1}_A}{P_A}\,.  
\label{series_expansion_lens_equation_2PN}
\end{eqnarray}
 
\noindent  
The above standing results in (\ref{higher_order_terms_1PN}) and (\ref{higher_order_terms_2PN}) coincide, in the limit of body at rest, 
with Eqs.~(\ref{series_expansion_lens_equation_1PN}) and (\ref{series_expansion_lens_equation_2PN}). 
The 3PN term of light deflection for body at rest has already been  
considered in Eq.~(27) in \cite{Zschocke_Lense_Equation} and reads:  
\begin{eqnarray}
\varphi_{\rm 3PN} \le 128\,\frac{m_A^3}{P_A^3}\,\left(\frac{r^{\,1}_A}{P_A}\right)^2\,.   
\label{higher_order_terms_3PN}
\end{eqnarray}

\noindent 
The same result has also been obtained within the Time Transfer Function approach in \cite{LinetTeyssandier2013_a,LinetTeyssandier2013_b}  
(cf. Eq.~(93) in \cite{LinetTeyssandier2013_a} or Eq.~(21) in \cite{LinetTeyssandier2013_b}). One might believe that  
(\ref{higher_order_terms_3PN}) could also be concluded from the second term on the r.h.s. of (\ref{series_arcsin}), but this would  
be incomplete as long as the transformation $\ve{n}$ to $\ve{k}$ is only known in the 2PN approximation, because the 
first term on the r.h.s. of (\ref{series_arcsin}) contributes to any order.  
Inserting numerical parameter of Table~\ref{Table1}  
one obtains for grazing rays at Jupiter and Saturn about $\varphi_{\rm 3PN} = 32\,{\rm nas}$ and $\varphi_{\rm 3PN} = 7\,{\rm nas}$, 
respectively, in light deflection, while  
in the field of earth-like planets or Sun at $45^{\circ}$  
they would contribute much less than $1\,{\rm nas}$; for grazing
ray at the Sun the 3PN term (\ref{higher_order_terms_3PN}) amounts to be about $12 \cdot 10^3\,{\rm nas}$ in light deflection, 
as already noticed by Eq.~(22) in \cite{LinetTeyssandier2013_b}. 

From these considerations it becomes  
certain, that {\it enhanced terms} in the third post-Newtonian (3PN) approximation have to be taken into account for
astrometry on the nas-level of accuracy. But it is clear that such calculation would be a rather ambitious assignment 
of a task for moving bodies. Therefore, in order to get the light trajectory $\ve{x}\left(t\right)$ in the 3PN approximation 
for moving bodies, one should consider the much simpler case of 3PN light trajectory in the field of one monopole at rest, 
$\ve{x}_A = {\rm const}$, and then just take the retarded position of the massive body, $\ve{x}_A = \ve{x}_A\left(s_1\right)$,  
in order to account for the body's motion.  

Finally, from very similar considerations it becomes clear that 3.5PN terms and 4PN terms will not be of relevance for nas-astrometry. 
For instance, we would obtain 
\begin{eqnarray}
\varphi_{\rm 4PN} \le 1280\,\frac{m_A^4}{P_A^4}\,\left(\frac{r^{\,1}_A}{P_A}\right)^3\,, 
\label{higher_order_terms_4PN}
\end{eqnarray}

\noindent
which is much less than $1\,{\rm nas}$ for Sun at $45^{\circ}$  
and any other Solar System body; but we notice that for grazing 
ray at the Sun the 4PN term (\ref{higher_order_terms_4PN}) amounts to be about $50\,{\rm nas}$ in light deflection.  
So the strict statement is that the impact of {\it enhanced terms} becomes smaller and smaller the higher the post-Newtonian order is,  
and can be neglected from the 3.5PN order on, even for ultra-high precision of the nas-level of accuracy in astrometry, except 
for grazing rays at the Sun where the 4PN order has to be accounted for.

\section{Summary}\label{Section8}

Todays precision in angular measurements of celestial objects has reached a level of a few micro-arcseconds.  
In fact, the very recent Data Release 2 of the ESA astrometry mission Gaia contains 
precise positions, proper motions, and parallaxes for more than $1300$ million stars and   
provides astrometric data for parallaxes having uncertainties of only about $30$ \muas $\;$ 
for bright stars with $V \!\! = \!\! 15\;{\rm mag}$ in stellar magnitude \cite{GAIA_DR2_1,Gaia_Archive,GAIA_DR2_2}.  

The impressive progress of the ESA cornerstone mission Gaia in astrometric precision has encouraged the astrometric science  
to further proceed in nearest future. Over the next coming years, the Gaia science community will embark on an
intense series of workshops to develop the key science themes which will scope the
requirements for a future astrometry mission. This will culminate in a detailed white paper
which will be published to coincide with the first releases of Gaia data.
Furthermore, among several astrometry missions proposed to ESA the
M-5 mission proposals Gaia-NIR \cite{Gaia_NIR}, Theia \cite{Theia}, and NEAT \cite{NEAT1,NEAT2,NEAT3}, are mentioned 
which in this order are designed for a highly precise measurement aiming at the \muas$\;$ level, 
sub-$\mu{\rm as}$ level and even nas level of precision.  
Also feasibility studies of Earth-bounded telescopes are presently under consideration which aim at an
accuracy of about $10\,{\rm nas}$ \cite{nas_telescopes}.

Such ultra-highly precise accuracies on the sub-\muas-level presuppose corresponding advancements in the theory of light propagation 
in the Solar System. In particular, at such level of precision it is necessary to describe the propagation of a light signal in the  
gravitational field of $N$ Solar System bodies described by their full set of mass-multipoles $M_L^A$ and spin-multipoles $S_L^A$, 
allowing the bodies to have arbitrary shape, inner structure, oscillations and rotational motion.    
A remarkable and impressive progress has been achieved during recent years in determining the light trajectories in the gravitational  
field of bodies with higher multipoles, as there are:  

$\bullet$ A general solution for the light-trajectory in the stationary gravitational field of a localized source at rest, $\ve{x}_A = {\rm const}$,  
with time-independent intrinsic multipoles, $M^A_L$ and $S^A_L$, has been determined in 1.5PN approximation in \cite{Kopeikin1997}. 
 
$\bullet$ Furthermore, the light-trajectory in the field of a localized source at rest with time-dependent intrinsic   
multipoles, $M^A_L\left(t\right)$ and $S^A_L\left(t\right)$, has been obtained
in \cite{KopeikinKorobkovPolnarev2006,KopeikinKorobkov2005} in 1PM approximation; see also \cite{KSGE}. Furthermore,  
the light trajectory in the field of an arbitrarily moving body with quadrupole-structure has been determined in \cite{KopeikinMakarov2007}.

$\bullet$ In the investigation \cite{moving_axisymmetric_body} the light propagation in the field of an uniformly moving axisymmetric body
has been determined in terms of the full mass-multipole structure of the body.  
Furthermore, an analytical formula for the time-delay caused by the gravitational field of a body in slow and
uniform motion with arbitrary multipoles has been derived in \cite{Soffel_Han}.

$\bullet$ A general solution for light trajectories in the field of arbitrarily moving bodies characterized by intrinsic multipoles 
has been determined in the 1PN approximation \cite{Zschocke1} where the moving bodies are endowed with time-dependent intrinsic 
mass-multipoles $M^A_L\left(t\right)$, as well as in the 1.5PN approximation \cite{Zschocke2} where the moving bodies are endowed with 
both time-dependent intrinsic mass-multipoles $M^A_L\left(t\right)$ and spin-multipoles $S^A_L\left(t\right)$.  

Moreover, it is clear that astrometry on  
the sub-micro-arcsecond level necessitates to determine the light trajectory in the second post-Newtonian approximation  
\cite{Conference_Cambridge,Deng_Xie,Deng_2015,Minazzoli2,Xu_Wu,Xu_Gong_Wu_Soffel_Klioner,Minazzoli1,2PN_Light_PropagationA,Xie_Huang}.
Thus far, the light trajectory in 2PN approximation has only been determined in the field of one monopole  
at rest \cite{Brumberg1991,Brumberg1987}, a result which has later been confirmed within several investigations  
\cite{Deng_Xie,Deng_2015,Minazzoli2,Article_Zschocke1,LLT2004,TL2008,Teyssandier,HBL2014b,AshbyBertotti2010,Moving_Kerr_Black_Hole1}.  
Very recently, an analytical solution in 2PN approximation for the light trajectory in the field of one arbitrarily moving pointlike monopole 
has been obtained in \cite{Zschocke3,Zschocke4}. That solution has solved the so-called initial-value problem (\ref{Initial_Boundary_Conditions}).  
The initial value problem is just the first step in the theory of light propagation, while practical modeling of astronomical observations 
needs to solve the boundary value problem (\ref{Boundary_Value_Conditions}), which is the primary topic of this investigation.  

The solution of the boundary value problem (\ref{Boundary_Value_Conditions}) comprises a set of altogether three transformations, which 
represent the first part of the main results of this investigation:  
\begin{enumerate} 
\item[1.] Transformation $\ve{k} \rightarrow \ve{\sigma}$ given by Eq.~(\ref{Transformation_k_to_sigma}),  
\item[2.] Transformation $\ve{\sigma} \rightarrow \ve{n}$ given by Eq.~(\ref{Transformation_sigma_to_n}),  
\item[3.] Transformation $\ve{k} \rightarrow \ve{n}$ given by Eq.~(\ref{Transformation_k_to_n}).  
\end{enumerate}

\noindent
These transformations are of rather involved structure which inherits two problems: (i) a highly effective algorithm 
in data reduction requires a simpler solution and (ii) a simplified solution reveals which terms are of relevance 
for a given goal accuracy in the sub-\muas$\;$ domain. Therefore, in this investigation we have determined upper limits 
for each individual term in these transformations.  

Furthermore, in meanwhile it has become a well-known fact that light propagation in second post-Newtonian approximation leads to the
occurrence of so-called {\it enhanced terms}. The occurrence of {\it enhanced terms} have been recovered at the first time for 
the case of light propagation in the field of bodies at rest \cite{Article_Zschocke1,Teyssandier,AshbyBertotti2010}.  
Such {\it enhanced terms}, despite that they are of second post-Newtonian order,  
contain a large factor proportional to $r^{\,1}_A\left(s_1\right)/P_A \gg 1$, where $r^{\,1}_A\left(s_1\right)$ is the
distance between body and observer and $P_A$ is the equatorial radius of the body.
These {\it enhanced 2PN terms} are: $\ve{\rho}_6$ in (\ref{Term_rho_6_1}), $\ve{\rho}_7$ in (\ref{Term_rho_7_1}),
$\ve{\varphi}_8$ in (\ref{Appendix_Estimation_phi_8_1}), $\ve{\varphi}_9$ in (\ref{Appendix_Estimation_phi_9_1}), and
$\ve{\varphi}_{10}$ in (\ref{Appendix_Estimation_phi_10_1}). The simplified transformations contain only those terms which are  
relevant for the given threshold in light deflection of at least $1.0\,{\rm nas}$, as there are: 1PN terms,  
1.5PN terms and the just mentioned enhanced 2PN terms. These simplified transformations represent the second part of the main results of  
this investigation: 
\begin{enumerate}
\item[1.] Simplified transformation $\ve{k} \rightarrow \ve{\sigma}$ given by Eq.~(\ref{Simplified_Transformation_k_to_sigma}),
\item[2.] Simplified transformation $\ve{\sigma} \rightarrow \ve{n}$ given by Eq.~(\ref{Simplified_Transformation_sigma_to_n}),
\item[3.] Simplified transformation $\ve{k} \rightarrow \ve{n}$ given by Eq.~(\ref{Simplified_Transformation_k_to_n}).
\end{enumerate}

\noindent
The simplified transformations $\ve{k} \rightarrow \ve{\sigma}$ and $\ve{\sigma} \rightarrow \ve{n}$ are valid 
with an accuracy of $1.0\,{\rm nas}$ and $1.2\,{\rm nas}$, respectively, while  
the simplified transformation $\ve{k} \rightarrow \ve{n}$ is valid with an accuracy of at least $1.3\,{\rm nas}$. 
These statements are valid for light deflection for Sun at $45^{\circ}$ (solar aspect angle adopted from the Gaia mission) 
and all the other Solar System bodies.  

But one has to take care about the fact that higher order terms may also significantly contribute on the nas-level of accuracy.  
Therefore, the impact of possible 2.5PN and 3PN {\it enhanced terms} to order ${\cal O}\left(c^{-5}\right)$ and ${\cal O}\left(c^{-6}\right)$ 
has been considered. While it might be that 2.5PN terms are relevant, it has turned out that 3PN terms will certainly have an 
impact on the nas-scale of accuracy, 
namely about $32\,{\rm nas}$ for grazing rays at Jupiter and about $7\,{\rm nas}$ for grazing rays at Saturn. 
That means, in order to arrive at a light propagation model having an accuracy of $1\,{\rm nas}$ in angular determination, 
the 3PN solution for the light trajectory needs to be determined. For such a sophisticated calculation it would be  
sufficient to consider the case of one monopole at rest and then to take just the retarded position of the body at $s_1$ in 
order to account for the motion of the body.  
Furthermore, we have argued that {\it enhanced terms} in 3.5PN and 4PN approximation contribute certainly less than $1\,{\rm nas}$ for Sun at $45^{\circ}$ 
and all the other Solar System bodies, except for grazing rays at the Sun, where 4PN terms amount to be about $50\,{\rm nas}$ in light deflection.  

The primary aim of our investigations is to develop a fully analytical model of light propagation in the gravitational field of the 
Solar System which allows for astrometry on the sub-\muas$\;$ and even nas-level of accuracy. Before this aim is in reach, 
further aspects of the theory of light propagation are of decisive importance, for instance:  
\begin{enumerate}
\item[(a)] 1PN \cite{Zschocke1} and 1.5PN \cite{Zschocke2} light trajectory needs further to be investigated.  
\item[(b)] 2PN light trajectory in the field of $N$ moving monopoles.  
\item[(c)] 2PN effects of light propagation in the field of finite sized bodies at rest.   
\item[(d)] Enhanced terms in 2.5PN approximation in the field of one moving monopole.   
\item[(e)] Enhanced terms in 3PN approximation in the field of one monopole at rest.   
\item[(f)] Impact of the motion of source and observer.   
\end{enumerate}

\noindent
Each of these and certainly further problems, for instance light propagation in the post-Minkowskian scheme 
(which allows for astrometry in the far-zone of the Solar System),  
need to be solved before light propagation models become feasible for astrometry on the sub-\muas-level  
or nas-level of accuracy.

\section{Acknowledgment}

This work was funded by the German Research Foundation (Deutsche Forschungsgemeinschaft DFG) under grant number 263799048.  
Sincere gratitude is expressed to Prof. S.A. Klioner and Prof. M.H. Soffel for kind encouragement and enduring support.  
The author also wish to thank Dr. A.G. Butkevich, Prof. R. Sch\"utzhold, Priv.-Doz. Dr. G. Plunien, Prof. B. K\"ampfer, and Prof. L.P. Csernai  
for inspiring discussions about general theory of relativity and astrometry during recent years.

\appendix

\section{Notation}\label{Appendix0}

Throughout the investigation the following notation is in use:  

\begin{itemize}
\item $G$ is the Newtonian constant of gravitation  
\item $c$ is the vacuum speed of light in Minkowskian space-time  
\item $M_A$ is the rest mass of the body A  
\item $m_A = G\,M_A/c^2$ is the Schwarzschild radius of the body A  
\item $P_A$ denotes the equatorial radius of the body A  
\item $v_A$ denotes the orbital velocity of the body A  
\item $a_A$ denotes the orbital acceleration of the body A  
\item Theta-function: $\Theta\left(x\right) = 0$ for $x < 0$ and $\Theta\left(x\right) = 1$ for $x \ge 0$.
\item Lower case Latin indices take values $1,2,3$  
\item $\delta_{ij} = \delta^{ij} = {\rm diag}\left(+1,+1,+1\right)$ is the Kronecker delta  
\item $\varepsilon_{ijk} = \varepsilon^{ijk}$ with $\varepsilon_{123} = + 1$ is the fully anti-symmetric Levi-Civita symbol  
\item Triplet of spatial coordinates (three-vectors) are in boldface: e.g. $\ve{a}$, $\ve{b}$, $\ve{k}$, $\ve{\sigma}$, $\ve{r}_A$ 
\item Contravariant components of three-vectors: $a^{i} = \left(a^{\,1},a^2,a^3\right)$
\item Absolute value of a three-vector: $a = |\ve{a}| = \sqrt{a^{\,1}\,a^{\,1}+a^2\,a^2+a^3\,a^3}$
\item Scalar product of three-vectors: $\ve{a}\,\cdot\,\ve{b}=\delta_{ij}\,a^i\,b^j$  
\item Vector product of two three-vectors: $\left(\ve{a}\times\ve{b}\right)^i=\varepsilon_{ijk}\,a^j\,b^k$  
\item Angle $\alpha$ between three-vectors $\ve{a}$ and $\ve{b}$ is determined by  
$\displaystyle \alpha = \arccos \frac{\ve{a} \cdot \ve{b}}{\left|\ve{a}\right|\,\left|\ve{b}\right|}$
\item Lower case Greek indices take values 0,1,2,3
\item $\eta_{\alpha\beta} = \eta^{\alpha \beta} = {\rm diag}\left(-1,+1,+1,+1\right)$ is the metric tensor of flat space-time
\item $g_{\alpha\beta}$ and $g^{\alpha\beta}$ are the covariant and contravariant components of the metric tensor 
\item the signature of the metric tensor is adopted to be $\left(-,+,+,+\right)$ 
\item Contravariant components of four-vectors: $a^{\mu} = \left(a^{\,0},a^{\,1},a^2,a^3\right)$
\item Scalar product of four-vectors:
$a \cdot b = \eta_{\mu \nu}\,a^{\mu}\,b^{\mu}$ in Minkowskian metric $\eta_{\mu\nu}$
\item $f_{,\mu} = \partial_{\mu}\,f = \frac{\displaystyle \partial f}{\displaystyle \partial x^{\mu}}$ denotes partial derivative of $f$ with respect to $x^{\mu}$ 
\item $A^{\alpha}_{\,;\,\mu} = A^{\alpha}_{\,,\,\mu} + \Gamma^{\alpha}_{\mu\nu}\,A^{\nu}$ is covariant derivative of first rank tensor.
\item $B^{\alpha\beta}_{\;\;\;\;;\,\mu} = B^{\alpha\beta}_{\;\;\;\;,\,\mu} + \Gamma^{\alpha}_{\mu\nu}\,B^{\nu\beta} + \Gamma^{\beta}_{\mu\nu}\,B^{\alpha\nu}$
is covariant derivative of second rank tensor.
\item Einstein's convention is used, i.e. repeated indices are implicitly summed over   
\item $\displaystyle 1\,{\rm mas}\; ({\rm milli-arcsecond}) = \frac{\pi}{180 \times 60 \times 60}\,10^{-3}\,{\rm rad} \simeq 4.85 \times 10^{-9}\,{\rm rad}$
\item $\displaystyle 1\,\muas\; ({\rm micro-arcsecond}) = \frac{\pi}{180 \times 60 \times 60}\,10^{-6}\,{\rm rad} \simeq 4.85 \times 10^{-12}\,{\rm rad}$
\item $\displaystyle 1\,{\rm nas}\; ({\rm nano-arcsecond}) = \frac{\pi}{180 \times 60 \times 60}\,10^{-9}\,{\rm rad} \simeq 4.85 \times 10^{-15}\,{\rm rad}$
\end{itemize}

\section{The vectorial functions for light propagation in 2PN approximation}\label{Appendix2}

\subsection{The vectorial functions for the coordinate velocity of a light signal}

The vectorial functions $\ve{A}_1$, $\ve{A}_2$, $\ve{A}_3$, and $\ve{\epsilon}_1$ are given by  
\begin{eqnarray}
\fl \ve{A}_1\left(\ve{x}\right) =
- 2\,\left(\frac{\ve{\sigma} \times \left(\ve{x} \times \ve{\sigma}\right)}
{x \left(x - \ve{\sigma} \cdot \ve{x}\right)}
+ \frac{\ve{\sigma}}{x} \right),
\label{Vectorial_Function_A1}
\\
\nonumber\\
\fl \ve{A}_2\left(\ve{x},\ve{v}\right) = + \,2\,\frac{\ve{\sigma} \times \left(\ve{x} \times \ve{\sigma}\right)}
{x \left(x - \ve{\sigma} \cdot \ve{x}\right)}\,
\frac{\ve{\sigma} \cdot \ve{v}}{c}
+ \frac{4}{x}\,\frac{\ve{v}}{c}
+ 2\,\frac{\ve{\sigma} \times \left(\ve{x} \times \ve{\sigma}\right)}
{x^2}\,\frac{\ve{\sigma} \cdot \ve{v}}{c}
- 2\,\frac{\ve{\sigma}}{x^2}\,
\frac{\ve{x} \cdot \ve{v}}{c}
\nonumber\\
\nonumber\\
\fl \hspace{1.8cm} -\,2\,\frac{\ve{\sigma} \times \left(\ve{x} \times \ve{\sigma}\right)}
{x^2\,\left(x - \ve{\sigma} \cdot \ve{x}\right)}\,
\frac{\left(\ve{\sigma} \times \left(\ve{x} \times \ve{\sigma}\right) \right) \cdot \ve{v}}{c}\,,
\label{Vectorial_Function_A2}
\\
\nonumber\\
\fl \ve{A}_3\left(\ve{x}\right) = - \frac{1}{2}\,\frac{\ve{\sigma} \cdot \ve{x}}{x^4}\,\ve{x}
+ 8\,\frac{\ve{\sigma} \times \left(\ve{x} \times \ve{\sigma}\right)}{x^2\,\left(x - \ve{\sigma} \cdot\ve{x} \right)}\,
+ 4\,\frac{\ve{\sigma} \times \left(\ve{x} \times \ve{\sigma}\right)}{x\,\left(x - \ve{\sigma} \cdot \ve{x} \right)^2}\,
- 4\,\frac{\ve{\sigma}}{x\,\left(x - \ve{\sigma} \cdot \ve{x} \right)}
+ \frac{9}{2}\,\frac{\ve{\sigma}}{x^2}
\nonumber\\
\nonumber\\
\fl \hspace{1.6cm} -\,\frac{15}{4}\,\left(\ve{\sigma} \cdot \ve{x}\right)\,
\frac{\ve{\sigma} \times \left(\ve{x} \times \ve{\sigma}\right)}{x^2\,\left|\ve{\sigma} \times \ve{x}\right|^2}
- \frac{15}{4}\,\frac{\ve{\sigma} \times \left(\ve{x} \times \ve{\sigma}\right)}{\left|\ve{\sigma} \times \ve{x}\right|^3}
\left(\arctan \frac{\ve{\sigma} \cdot \ve{x}}{\left|\ve{\sigma} \times \ve{x}\right|} + \frac{\pi}{2}\right),
\label{Vectorial_Function_A3}
\end{eqnarray}

\noindent
and the vectorial function $\ve{\epsilon}_1$ is given as follows,
\begin{eqnarray}
\fl \ve{\epsilon}_1\left(\ve{x},\ve{v}\right) =
- \frac{v^2}{c^2}\,\frac{\ve{\sigma} \times \left(\ve{x} \times \ve{\sigma}\right)}{x - \ve{\sigma} \cdot \ve{x}}\,\frac{1}{x}
- 2\,\left(\frac{\ve{v} \cdot \ve{x}}{c\,x}\right)^2\,
\frac{\ve{\sigma} \times \left(\ve{x} \times \ve{\sigma}\right)}{x - \ve{\sigma} \cdot \ve{x}}\,\frac{1}{x}
\nonumber\\
\fl \hspace{0.35cm} -\, 2\,
\left(\frac{\ve{\sigma} \cdot \ve{v}}{c}\right)^2\,\frac{\ve{\sigma} \times \left(\ve{x} \times \ve{\sigma}\right)}{x - \ve{\sigma} \cdot \ve{x}}\,\frac{1}{x}
+\, 4\,\left(\frac{\ve{\sigma} \cdot \ve{v}}{c}\right) \, \left(\frac{\ve{v} \cdot \ve{x}}{c\,x}\right) \,
\frac{\ve{\sigma} \times \left(\ve{x} \times \ve{\sigma}\right)}{x - \ve{\sigma} \cdot \ve{x}}\,\frac{1}{x}
\nonumber\\
\fl  \hspace{0.35cm} +\, 4\,\frac{\ve{v}}{c}\,\left(\frac{\ve{v} \cdot \ve{x}}{c\,x}\right) \,\frac{1}{x}
- 4\,\frac{\ve{v}}{c}\,\left(\frac{\ve{\sigma} \cdot \ve{v}}{c}\right)\,\frac{1}{x}
 - \, \frac{v^2}{c^2}\,\frac{\ve{\sigma}}{x}
- 2\, \left(\frac{\ve{v} \cdot \ve{x}}{c\,x}\right)^2\,\frac{\ve{\sigma}}{x}
+ 2\, \left(\frac{\ve{\sigma} \cdot \ve{v}}{c}\right)^2\,\frac{\ve{\sigma}}{x}\,.
\label{epsilon_1}
\end{eqnarray}

\subsection{The vectorial functions for the trajectory of a light signal}

The vectorial functions for the second integration of geodesic equation (\ref{Second_Integration}) are given as follows:
\begin{eqnarray}
\fl \ve{B}_1\left(\ve{x}\right) = -\,2\, \frac{\ve{\sigma} \times \left(\ve{x} \times \ve{\sigma}\right)}{x - \ve{\sigma} \cdot \ve{x}}
+\,2\,\ve{\sigma}\,\ln \left(x - \ve{\sigma} \cdot \ve{x}\right),
\label{Vectorial_Function_C1}
\\
\nonumber\\
\fl \ve{B}^A_2\left(\ve{x},\ve{v}\right) = - 2\,\frac{\ve{v}}{c}\,\ln \left(x - \ve{\sigma} \cdot \ve{x}\right),
\label{Vectorial_Function_C2_A}
\\
\nonumber\\
\fl \ve{B}^B_2\left(\ve{x},\ve{v}\right) =
+ 2\,\frac{\ve{\sigma} \times \left(\ve{x} \times \ve{\sigma}\right)}{x - \ve{\sigma} \cdot \ve{x}}\,\frac{\ve{\sigma} \cdot \ve{v}}{c}
+ 2\,\frac{\ve{v}}{c}\,,
\label{Vectorial_Function_C2_B}
\\
\nonumber\\
\fl \ve{B}_3\left(\ve{x}\right) = + 4\,\frac{\ve{\sigma}}{x - \ve{\sigma} \cdot \ve{x}}
+\,4\,\frac{\ve{\sigma} \times \left(\ve{x} \times \ve{\sigma}\right)}{\left(x - \ve{\sigma} \cdot \ve{x}\right)^2}
+\,\frac{1}{4}\,\frac{\ve{x}}{x^2}
-\,\frac{15}{4}\,\frac{\ve{\sigma}}{\left|\ve{\sigma} \times \ve{x}\right|} \,
\arctan \frac{\ve{\sigma} \cdot \ve{x}}{\left|\ve{\sigma} \times \ve{x}\right|}
\nonumber\\
\nonumber\\
\fl \hspace{1.55cm} -\,\frac{15}{4}\,\left(\ve{\sigma} \cdot \ve{x}\right) \frac{\ve{\sigma} \times \left(\ve{x} \times \ve{\sigma}\right)}
{\left|\ve{\sigma} \times \ve{x}\right|^3} \left(\arctan \frac{\ve{\sigma} \cdot \ve{x}}{\left|\ve{\sigma} \times \ve{x}\right|}
+ \frac{\pi}{2}\right).
\label{Vectorial_Function_C3}
\end{eqnarray}

\noindent
We notice that the second term in the vectorial function $\ve{B}^B_2$ would vanish in case of $N$ bodies;  
cf.~relation (C.20) in \cite{Zschocke4}.
The vectorial function $\ve{\epsilon}_2$ with well-defined logarithms is given as follows:  
\begin{eqnarray}
\fl \ve{\epsilon}_2\left(s,s_0\right) = \ve{\epsilon}^{\rm A}_2\left(s,s_0\right) + \ve{\epsilon}^{\rm B}_2\left(s,s_0\right),
\label{epsilon_3}
\\
\nonumber\\
\fl \ve{\epsilon}^{\rm A}_2\left(s,s_0\right) = - \frac{v_A^2\left(s\right)}{c^2}\,
\frac{\ve{\sigma} \times \left(\ve{r}_A\left(s\right) \times \ve{\sigma}\right)}{r_A\left(s\right) - \ve{\sigma} \cdot \ve{r}_A\left(s\right)}
+ \frac{v_A^2\left(s_0\right)}{c^2}\,
\frac{\ve{\sigma} \times \left(\ve{r}_A\left(s_0\right) \times \ve{\sigma}\right)}{r_A\left(s_0\right) 
- \ve{\sigma} \cdot \ve{r}_A\left(s_0\right)}  
\nonumber\\
\nonumber\\
\fl \hspace{1.9cm} + \frac{v_A^2\left(s_0\right)}{c^2}\, \ve{\sigma}\,
\ln \frac{r_A\left(s\right) - \ve{\sigma} \cdot \ve{r}_A\left(s\right)}{r_A\left(s_0\right) - \ve{\sigma} \cdot \ve{r}_A\left(s_0\right)}\,, 
\label{epsilon_3a}
\\
\nonumber\\
\fl \ve{\epsilon}^{\rm B}_2\left(s,s_0\right) = + 2\,\ve{d}_A\left(s_0\right)\,\frac{\ve{\sigma} \cdot \ve{a}_A\left(s_0\right)}{c^2}\,
\,\ln \frac{r_A\left(s\right) - \ve{\sigma} \cdot \ve{r}_A\left(s\right)}{r_A\left(s_0\right) - \ve{\sigma} \cdot \ve{r}_A\left(s_0\right)}
\nonumber\\
\nonumber\\
\fl \hspace{0.5cm} + 2\,\frac{\ve{a}_A\left(s_0\right)}{c^2}
\left[r_A\left(s\right) - \ve{\sigma} \cdot \ve{r}_A\left(s\right)
- r_A\left(s_0\right) + \ve{\sigma} \cdot \ve{r}_A\left(s_0\right) \right]
\nonumber\\
\nonumber\\
\fl \hspace{0.5cm} - 2\,\frac{\ve{a}_A\left(s_0\right)}{c^2}
\left(r_A\left(s_0\right) - \ve{\sigma} \cdot \ve{r}_A\left(s_0\right) \right)
\ln \frac{r_A\left(s\right) - \ve{\sigma} \cdot \ve{r}_A\left(s\right)}{r_A\left(s_0\right) - \ve{\sigma} \cdot \ve{r}_A\left(s_0\right)}  
\nonumber\\
\nonumber\\
\fl \hspace{0.5cm} + 2\,\frac{\ve{a}_A\left(s\right)}{c^2}
\left(r_A\left(s_0\right) - \ve{\sigma} \cdot \ve{r}_A\left(s_0\right) - r_A\left(s\right) 
+ \ve{\sigma} \cdot \ve{r}_A\left(s\right) \right)
\ln \frac{r_A\left(s\right) - \ve{\sigma} \cdot \ve{r}_A\left(s\right)}{r_A\left(s_0\right) - \ve{\sigma} \cdot \ve{r}_A\left(s_0\right)}\,.
\nonumber\\
\label{epsilon_3b}
\end{eqnarray}

\noindent
As mentioned above (cf. text below Eq.~(\ref{Series_A}))  
the last term in (\ref{epsilon_3b}) is caused by the replacement of $\ve{v}_A\left(s_0\right)$ in Eq.~(128) in \cite{Zschocke4} 
by $\ve{v}_A\left(s\right)$ according to the series expansion (\ref{Series_A}) which, however, implies to account just 
for the last term in (\ref{epsilon_3b}). Of course, since $\ve{a}_A\left(s\right) = \ve{a}_A\left(s_0\right) + {\cal O}\left(c^{-1}\right)$, 
the last two terms in (\ref{epsilon_3b}) can be combined to simplify the expression (\ref{epsilon_3b}); cf. the vectorial function (\ref{Transformation_k_to_sigma_epsilon}) 
where the last two terms in (\ref{epsilon_3b}) have been combined.

\section{Some useful relations for the transformations}\label{Appendix5}

First of all, we notice two important relations between $\ve{\sigma}$ and $\ve{k}$, namely 
\begin{eqnarray}
\fl \hspace{2.0cm} \ve{\sigma} = \ve{k} - 2\,\frac{m_A}{R}\,
\left(\frac{\ve{d}^k_A\left(s_1\right)}{k \cdot r^{\,1}_A\left(s_1\right)}
- \frac{\ve{d}^k_A\left(s_0\right)}{k \cdot r_A^{\,0}\left(s_0\right)}\right)
+ {\cal O}\left(c^{-3}\right), 
\label{appendix_E_10}
\end{eqnarray}

\noindent
which is just the term in the second line in (\ref{Transformation_k_to_sigma}), and  
\begin{eqnarray}
\fl \hspace{2.0cm} \ve{\sigma} \cdot \ve{k} = 1 - \frac{1}{2}\,\frac{m_A^2}{R^2}\,
\bigg|\ve{k} \times \left( \ve{B}_1 \left(\ve{r}^{\,1}_A\left(s_1\right)\right) - \ve{B}_1 \left(\ve{r}^{\,0}_A\left(s_0\right)\right) \right)\bigg|^2
+ {\cal O}\left(c^{-5}\right),
\label{appendix_E_5}
\end{eqnarray}

\noindent
which has already been given by Eq.~(157) in \cite{Zschocke4} and which  
is needed in order to obtain the formal expression in (\ref{Transformation_k_to_sigma_5}).

According to (\ref{appendix_E_10}), up to the 1.5PN approximation there is no need to distinguish between the 
impact vectors (\ref{Impact_Vector_Sigma_s}), (\ref{Impact_Vector_Sigma_s0}) and (\ref{Impact_Vector_k0}), (\ref{Impact_Vector_k1}), 
simply because of  
$\ve{\sigma} = \ve{k} + {\cal O}\left(c^{-2}\right)$. However, beyond the 1.5PN approximation one has carefully  
to distinguish between these impact vectors. These impact vectors are related to each other,  
\begin{eqnarray}
\fl \ve{d}_A\left(s\right) = \ve{d}^k_A\left(s\right)
+ 2\,\frac{m_A}{R}\,\ve{k}\,\left(
\frac{\ve{d}^k_A\left(s\right) \cdot \ve{d}^k_A\left(s_1\right)}{k \cdot r^{\,1}_A\left(s_1\right)}
- \frac{\ve{d}^k_A\left(s\right) \cdot \ve{d}^k_A\left(s_0\right)}{k \cdot r_A^{\,0}\left(s_0\right)} \right)
\nonumber\\
\nonumber\\
\fl \hspace{2.5cm} + 2\,\frac{m_A}{R}\,\ve{k} \cdot \ve{r}_A\left(s\right)\,\left(
\frac{\ve{d}_A^k\left(s_1\right)}{k \cdot r^{\,1}_A\left(s_1\right)}
- \frac{\ve{d}_A^k\left(s_0\right)}{k \cdot r_A^{\,0}\left(s_0\right)}\right)
+ {\cal O}\left(c^{-3}\right),
\label{Relation_Impact_Vectors_1}
\end{eqnarray}

\noindent
which follows from (\ref{appendix_E_10}). 
Actually, what we need is the relation between these impact vectors for the specific case of the retarded moment of emission $s_0$
and the retarded moment of reception $s_1$ of the light signal, which is easily obtained from (\ref{Relation_Impact_Vectors_1})
just by specifying either $s=s_0$ or $s=s_1$.  
Furthermore, we notice the following relation which follows from (\ref{appendix_E_10}),  
\begin{eqnarray}
\fl \frac{1}{\sigma \cdot r_A\left(s\right)} \! = \! \frac{1}{k \cdot r_A\left(s\right)}
+ \frac{1}{R}\, \frac{2\,m_A}{\left(k \cdot r_A\left(s\right)\right)^2}
\!\left(\!\frac{\ve{d}^k_A\left(s\right) \cdot \ve{d}^k_A\left(s_1\right)}{k \cdot r^{\,1}_A\left(s_1\right)}
- \frac{\ve{d}^k_A\left(s\right) \cdot \ve{d}^k_A\left(s_0\right)}{k \cdot r_A^{\,0}\left(s_0\right)}\!\right)\! 
+ {\cal O}\left(c^{-3}\right),
\nonumber\\ 
\label{appendix_E_15}
\end{eqnarray}

\noindent
from which one may deduce the expressions for the specific cases $s=s_0$ or $s=s_1$.

\section{Parameters for massive Solar System bodies}  

In order to quantify the numerical magnitude of the upper limits  
we will use the parameters of the most massive bodies of the Solar System as presented in Table~\ref{Table1}.
\begin{table}[h!] 
\caption{\label{Table1}The numerical parameters Schwarzschild radius $m_A$, equatorial radius $P_A$, orbital velocity $v_A$, and orbital acceleration
$a_A$ of Solar System bodies \cite{JPL}.
For the distance $r^{\,1}_A\left(s_1\right)$ between massive body and observer we take the maximal possible distance, which
is computed under the assumption that the observer
is located at Lagrange point $L_2$, that is $1.5 \cdot 10^9\,{\rm m}$ from the Earth's orbit.  
For Sun at $45^{\circ}$  (solar aspect angle adopted from the Gaia mission) 
one has to replace $P_A \rightarrow \sin \left(\pi/4\right)\,r_A^{\,1}\left(s_1\right) = 0.105 \cdot 10^{12}\,{\rm m}$.}  
\footnotesize
\begin{tabular}{@{}cccccc}
\br
Object & $m_A\,[{\rm m}]$ & $P_A\,[{\rm 10^6\,m}]$ & $v_A/c$ & $a_A\,[{\rm 10^{-3}\,m/s^2}]$ & $r^{\,1}_A\left(s_1\right)\,[{\rm 10^{12}\,m}]$ \\
\mr
Sun & $1476$ & $ 696$ & $4.0 \cdot 10^{-8}$ & $ - $ & $0.149$ \\
Mercury & $ 0.245 \cdot 10^{-3}$ & $ 2.440$ & $15.8 \cdot 10^{-5}$ & $ 38.73 $ & $0.208$ \\
Venus & $ 3.615 \cdot 10^{-3} $& $ 6.052$ & $11.7 \cdot 10^{-5}$ & $ 11.34 $ & $0.258$ \\
Earth & $ 4.438 \cdot 10^{-3}$ & $ 6.378$ & $9.9 \cdot 10^{-5}$ & $ 5.93$ & $0.0015$ \\
Mars & $ 0.477 \cdot 10^{-3}$ & $ 3.396$ & $8.0 \cdot 10^{-5}$ & $ 2.55$ & $0.399$ \\
Jupiter & $1.410$ & $ 71.49$ & $4.4 \cdot 10^{-5}$ & $0.21$ & $ 0.898$ \\
Saturn & $ 0.422 $ & $ 60.27$ & $3.2 \cdot 10^{-5}$ & $ 0.06$ & $ 1.646$ \\
Uranus & $ 0.064 $ & $ 25.56$ & $2.3 \cdot 10^{-5}$ & $ 0.016$ & $ 3.142$ \\
Neptune & $ 0.076 $ & $ 24.76$ & $1.8 \cdot 10^{-5}$ & $ 0.0065$ & $4.638$ \\
\br
\end{tabular}\\
\end{table}
\normalsize

\section{The approach for the estimation of the upper limits}\label{Appendix_Estimation1} 

\subsection{Preliminary remarks} 

The transformations $\ve{k}$ to $\ve{\sigma}$ and $\ve{\sigma}$ to $\ve{n}$ were given by Eqs.~(\ref{Transformation_k_to_sigma})  
and (\ref{Transformation_sigma_to_n}), respectively, and the transformation $\ve{k}$ to $\ve{n}$ was given by Eq.~(\ref{Transformation_k_to_n}).  
These formulae are of rather involved algebraic structure and 
it is necessary to simplify these expressions by estimations of the upper limit of each individual term which allows to neglect  
all those terms which contribute less than $1\,{\rm nas}$.  
The estimation of the terms for light propagation in the gravitational field of moving bodies is considerably more  
complicated than in case of bodies at rest as presented in our article \cite{Article_Zschocke1}. This fact is mainly caused  
by the circumstance that the impact vectors do not coincide for moving bodies, $\ve{d}_A^k\left(s_0\right) \neq \ve{d}_A^k\left(s_1\right)$,  
while in case of bodies at rest the impact vector $\ve{d}_A^k$ is constant. 
In the following the approach is described, while in a subsequent \ref{Appendix_Example} an example is considered in more detail. 
\begin{figure}[!ht]
\begin{indented}
\item[]
\includegraphics[scale=0.13]{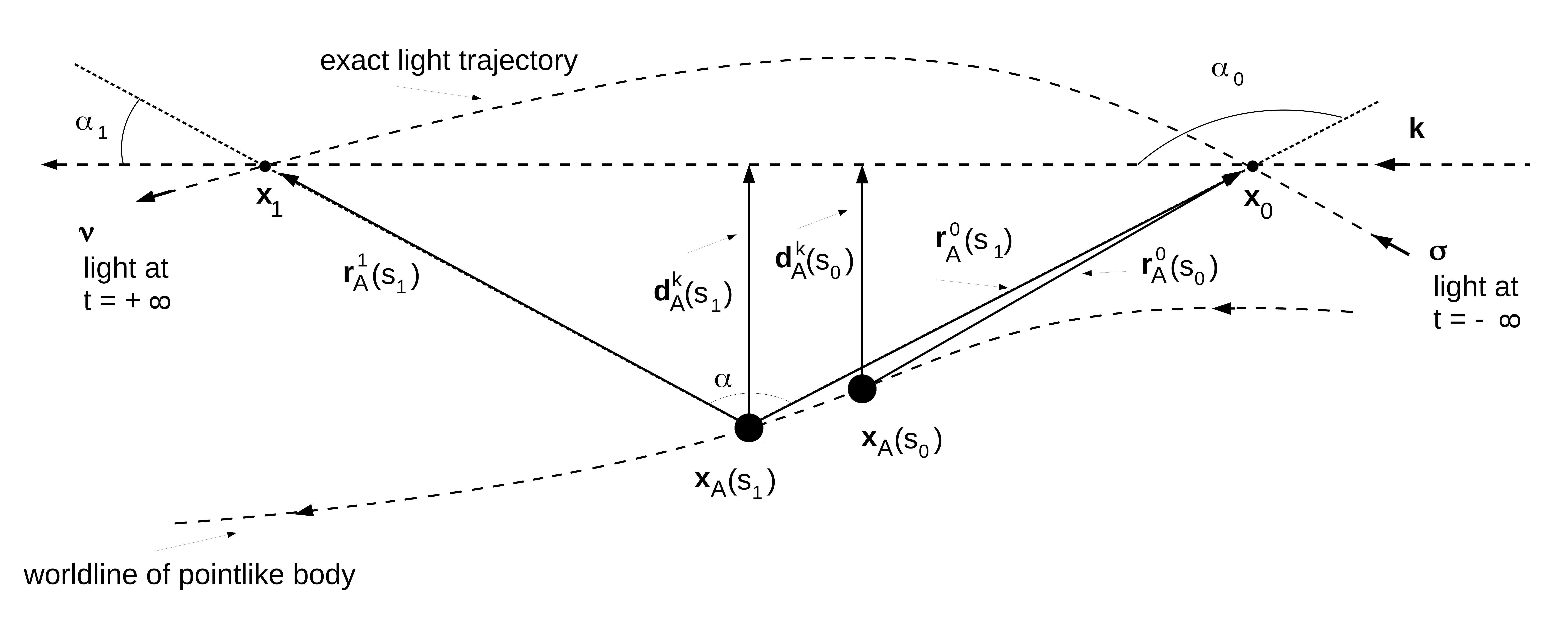}
\end{indented}
\caption{A geometrical illustration of a configuration of region A (Eq.~(\ref{Config_k1})), where the massive body is located between
the observer at $\ve{x}_1$ and the light source at $\ve{x}_0$, i.e. $\frac{\pi}{2} \le \alpha_0 \le \pi$ and
$0 \le \alpha_1 \le \frac{\pi}{2}$.}
\label{Diagram2}
\end{figure}
\begin{figure}[!ht]
\begin{indented}
\item[]
\includegraphics[scale=0.13]{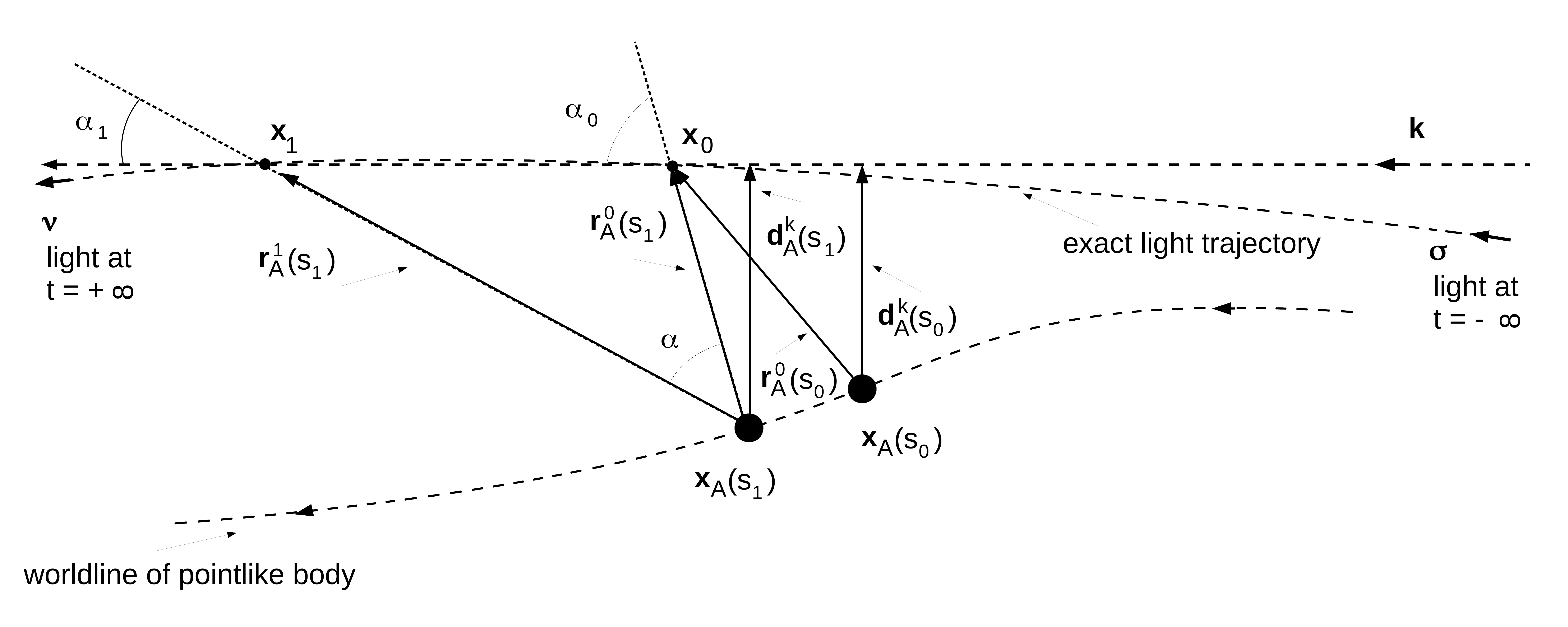}
\end{indented}
\caption{A geometrical illustration of a configuration of region B (Eq.~(\ref{Config_k2})), where the light source at $\ve{x}_0$ is located
between the massive body and observer at $\ve{x}_1$, i.e. $0 \le \alpha_0 \le \frac{\pi}{2}$ and
$0 \le \alpha_1 \le \frac{\pi}{2}$ and the condition $0 \le x \le 1$.}
\label{Diagram3}
\end{figure}
\begin{figure}[!ht]
\begin{indented}
\item[]
\includegraphics[scale=0.13]{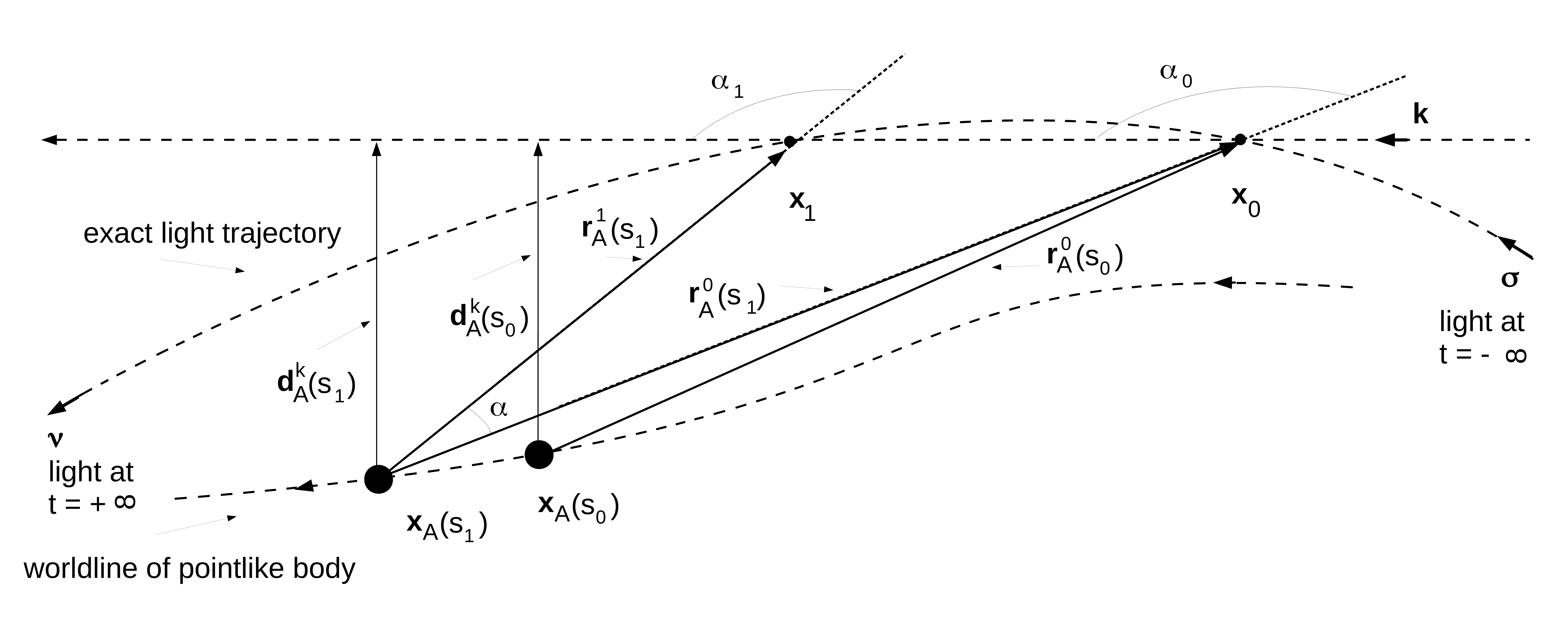}
\end{indented}
\caption{A geometrical illustration of a configuration of region C (Eq.~(\ref{Config_k3})), where the observer at $\ve{x}_1$ is located
between the massive body and light source at $\ve{x}_0$, i.e. $\frac{\pi}{2} \le \alpha_0 \le \pi$ and
$\frac{\pi}{2} \le \alpha_1 \le \pi$ and the condition $x \ge 1$.}
\label{Diagram4}
\end{figure}

\subsection{The distance vector $\ve{R}$}  

The following notation for the angles is introduced,
\begin{eqnarray}
\fl \hspace{1.0cm} 0 \le \alpha_0 = \angle \left(\ve{k},\ve{r}^{\,0}_A\left(s_1\right)\right) \le \pi \quad {\rm and} \quad
0 \le \alpha_1 = \angle \left(\ve{k},\ve{r}^{\,1}_A\left(s_1\right)\right) \le \pi\;.
\label{alpha0_alpha1}
\end{eqnarray}

\noindent 
For the estimations it is reasonable to express the distance vector $\ve{R}$ in (\ref{Boundary_3})  
in terms of $\ve{r}^{\,0}_A\left(s_1\right)$ and $\ve{r}^{\,1}_A\left(s_1\right)$ as follows,  
\begin{eqnarray}
\ve{R} = \ve{r}^{\,1}_A\left(s_1\right) - \ve{r}^{\,0}_A\left(s_1\right),  
\label{vector_R}
\end{eqnarray}

\noindent
where $\ve{r}^{\,1}_A\left(s_1\right)$ and $\ve{r}^{\,0}_A\left(s_1\right)$ are given by Eqs.~(\ref{vector_rA_1}) and (\ref{vector_rA_0_s1}),  
respectively. The three-vector $\ve{R}$ in Eq.~(\ref{Boundary_3}) is time-independent.   
The vector $\ve{R}$ in Eq.~(\ref{vector_R}) is identical to vector $\ve{R}$ in Eq.~(\ref{Boundary_3}), hence also time-independent.  
The absolute value is  
\begin{eqnarray}
R = \sqrt{\left(r^{\,0}_A\left(s_1\right)\right)^2 + \left(r^{\,1}_A\left(s_1\right)\right)^2 
- 2\,r_A^{\,0}\left(s_1\right) r^{\,1}_A\left(s_1\right)\,\cos \left(\alpha_0 - \alpha_1\right)}\;,  
\label{vector_R_series_6}
\end{eqnarray}

\noindent
where 
\begin{eqnarray}
\angle \left(\ve{r}^{\,1}_A\left(s_1\right)\,,\,\ve{r}^{\,0}_A\left(s_1\right)\right) 
= \angle \left(\ve{k}\,,\,\ve{r}^{\,0}_A\left(s_1\right)\right) - \angle \left(\ve{k}\,,\,\ve{r}^{\,1}_A\left(s_1\right)\right)  
\label{Relation_alpha} 
\end{eqnarray}

\noindent   
has been used; cf. Eq.~(69) in \cite{Article_Zschocke1} for the same angular relation in case of body at rest 
and notice that $\alpha_0 \ge \alpha_1$ in any astrometric configuration. In order to show the validity of  
the angular relation (\ref{Relation_alpha}) one should keep in mind that the usual vector operations of Euclidean space  
can be applied to three-vectors like $\ve{k}$, $\ve{r}^{\,1}_A\left(s_1\right)$, $\ve{r}^{\,0}_A\left(s_1\right)$  
\cite{Kopeikin_Efroimsky_Kaplan,Brumberg1991,Thorne,Poisson_Lecture_Notes,Poisson_Will}; 
e.g. text below Eq.~(3.1.45) in \cite{Brumberg1991}. Accordingly, relation (\ref{Relation_alpha}) asserts the following,  
\begin{eqnarray}
\arccos \frac{\ve{r}^{\,1}_A\left(s_1\right) \cdot \ve{r}^{\,0}_A\left(s_1\right)}{r^{\,1}_A\left(s_1\right)\,r^{\,0}_A\left(s_1\right)}  
\! = \! \arccos \frac{\ve{k} \cdot \ve{r}^{\,0}_A\left(s_1\right)}{r^{\,0}_A\left(s_1\right)} 
- \arccos \frac{\ve{k} \cdot \ve{r}^{\,1}_A\left(s_1\right)}{r^{\,1}_A\left(s_1\right)}\,. 
\label{Proof_Relation_alpha_1}
\end{eqnarray}
 
\noindent
Using $\arccos x - \arccos y = \arccos\left(x\,y + \sqrt{1 - x^2}\,\sqrt{1 - y^2} \right)$ 
(cf. Eq.~(4.4.33) on p. 80 in \cite{Abramowitz_Stegun}) as well as 
Eqs.~(\ref{Boundary_3}) and (\ref{vector_R}), one may demonstrate the validity of the angular relation (\ref{Relation_alpha}).  
 
Furthermore, for the estimations it is convenient to introduce the ratio 
\begin{eqnarray}
x = \frac{r_A^{\,0}\left(s_1\right)}{r^{\,1}_A\left(s_1\right)}  \quad {\rm with} \quad x \ge 0\,. 
\label{ratio_x_1}
\end{eqnarray}

\noindent 
Formally, the astrometric configurations allow all individual values for angles, that means $0 \le \alpha_0 \le \pi$ and $0 \le \alpha_1 \le \pi$ 
as already noticed in their definitions (\ref{alpha0_alpha1}). However, from (\ref{Impact_Parameter_k1}) we get 
$r^{\,0}_A\left(s_1\right)\,\sin \alpha_0 = r^{\,1}_A\left(s_1\right)\,\sin \alpha_1$, hence as soon as the parameter $x$ 
in (\ref{ratio_x_1}) is fixed, the combinations of these both angles are not arbitrary anymore,  
but restricted by the relation (note that $\sin \alpha_0 \ge 0$ as well as $\sin \alpha_1 \ge 0$)  
\begin{eqnarray}
x = \frac{\sin \alpha_1}{\sin \alpha_0}\,.  
\label{ratio_x_2} 
\end{eqnarray}

\noindent
So, relation (\ref{ratio_x_1}) is the exact definition of parameter $x$, while (\ref{ratio_x_2}) follows from (\ref{Impact_Parameter_k1}).  
Finally, from (\ref{ratio_x_2}) we deduce  
\begin{equation}
\fl \begin{array}[c]{l}
\displaystyle
\alpha_0 = \left \{ \begin{array}[c]{l}
\displaystyle
\pi - \arcsin\left(\frac{\sin \alpha_1}{x}\right)   
\quad \mbox{for} \quad \frac{\pi}{2} \le \alpha_0 \le \pi \quad {\rm and} \quad x \ge 0 \\
\\
\displaystyle
\arcsin\left(\frac{\sin \alpha_1}{x}\right)   
\quad \mbox{for} \quad 0 \le \alpha_0 \le \frac{\pi}{2} \quad {\rm and} \quad 1 \ge x \ge 0  
\end{array} \right \} \;, 
\end{array}
\label{ratio_x_3}
\end{equation}

\noindent
while the region $0 \le \alpha_0 \le \pi/2$ and $x > 1$ is not possible, because in such configurations the light signal 
would not be received by an observer; see also comment below Eqs.~(\ref{Config_k1}) - (\ref{Config_k3}).  
The relations in (\ref{ratio_x_3}) are needed if one evaluates the 
term $\cos \left(\alpha_0 - \alpha_1\right)$ in Eq.~(\ref{vector_R_series_6}) for the distance $R$.
In the estimations, the variables $x$ and $\alpha_1$ are considered as independent of each other, but restricted 
by the possible configurations as defined in Eqs.~(\ref{Config_k1}) - (\ref{Config_k3}). 
That means, in region A and C, as defined below by Eqs.~(\ref{Config_k1}) and (\ref{Config_k3}), one has to use the 
first line of (\ref{ratio_x_3}), while in region B, as defined below by Eq.~(\ref{Config_k2}), one has to 
use the relation in the second line of (\ref{ratio_x_3}). We also notice that (\ref{ratio_x_2}) implies   
$\displaystyle 0 \le \frac{\sin \alpha_1}{x} \le 1$, so that the relations in (\ref{ratio_x_3}) are uniquely defined.

\subsection{The possible configurations} 

The approach for the estimation of the upper limit of each individual term is the following. 
We separate all possible configurations into three angular areas,   
\begin{eqnarray}
{\rm A:}\; \frac{\pi}{2} \le \alpha_0 \le \pi \;,\; 0 \le \alpha_1 \le \frac{\pi}{2} \;: \quad  
d^k_A\left(s_1\right) \ge P_A\;, 
x \ge 0\,,  
\label{Config_k1}
\\
{\rm B:}\;\; 0 \le \alpha_0 < \frac{\pi}{2} \;,\; 0 \le \alpha_1 \le \frac{\pi}{2} \;: \quad  
d^k_A\left(s_1\right) \ge P_A\;,\; 1 \ge x \ge 0\,,  
\label{Config_k2}
\\
{\rm C:}\; \frac{\pi}{2} \le \alpha_0 \le \pi \;,\; \frac{\pi}{2} < \alpha_1 \le \pi \;: \quad  
d^k_A\left(s_1\right) \ge 0\;\; \;,\; x \ge 1\,.   
\label{Config_k3}
\end{eqnarray}

\noindent 
A graphical representation of a typical configuration belonging to region A, B, and C, is given by the 
Figures~\ref{Diagram2}, \ref{Diagram3}, and \ref{Diagram4}, respectively.  
The constraints for the impact parameter were given by Eqs.~(\ref{Impact_Vector_k_Constraint_1}) and (\ref{Impact_Vector_k_Constraint_2}), while  
the constraints $x \le 1$ in (\ref{Config_k2}) and $x \ge 1$ in (\ref{Config_k3}) are necessary because otherwise the light signal cannot  
be received by the observer.

\subsection{The approach for the estimations}\label{Approach_Appendix}  

The determination of the upper limit of each individual term in the transformations  
$\ve{k}$ to $\ve{\sigma}$ in (\ref{Transformation_k_to_sigma}) and $\ve{\sigma}$ to $\ve{n}$ in  
(\ref{Transformation_sigma_to_n}) proceeds as follows:  
\begin{enumerate}
\item[1.] Series expansion of the individual expression in terms of $s_1$,  
\item[2.] Inserting (\ref{vector_R_series_6}) for the absolute value $R$ of the distance vector,  
\item[3.] Rewriting the expression in terms of the variables $x$ (\ref{ratio_x_1}) and $\alpha_0, \alpha_1$ (\ref{alpha0_alpha1}),  
\item[4.] Using relations (\ref{ratio_x_3}) in line with the regions (\ref{Config_k1}) - (\ref{Config_k3}), 
\item[5.] Estimation of the term for each possible region separately.  
\end{enumerate}

\section{An example: the estimation of $\rho_1$}\label{Appendix_Example}

The estimation of the upper limit of each individual term implies some algebraic effort. 
So an example is considered in some more detail, which comprehensively elucidates the basic steps about how the approach runs.  
Accordingly, we shall consider the determination of the upper  
limit of the term in the second line of (\ref{Transformation_k_to_sigma}), which reads  
\begin{eqnarray}
\ve{\rho}_1\left(s_1,s_0\right) = - 2\,\frac{m_A}{R}
\left(\frac{\ve{d}^k_A\left(s_1\right)}{k \cdot r^{\,1}_A\left(s_1\right)} 
- \frac{\ve{d}^k_A\left(s_0\right)}{k \cdot r_A^{\,0}\left(s_0\right)}\right). 
\label{Example_rho_1_1}
\end{eqnarray}

\noindent
In what follows an upper limit of this expression will be given by means of the approach as just described in the previous section.

\subsection{Series expansion of $\ve{\rho}_1$} 

For the impact vector $\ve{d}^k_A\left(s_0\right)$ in (\ref{Example_rho_1_1}) the 
series expansion (\ref{Impact_Vector_Relation1}) is used. For the four-scaler product  
$k \cdot r_A^{\,0}\left(s_0\right) = - \left(r^{\,0}_A\left(s_0\right) - \ve{k} \cdot \ve{r}^{\,0}_A\left(s_0\right)\right)$ 
in (\ref{Example_rho_1_1}) the series expansions  
\begin{eqnarray}
\fl \ve{r}^{\,0}_A\left(s_0\right) = \ve{r}^{\,0}_A\left(s_1\right) + \frac{\ve{v}_A\left(s_1\right)}{c}\,
\bigg(r_A^{\,0}\left(s_1\right) - \ve{k} \cdot \ve{r}^{\,0}_A\left(s_1\right)
- r^{\,1}_A\left(s_1\right) + \ve{k} \cdot \ve{r}^{\,1}_A\left(s_1\right)\bigg)  
\nonumber\\
\fl \hspace{1.5cm} + \,{\cal O}\left(c^{-2}\right),  
\label{series_expansion_r0}
\\
\nonumber\\
\fl r^{\,0}_A\left(s_0\right) = r^{\,0}_A\left(s_1\right) 
+ \frac{\ve{r}^{\,0}_A\left(s_1\right)}{r^{\,0}_A\left(s_1\right)} \cdot \frac{\ve{v}_A\left(s_1\right)}{c} 
\bigg(r_A^{\,0}\left(s_1\right) - \ve{k} \cdot \ve{r}^{\,0}_A\left(s_1\right)
- r^{\,1}_A\left(s_1\right) + \ve{k} \cdot \ve{r}^{\,1}_A\left(s_1\right)\bigg) 
\nonumber\\
\fl \hspace{1.5cm} + \,{\cal O}\left(c^{-2}\right),  
\label{series_expansion_r0_B}
\end{eqnarray}

\noindent
are applied, which follow by inserting the expansion (\ref{series_expansion_body_1}) into the definition (\ref{vector_rA_0})   
by keeping in mind relation (\ref{relation_k_2}); recall that  
$r^{\,0}_A\left(s_0\right) = \left|\ve{r}^{\,0}_A\left(s_0\right)\right|$ and 
$r^{\,0}_A\left(s_1\right) = \left|\ve{r}^{\,0}_A\left(s_1\right)\right|$ in (\ref{series_expansion_r0_B}) are  
absolute values of three-vectors. Using these relations, one obtains the following expansion for the term  
$\ve{\rho}_1$ in (\ref{Example_rho_1_1}),  
\begin{eqnarray}
\ve{\rho}_1\left(s_1,s_0\right) = \ve{\rho}_1\left(s_1,s_1\right) + \Delta \ve{\rho}_1\left(s_1,s_1\right) + {\cal O}\left(c^{-5}\right),  
\label{series_expansion_rho_1}
\end{eqnarray}

\noindent
where 
\begin{eqnarray}
\ve{\rho}_1\left(s_1,s_1\right) = - 2\,\frac{m_A}{R}
\left(\frac{\ve{d}^k_A\left(s_1\right)}{k \cdot r^{\,1}_A\left(s_1\right)}
- \frac{\ve{d}^k_A\left(s_1\right)}{k \cdot r_A^{\,0}\left(s_1\right)}\right), 
\label{series_expansion_rho_1_A}
\\
\nonumber\\
\fl \Delta \ve{\rho}_1\left(s_1,s_1\right) = + 2\,\frac{m_A}{R}  
\frac{\ve{d}^k_A\left(s_1\right)}{k \cdot r^{\,0}_A\left(s_1\right)}
\left(\ve{k} - \frac{\ve{r}^{\,0}_A\left(s_1\right)}{r^{\,0}_A\left(s_1\right)}\right)  
\cdot \frac{\ve{v}_A\left(s_1\right)}{c}\; 
\frac{k \cdot r^{\,0}_A\left(s_1\right) - k \cdot r^{\,1}_A\left(s_1\right)}{k \cdot r^{\,0}_A\left(s_1\right)}
\nonumber\\
\nonumber\\
\fl \hspace{2.25cm} - 2\,\frac{m_A}{R}\; \ve{k} \times \left(\frac{\ve{v}_A\left(s_1\right)}{c} \times \ve{k}\right)
\frac{k \cdot r^{\,0}_A\left(s_1\right) - k \cdot r^{\,1}_A\left(s_1\right)}{k \cdot r^{\,0}_A\left(s_1\right)}
\nonumber\\
\nonumber\\
\fl \hspace{2.25cm} + {\cal O}\left(\frac{v_A^2\left(s_1\right)}{c^2}\right)  
+ {\cal O}\left(\frac{a_A\left(s_1\right)}{c^2}\right),  
\label{series_expansion_rho_1_B}
\end{eqnarray}

\noindent
where the absolute value $R$ is still given by the exact expression (\ref{Boundary_3}).  
In what follows we show that  
\begin{eqnarray}
\fl \hspace{1.0cm} \rho_1\left(s_1,s_1\right) \le 4\,\frac{m_A}{P_A}\,,
\label{Estimation_rho1}
\\
\nonumber\\
\fl \hspace{0.75cm} \Delta \rho_1\left(s_1,s_1\right) \le 6\,\frac{m_A}{r^{\,1}_A\left(s_1\right)}\,\frac{v_A\left(s_1\right)}{c}
+ {\cal O}\left(\frac{v_A^2\left(s_1\right)}{c^2}\right) + {\cal O}\left(\frac{a_A\left(s_1\right)}{c^2}\right), 
\label{Example_rho1_Delta}
\end{eqnarray}

\noindent
for any kind of astrometric configuration. Because of  
\begin{eqnarray}
\Delta \rho_1\left(s_1,s_1\right) \le 1\,{\rm nas}\,,  
\label{Example_rho_1_3}
\end{eqnarray}

\noindent
for all Solar System bodies, only the term (\ref{series_expansion_rho_1_A}) is taken into  
account in the simplified transformation (\ref{Simplified_Transformation_k_to_sigma}).

First of all, we continue the exemplifying considerations with the expression (\ref{series_expansion_rho_1_A}), while the estimation of   
the term (\ref{series_expansion_rho_1_B}) proceeds in similar manner and is considered afterwards.

\subsection{Estimation of (\ref{series_expansion_rho_1_A})} 

\subsubsection{\underline{Region A: $\pi/2 \le \alpha_0 \le \pi \;,\; 0 \le \alpha_1 \le \pi/2$:}} 

Using the relations  
\begin{eqnarray}
\fl \frac{1}{k \cdot r^{\,1}_A\left(s_1\right)} 
= - \frac{r^{\,1}_A\left(s_1\right) + \ve{k} \cdot \ve{r}^{\,1}_A\left(s_1\right)}{\left(d^k_A\left(s_1\right)\right)^2} \;\; {\rm and} \;\;
\frac{1}{k \cdot r^{\,0}_A\left(s_1\right)} 
= - \frac{r^{\,0}_A\left(s_1\right) + \ve{k} \cdot \ve{r}^{\,0}_A\left(s_1\right)}{\left(d^k_A\left(s_1\right)\right)^2}\,, 
\label{Example_rho_1_4}
\end{eqnarray}

\noindent
we get for the absolute value of (\ref{series_expansion_rho_1_A}),  
\begin{eqnarray}
\fl \rho_1\left(s_1,s_1\right) = 2\,\frac{m_A}{d^k_A\left(s_1\right)}\,
\left|\frac{r^{\,1}_A\left(s_1\right) + \ve{k} \cdot \ve{r}^{\,1}_A\left(s_1\right) 
- r^{\,0}_A\left(s_1\right) - \ve{k} \cdot \ve{r}^{\,0}_A\left(s_1\right)}{R}\right|\,.  
\label{Example_rho_1_6}
\end{eqnarray}

\noindent
Now we insert for the absolute value $R$ the expression (\ref{vector_R_series_6}), and then we can  
rewrite (\ref{Example_rho_1_6}) in terms of the variables $x$ (\ref{ratio_x_1}) as well as 
the angles $\alpha_0$, $\alpha_1$ (\ref{alpha0_alpha1}) as follows,
\begin{eqnarray}
\fl \hspace{1.0cm} \rho_1\left(s_1,s_1\right) = 2\,\frac{m_A}{d^k_A\left(s_1\right)}\,
\left|\frac{1 + \cos \alpha_1 - x - x\,\cos \alpha_0}{\sqrt{1 + x^2 - 2\,x\,\cos \left(\alpha_0 - \alpha_1\right)}}\right|\,.  
\label{Example_rho_1_7_A}
\end{eqnarray}

\noindent
Keeping in mind that in region A the first line of the angular relations (\ref{ratio_x_3}) is valid, we find that  
(\ref{Example_rho_1_7_A}) depends on two variables only, namely $x$ and $\alpha_1$.  
One may demonstrate with the aid of the computer algebra system {\it Maple} \cite{Maple} that  
\begin{eqnarray}
\fl f_1 = \left|\frac{1 + \cos \alpha_1 - x - x\,\cos \alpha_0}{\sqrt{1 + x^2 - 2\,x\,\cos \left(\alpha_0 - \alpha_1\right)}}\right| 
\le 2\,,\quad {\rm for} \quad 0 \le \alpha_1 \le \frac{\pi}{2} \quad {\rm and} \quad x \ge 0\,.   
\label{Term_rho_1_6_B}
\end{eqnarray}

\noindent
Inserting (\ref{Term_rho_1_6_B}) into (\ref{Example_rho_1_7_A}) yields for the upper limit 
\begin{eqnarray}
\rho_1\left(s_1,s_1\right) \le 4\,\frac{m_A}{d^k_A\left(s_1\right)}\,,  
\label{Example_rho_1_7}
\end{eqnarray}
 
\noindent
which validates the upper limit (\ref{Estimation_rho1}) for region A, because $d^k_A\left(s_1\right) \ge P_A$ in region A.

\subsubsection{\underline{Region B: $0 \le \alpha_0 \le \pi/2 \;,\; 0 \le \alpha_1 \le \pi/2$:}} 

The same steps as in the previous Section yield the same result as given by Eq.~(\ref{Example_rho_1_7_A}). 
Keeping in mind that in region B the second line of the angular relations (\ref{ratio_x_3}) is valid, one may show that  
the inequality (\ref{Term_rho_1_6_B}) is also valid for region B, 
\begin{eqnarray}
\fl f_1 = \left|\frac{1 + \cos \alpha_1 - x - x\,\cos \alpha_0}{\sqrt{1 + x^2 - 2\,x\,\cos \left(\alpha_0 - \alpha_1\right)}}\right|
\le 2 \quad {\rm for} \quad 0 \le \alpha_1 \le \frac{\pi}{2} \quad {\rm and} \quad 1 \ge x \ge 0\,. 
\label{Term_rho_1_6_C}
\end{eqnarray}

\noindent 
Hence, one obtains that (\ref{Example_rho_1_7}) is also valid in region B,  
\begin{eqnarray}
\rho_1\left(s_1,s_1\right) \le 4\,\frac{m_A}{d^k_A\left(s_1\right)}\,,  
\label{Example_rho_1_8}
\end{eqnarray}

\noindent
which confirms the validity of the upper limit (\ref{Estimation_rho1}) for region B, because $d^k_A\left(s_1\right) \ge P_A$ in region B.

\subsubsection{\underline{Region C: $\pi/2 \le \alpha_0 \le \pi \;,\; \pi/2 \le \alpha_1 \le \pi$:}} 

In this angular region the impact parameter $d^k_A\left(s_1\right)$ can be arbitrarily small. Therefore, an estimation  
for the expression in the first line of (\ref{series_expansion_rho_1}) is only meaningful if $d^k_A\left(s_1\right)$ does  
not appear in the denominator. But due to $r_A\left(s_1\right) \gg P_A$, we may get an upper limit 
where $r^{\,1}_A\left(s_1\right)$ appears in the denominator rather than $d^k_A\left(s_1\right)$.  
Hence, we reshape identically the expression in (\ref{Example_rho_1_6}), which is also valid for region C, as follows,  
\begin{eqnarray}
\fl \rho_1\left(s_1,s_1\right) = 2\,\frac{m_A}{r^{\,1}_A\left(s_1\right)}\,\frac{r^{\,1}_A\left(s_1\right)}{d^k_A\left(s_1\right)}\,
\left|\frac{r^{\,1}_A\left(s_1\right) + \ve{k} \cdot \ve{r}^{\,1}_A\left(s_1\right)
- r^{\,0}_A\left(s_1\right) - \ve{k} \cdot \ve{r}^{\,0}_A\left(s_1\right)}{R}\right|\,. 
\nonumber\\ 
\label{Example_rho_1_9}
\end{eqnarray}

\noindent 
Inserting the expression (\ref{vector_R_series_6}) for the distance $R$ and using the notation (\ref{alpha0_alpha1}) 
for the angles $\alpha_0,\alpha_1$, one obtains  
\begin{eqnarray}
\fl \rho_1\left(s_1,s_1\right) = 2\,\frac{m_A}{r^{\,1}_A\left(s_1\right)}\,\left|\frac{1}{\sin \alpha_1}\,
\frac{1 + \cos \alpha_1 - x - x\,\cos\alpha_0}{\sqrt{1 + x^2 - 2\,x\,\cos\left(\alpha_0 - \alpha_1\right)}} \right|\,.  
\label{Example_rho_1_10}
\end{eqnarray}

\noindent
Keeping in mind that in region C the first line of the angular relations (\ref{ratio_x_3}) is valid, 
relation (\ref{Example_rho_1_10}) depends on two variables only, namely $x$ and $\alpha_1$. Then, one may show that  
\begin{eqnarray}
\fl f_2 = \left|\frac{1}{\sin \alpha_1}\, 
\frac{1 + \cos \alpha_1 - x - x\,\cos\alpha_0}{\sqrt{1 + x^2 - 2\,x\,\cos\left(\alpha_0 - \alpha_1\right)}} \right| \le 1  
\quad {\rm for} \quad \frac{\pi}{2} \le \alpha_1 \le \pi \quad {\rm and} \quad x \ge 1\,,  
\nonumber\\ 
\label{Example_rho_1_11}
\end{eqnarray}

\noindent
which can demonstrated with the aid of the computer algebra system {\it Maple} \cite{Maple}.  
Hence, by inserting (\ref{Example_rho_1_11}) into (\ref{Example_rho_1_10}) we get  
\begin{eqnarray}
\rho_1\left(s_1,s_1\right) \le \frac{2\,m_A}{r^{\,1}_A\left(s_1\right)}\,, 
\label{Example_rho_1_12}
\end{eqnarray}

\noindent 
which also confirms (\ref{Estimation_rho1}) because $r^{\,1}_A\left(s_1\right) \gg P_A$.  
The upper limits (\ref{Example_rho_1_7}), (\ref{Example_rho_1_8}), and (\ref{Example_rho_1_12}) confirm the  
estimation given by Eq.~(\ref{Estimation_rho1}) for any astrometric configuration.

\subsection{Estimation of (\ref{series_expansion_rho_1_B})} 

The expression (\ref{series_expansion_rho_1_B}) is separated into two pieces, 
\begin{eqnarray}
\fl \Delta \ve{\rho}_1\left(s_1,s_1\right) = \Delta \ve{\rho}^A_1\left(s_1,s_1\right) + \Delta \ve{\rho}^B_1\left(s_1,s_1\right)  
+ {\cal O}\left(\frac{v_A^2\left(s_1\right)}{c^2}\right) + {\cal O}\left(\frac{a_A\left(s_1\right)}{c^2}\right), 
\label{series_expansion_rho_1_B_0}
\end{eqnarray}

\noindent 
where 
\begin{eqnarray}
\fl \Delta \ve{\rho}^A_1\left(s_1,s_1\right) = + 2\,\frac{m_A}{R}
\frac{\ve{d}^k_A\left(s_1\right)}{k \cdot r^{\,0}_A\left(s_1\right)}
\left(\ve{k} - \frac{\ve{r}^{\,0}_A\left(s_1\right)}{r^{\,0}_A\left(s_1\right)}\right) \cdot \frac{\ve{v}_A\left(s_1\right)}{c}\;
\frac{k \cdot r^{\,0}_A\left(s_1\right) - k \cdot r^{\,1}_A\left(s_1\right)}{k \cdot r^{\,0}_A\left(s_1\right)}\,, 
\nonumber\\ 
\label{series_expansion_rho_1_B_1}
\\
\fl \Delta \ve{\rho}^B_1\left(s_1,s_1\right) = - 2\,\frac{m_A}{R}\; \ve{k} \times \left(\frac{\ve{v}_A\left(s_1\right)}{c} \times \ve{k}\right)
\frac{k \cdot r^{\,0}_A\left(s_1\right) - k \cdot r^{\,1}_A\left(s_1\right)}{k \cdot r^{\,0}_A\left(s_1\right)}\;.
\label{series_expansion_rho_1_B_2}
\end{eqnarray}

\noindent
The estimation of these terms proceeds in the same way as the estimation of (\ref{series_expansion_rho_1_A}). 
Inserting the expression (\ref{vector_R_series_6}) for the distance $R$ and using the notation (\ref{alpha0_alpha1})
for the angles $\alpha_0,\alpha_1$, one obtains  
\begin{eqnarray}
\fl \Delta \ve{\rho}^A_1\left(s_1,s_1\right) = 2\,\frac{m_A}{r^{\,1}_A\left(s_1\right)}\,\frac{v_A\left(s_1\right)}{c}\,
\left|\frac{\sqrt{2}}{x^2}\,\frac{1 - \cos \alpha_1 - x + x\,\cos\alpha_0}{\sqrt{1 + x^2 - 2\,x\,\cos \left(\alpha_0 - \alpha_1\right)}}\,
\frac{\sin \alpha_1}{\left(1 - \cos \alpha_0\right)^{3/2}}\right|, 
\nonumber\\
\label{Term_rho_10_1_A}
\\
\fl \Delta \ve{\rho}^B_1\left(s_1,s_1\right) = 2\,\frac{m_A}{r^{\,1}_A\left(s_1\right)}\,\frac{v_A\left(s_1\right)}{c}\,
\left|\frac{1}{x}\,\frac{1 - \cos \alpha_1 - x + x\,\cos\alpha_0}{\sqrt{1 + x^2 - 2\,x\,\cos \left(\alpha_0 - \alpha_1\right)}}\,
\frac{1}{1 - \cos \alpha_0}\right|.
\label{Term_rho_11_1_A}
\end{eqnarray}

\noindent
where in (\ref{Term_rho_10_1_A}) we have used 
$\left|\ve{k} - \ve{r}^{\,0}_A\left(s_1\right)/r^{\,0}_A\left(s_1\right)\right| = \sqrt{2\,\left(1 - \cos \alpha_0\right)}$.  
For each region A, B, and C one obtains the following inequality,
\begin{eqnarray}
\fl \hspace{1.0cm} f_3 = \left|\frac{\sqrt{2}}{x^2}\,
\frac{1 - \cos \alpha_1 - x + x\,\cos\alpha_0}{\sqrt{1 + x^2 - 2\,x\,\cos \left(\alpha_0 - \alpha_1\right)}}\,
\frac{\sin \alpha_1}{\left(1 - \cos \alpha_0\right)^{3/2}}\right| \le 2\,, 
\label{Term_rho_10_1_B}
\\
\nonumber\\ 
\fl \hspace{1.0cm} f_4 = \left|\frac{1}{x}\,
\frac{1 - \cos \alpha_1 - x + x\,\cos\alpha_0}{\sqrt{1 + x^2 - 2\,x\,\cos \left(\alpha_0 - \alpha_1\right)}}\,
\frac{1}{1 - \cos \alpha_0}\right| \le 1\,.
\label{Term_rho_11_1_B}
\end{eqnarray}

\noindent
Hence, inserting (\ref{Term_rho_10_1_B}) into (\ref{Term_rho_10_1_A}) and (\ref{Term_rho_11_1_B}) into (\ref{Term_rho_11_1_A})  
yields for the upper limit of (\ref{series_expansion_rho_1_B_0})  
\begin{eqnarray}
\Delta \rho_1\left(s_1,s_1\right) \le 6\,\frac{m_A}{r^{\,1}_A\left(s_1\right)}\,\frac{v_A\left(s_1\right)}{c}\,,
\label{Term_rho_10_2}
\end{eqnarray}

\noindent
which is less than $1\,{\rm nas}$ for any Solar System body, as already stated in Eq.~(\ref{Example_rho_1_3}).

The calculation of the remaining terms of order ${\cal O}\left(v_A^2/c^2\right)$ and ${\cal O}\left(a_A/c^2\right)$ 
in (\ref{series_expansion_rho_1_B_0}) involves a considerable algebraic effort which we believe cannot be of much interest.  
To present all these detailed calculations explicitly here would be disadvantageous to the clarity. So we are obliged to confine 
ourselves here by the statement that these terms turn out to be even smaller than the terms (\ref{series_expansion_rho_1_B_1})  
and (\ref{series_expansion_rho_1_B_2}) and will, also for this reason, not be presented here in their explicit form.

\section{Estimation of the terms in the transformation $\ve{k}$ to $\ve{\sigma}$}\label{Appendix_EstimationA} 

\subsection{Estimation of $\rho_2$}\label{Estimation_rho2}  

The term in the third line of (\ref{Transformation_k_to_sigma}) is denoted as $\ve{\rho}_2$ and reads  
\begin{eqnarray}
\ve{\rho}_2\left(s_1,s_0\right) = 2\,\frac{m_A}{R}\,\ve{k} \times \left(\frac{\ve{v}_A\left(s_1\right)}{c} \times \ve{k}\right)
\ln \frac{k \cdot r^{\,1}_A\left(s_1\right)}{k \cdot r_A^{\,0}\left(s_0\right)} 
\label{Term_rho_2_1_A}
\\
\nonumber\\
\hspace{1.5cm} = \ve{\rho}_2\left(s_1,s_1\right) + \Delta \ve{\rho}_2\left(s_1,s_1\right) + {\cal O}\left(c^{-5}\right),  
\label{Term_rho_2_1_B}
\end{eqnarray}

\noindent
where  
\begin{eqnarray}
\fl \hspace{0.25cm} \ve{\rho}_2\left(s_1,s_1\right) = 2\,\frac{m_A}{R}\,\ve{k} \times \left(\frac{\ve{v}_A\left(s_1\right)}{c} \times \ve{k}\right)
\ln \frac{k \cdot r^{\,1}_A\left(s_1\right)}{k \cdot r_A^{\,0}\left(s_1\right)}\,, 
\label{Term_rho_2_1_C}
\\
\nonumber\\ 
\fl \Delta \ve{\rho}_2\left(s_1,s_1\right)  
\!=\! 2 \frac{m_A}{R} \ve{k} \times \left(\!\frac{\ve{v}_A\left(s_1\right)}{c} \times \ve{k}\!\right)\!\! 
\left(\!\ve{k} - \frac{\ve{r}^0_A\left(s_1\right)}{r^{\,0}_A\left(s_1\right)}\!\right) \cdot \frac{\ve{v}_A\left(s_1\right)}{c}\,   
\frac{k \cdot r^{\,0}_A\left(s_1\right) - k \cdot r^{\,1}_A\left(s_1\right)}{k \cdot r^{\,0}_A\left(s_1\right)}\,.  
\nonumber\\
\label{Term_rho_2_1_D}
\end{eqnarray}

\noindent
The upper limits of the absolute values are given by  
\begin{eqnarray}
\rho_2\left(s_1,s_1\right) \le 2\,\frac{m_A}{\sqrt{P_A\,r^{\,0}_A\left(s_1\right)}}\,\frac{v_A\left(s_1\right)}{c} 
\le 2\,\frac{m_A}{P_A}\,\frac{v_A\left(s_1\right)}{c}\,,  
\label{Term_rho_2_2}
\\
\nonumber\\
\Delta \rho_2\left(s_1,s_1\right) \le 4\,\frac{m_A}{P_A}\,\frac{v^2_A\left(s_1\right)}{c^2} \ll 1\,{\rm nas}\,. 
\label{Term_rho_2_1_E}
\end{eqnarray}

\noindent 
For the second estimation in (\ref{Term_rho_2_2}) we have taken into account that $r^{\,0}_A\left(s_1\right) \simeq P_A$ is quite possible, for  
instance by a moon orbiting around a planet.  
Hence, for all Solar System bodies only the term (\ref{Term_rho_2_1_C})  
is taken into account in the simplified transformation (\ref{Simplified_Transformation_k_to_sigma}).

\subsection{Estimation of $\rho_3$}

The term in the fourth line of (\ref{Transformation_k_to_sigma}) is denoted as $\ve{\rho}_3$ and reads 
\begin{eqnarray}
\ve{\rho}_3\left(s_1,s_0\right) =  - 2\,\frac{m_A}{R}\,
\left( \ve{k} \times \left(\frac{\ve{v}_A\left(s_1\right)}{c} - \frac{\ve{v}_A\left(s_0\right)}{c}\right) \times \ve{k} \right)  
\label{Term_rho_3_1}
\\
\nonumber\\ 
\hspace{1.6cm} = \ve{\rho}_3\left(s_1,s_1\right) + \Delta \ve{\rho}_3\left(s_1,s_1\right) + {\cal O}\left(c^{-5}\right),  
\label{Term_rho_3_2}
\end{eqnarray} 

\noindent
where 
\begin{eqnarray}
\fl \ve{\rho}_3\left(s_1,s_1\right) = 0\,,
\label{Term_rho_3_3_A}
\\
\fl \Delta \ve{\rho}_3\left(s_1,s_1\right) = - 2\,m_A\,\ve{k} \times \left(\frac{\ve{a}_A\left(s_1\right)}{c^2} \times \ve{k}\right) 
\frac{k \cdot r^{\,0}_A\left(s_1\right) - k \cdot r^{\,1}_A\left(s_1\right)}{R}\,.  
\label{Term_rho_3_3}
\end{eqnarray} 

\noindent
The upper limit of the absolute value $\rho_3$ is given by
\begin{eqnarray}
\Delta \rho_3\left(s_1,s_1\right) \le 4\,m_A\,\frac{a_A\left(s_1\right)}{c^2} \ll 1\,{\rm nas}\,,  
\label{Term_rho_3_4}
\end{eqnarray} 

\noindent
hence the term $\ve{\rho}_3$ is not taken into account in the simplified transformation (\ref{Simplified_Transformation_k_to_sigma}).  
As already mentioned above (see text below Eq.~(\ref{Vectorial_Function_C3})), the term (\ref{Term_rho_3_1}) actually vanishes in 
case of $N$ moving monopoles; cf. Eq.~(C.20) in \cite{Zschocke4}.  

\subsection{Estimation of $\rho_4$}\label{Estimation_rho4} 

The term in the fifth line of (\ref{Transformation_k_to_sigma}) is denoted as $\ve{\rho}_4$ and reads 
\begin{eqnarray}
\fl \hspace{1.0cm} \ve{\rho}_4\left(s_1,s_0\right) = 2\,\frac{m_A}{R}\,\left(\frac{\ve{k} \cdot \ve{v}_A\left(s_1\right)}{c}\,
\frac{\ve{d}^k_A\left(s_1\right)}{k \cdot r^{\,1}_A\left(s_1\right)}
- \frac{\ve{k} \cdot \ve{v}_A\left(s_0\right)}{c}\,
\frac{\ve{d}^k_A\left(s_0\right)}{k \cdot r_A^{\,0}\left(s_0\right)}\right)  
\label{Term_rho_4_1}
\\
\nonumber\\ 
\fl \hspace{2.6cm} = \ve{\rho}_4\left(s_1,s_1\right) + \Delta \ve{\rho}_4\left(s_1,s_1\right) + {\cal O}\left(c^{-5}\right),  
\label{Term_rho_4_2}
\end{eqnarray}

\noindent
where 
\begin{eqnarray}
\fl \ve{\rho}_4\left(s_1,s_1\right) = 2\,\frac{m_A}{R}\,\frac{\ve{k} \cdot \ve{v}_A\left(s_1\right)}{c}\, 
\left(\frac{\ve{d}^k_A\left(s_1\right)}{k \cdot r^{\,1}_A\left(s_1\right)}
 - \frac{\ve{d}^k_A\left(s_1\right)}{k \cdot r_A^{\,0}\left(s_1\right)}\right),  
\label{Term_rho_4_3} 
\\
\nonumber\\
\fl \Delta \ve{\rho}_4\left(s_1,s_1\right) = + 2\,\frac{m_A}{R}\,\frac{\ve{k} \cdot \ve{v}_A\left(s_1\right)}{c}\,
\ve{k} \times \left(\frac{\ve{v}_A\left(s_1\right)}{c} \times \ve{k}\right)
\frac{k \cdot r^{\,0}_A\left(s_1\right) - k \cdot r^{\,1}_A\left(s_1\right)}{k \cdot r^{\,0}_A\left(s_1\right)}
\nonumber\\
\nonumber\\
\fl \hspace{0.75cm} - 2\,\frac{m_A}{R}\,\frac{\ve{k} \cdot \ve{v}_A\left(s_1\right)}{c}\, 
\frac{\ve{d}^k_A\left(s_1\right)}{k \cdot r^{\,0}_A\left(s_1\right)}
\left(\ve{k} - \frac{\ve{r}^{\,0}_A\left(s_1\right)}{r^{\,0}_A\left(s_1\right)}\right) \cdot \frac{\ve{v}_A\left(s_1\right)}{c}\;  
\frac{k \cdot r^{\,0}_A\left(s_1\right) - k \cdot r^{\,1}_A\left(s_1\right)}{k \cdot r^{\,0}_A\left(s_1\right)}
\nonumber\\
\nonumber\\
\fl \hspace{0.75cm} - 2\,\frac{m_A}{R}\,\frac{\ve{k} \cdot \ve{a}_A\left(s_1\right)}{c^2}\, 
\ve{d}^k_A\left(s_1\right)\, 
\frac{k \cdot r^{\,0}_A\left(s_1\right) - k \cdot r^{\,1}_A\left(s_1\right)}{k \cdot r^{\,0}_A\left(s_1\right)}\,.  
\label{Term_rho_4_5}
\end{eqnarray}

\noindent
The upper limit of the absolute value $\rho_4$ is given by
\begin{eqnarray}
\rho_4\left(s_1,s_1\right) \le 4\,\frac{m_A}{P_A}\,\frac{v_A\left(s_1\right)}{c}\,, 
\label{Estimation_rho_4_1}
\\
\nonumber\\
\hspace{-0.25cm} \Delta \rho_4\left(s_1,s_1\right) \le 6\,\frac{m_A}{r^{\,1}_A\left(s_1\right)}\,\frac{v_A^2\left(s_1\right)}{c^2}  
+ 4\,m_A\,\frac{a_A\left(s_1\right)}{c^2} \ll 1\,{\rm nas}\,,  
\label{Estimation_rho_4_2}
\end{eqnarray}

\noindent 
for all Solar System bodies. Hence, only the term (\ref{Term_rho_4_3}) is taken into 
account in the simplified transformation (\ref{Simplified_Transformation_k_to_sigma}).

\subsection{Estimation of $\rho_5$}

The term in the sixth line of (\ref{Transformation_k_to_sigma}) is denoted as $\ve{\rho}_5$ and reads 
\begin{eqnarray}
\ve{\rho}_5\left(s_1,s_0\right) = - 2\,\frac{m_A^2}{R^2}\,\ve{k}\,  
\left|\frac{\ve{d}^k_A\left(s_1\right)}{k \cdot r^{\,1}_A\left(s_1\right)}
- \frac{\ve{d}^k_A\left(s_0\right)}{k \cdot r_A^{\,0}\left(s_0\right)}\right|^2   
\label{Term_rho_5_1}
\\
\nonumber\\
\hspace{1.6cm} = \ve{\rho}_5\left(s_1,s_1\right) + {\cal O}\left(c^{-5}\right),
\label{Term_rho_5_1_A}  
\end{eqnarray}

\noindent
where 
\begin{eqnarray}
\ve{\rho}_5\left(s_1,s_1\right)  = - 2\,\frac{m_A^2}{R^2}\,\ve{k}\,
\left|\frac{\ve{d}^k_A\left(s_1\right)}{k \cdot r^{\,1}_A\left(s_1\right)}
- \frac{\ve{d}^k_A\left(s_1\right)}{k \cdot r_A^{\,0}\left(s_1\right)}\right|^2\,. 
\label{Term_rho_5_1_B}  
\end{eqnarray}

\noindent 
The upper limit of the absolute value is given by
\begin{eqnarray}
\rho_5\left(s_1,s_1\right) \le 8\,\frac{m_A^2}{P^2_A}\,,  
\label{Term_rho_5_2}
\end{eqnarray}

\noindent
which is less than $1\,{\rm nas}$ for all Solar System bodies. Furthermore, (\ref{Term_rho_5_1}) is a {\it scaling term}.

\subsection{Estimation of $\rho_6$}

The term in the seventh line of (\ref{Transformation_k_to_sigma}) is denoted as $\ve{\rho}_6$ and reads 
\begin{eqnarray}
\fl \hspace{0.25cm} \ve{\rho}_6\left(s_1,s_0\right) = - 2\,\frac{m_A^2}{R^2}
\left(\frac{\ve{d}^k_A\left(s_1\right)}{k \cdot r^{\,1}_A\left(s_1\right)} 
+ \frac{\ve{d}^k_A\left(s_0\right)}{k \cdot r_A^{\,0}\left(s_0\right)}\right) 
\left|\frac{\ve{d}^k_A\left(s_1\right)}{k \cdot r^{\,1}_A\left(s_1\right)}
- \frac{\ve{d}^k_A\left(s_0\right)}{k \cdot r_A^{\,0}\left(s_0\right)}\right|^2 
\label{Term_rho_6_1}
\\
\nonumber\\
\fl \hspace{1.85cm} = \ve{\rho}_6\left(s_1,s_1\right) + {\cal O}\left(c^{-5}\right),
\label{Term_rho_6_1_B}
\end{eqnarray}

\noindent
where 
\begin{eqnarray}
\fl \ve{\rho}_6\left(s_1,s_1\right) = - 2\,\frac{m_A^2}{R^2}
\left(\frac{\ve{d}^k_A\left(s_1\right)}{k \cdot r^{\,1}_A\left(s_1\right)} 
+ \frac{\ve{d}^k_A\left(s_1\right)}{k \cdot r_A^{\,0}\left(s_1\right)}\right)
\left|\frac{\ve{d}^k_A\left(s_1\right)}{k \cdot r^{\,1}_A\left(s_1\right)}
- \frac{\ve{d}^k_A\left(s_1\right)}{k \cdot r_A^{\,0}\left(s_1\right)}\right|^2\,.  
\label{Term_rho_6_1_A}
\end{eqnarray}

\noindent
The upper limit of the absolute value is given by
\begin{eqnarray}
\rho_6\left(s_1,s_1\right) \le 16\,\frac{m_A^2}{P_A^2}\,\frac{r^{\,1}_A\left(s_1\right)}{P_A}\,,  
\label{Term_rho_6_2}
\end{eqnarray}

\noindent
which contains the large factor $r^{\,1}_A\left(s_1\right)/P_A$ and, therefore, is an {\it enhanced term}, hence   
(\ref{Term_rho_6_1_A}) has necessarily to be taken into account in the simplified transformation (\ref{Simplified_Transformation_k_to_sigma}).

\subsection{Estimation of $\rho_7$}

The term in the eighth line of (\ref{Transformation_k_to_sigma}) is denoted as $\ve{\rho}_7$ and reads 
\begin{eqnarray}
\ve{\rho}_7\left(s_1,s_0\right) = - 4\,\frac{m_A^2}{R}\,\left(\frac{\ve{d}^k_A\left(s_1\right)}{\left(k \cdot r^{\,1}_A\left(s_1\right)\right)^2}
- \frac{\ve{d}^k_A\left(s_0\right)}{\left(k \cdot r_A^{\,0}\left(s_0\right)\right)^2}\right)  
\label{Term_rho_7_1}
\\
\nonumber\\ 
\hspace{1.6cm} = \ve{\rho}_7\left(s_1,s_1\right) + {\cal O} \left(c^{-5}\right),  
\label{Term_rho_7_1_A}
\end{eqnarray}

\noindent
where 
\begin{eqnarray}
\ve{\rho}_7\left(s_1,s_1\right) = - 4\,\frac{m_A^2}{R}\,\left(\frac{\ve{d}^k_A\left(s_1\right)}{\left(k \cdot r^{\,1}_A\left(s_1\right)\right)^2}
- \frac{\ve{d}^k_A\left(s_1\right)}{\left(k \cdot r_A^{\,0}\left(s_1\right)\right)^2}\right).  
\label{Term_rho_7_1_B}
\end{eqnarray}

\noindent 
The upper limit of the absolute value is given by  
\begin{eqnarray}
\rho_7\left(s_1,s_1\right) \le 16\,\frac{m_A^2}{P_A^2}\,\frac{r^{\,1}_A\left(s_1\right)}{P_A}\,,  
\label{Term_rho_7_2}
\end{eqnarray}

\noindent
which is an {\it enhanced term} because of the large factor $r^{\,1}_A\left(s_1\right)/P_A$, hence (\ref{Term_rho_7_1_B}) must be taken 
into account in the simplified transformation (\ref{Simplified_Transformation_k_to_sigma}).

\subsection{Estimation of $\rho_8$} 

The terms in the ninth and tenth line of (\ref{Transformation_k_to_sigma}) are combined to a term $\ve{\rho}_8$,  
\begin{eqnarray}
\fl \ve{\rho}_8\left(s_1,s_0\right) = \ve{\rho}^A_8\left(s_1\right) + \ve{\rho}^B_8\left(s_0\right),  
\label{Term_rho_8_1}
\\
\nonumber\\ 
\fl \ve{\rho}^A_8\left(s_1\right) = + \frac{15}{4} \frac{m_A^2}{R}
\frac{\ve{d}^k_A\left(s_1\right)}{\left|\ve{k} \times \ve{r}^{\,1}_A\left(s_1\right)\right|^3} \, 
\left(\ve{k} \cdot \ve{r}^{\,1}_A\left(s_1\right)\right) 
\left(\arctan \frac{\ve{k} \cdot \ve{r}^{\,1}_A\left(s_1\right)}{\left|\ve{k} \times \ve{r}^{\,1}_A\left(s_1\right)\right|} + \frac{\pi}{2} \right),  
\label{Term_rho_8_1_A}
\\
\nonumber\\ 
\fl \ve{\rho}^B_8\left(s_0\right) = - \frac{15}{4} \frac{m_A^2}{R}  
\frac{\ve{d}^k_A\left(s_0\right)}{\left|\ve{k} \times \ve{r}^{\,0}_A\left(s_0\right)\right|^3} \, 
\left(\ve{k} \cdot \ve{r}^{\,0}_A\left(s_0\right)\right)
\left( \arctan \frac{\ve{k} \cdot \ve{r}^{\,0}_A\left(s_0\right)}{\left|\ve{k} \times \ve{r}^{\,0}_A\left(s_0\right)\right|} 
+ \frac{\pi}{2} \right).    
\label{Term_rho_8_1_B}
\end{eqnarray}

\noindent 
The series expansion of $\ve{\rho}_8$ in (\ref{Term_rho_8_1}) reads 
\begin{eqnarray}
\ve{\rho}_8\left(s_1,s_0\right) = \ve{\rho}_8\left(s_1,s_1\right) + {\cal O}\left(c^{-5}\right),   
\label{Term_rho_8_1_C}
\end{eqnarray}

\noindent
where 
\begin{eqnarray}
\fl \ve{\rho}_8\left(s_1,s_1\right) = \frac{15}{4} \frac{m_A^2}{R}\,
\frac{\ve{d}^k_A\left(s_1\right)}{\left|\ve{k} \times \ve{r}^{\,1}_A\left(s_1\right)\right|^3}\,  
\bigg[\left(\ve{k} \cdot \ve{r}^{\,1}_A\left(s_1\right)\right)
\left(\arctan \frac{\ve{k} \cdot \ve{r}^{\,1}_A\left(s_1\right)}{\left|\ve{k} \times \ve{r}^{\,1}_A\left(s_1\right)\right|} + \frac{\pi}{2} \right)
\nonumber\\
\fl \hspace{5.0cm} - \left(\ve{k} \cdot \ve{r}^{\,0}_A\left(s_1\right)\right)  
\left( \arctan \frac{\ve{k} \cdot \ve{r}^{\,0}_A\left(s_1\right)}{\left|\ve{k} \times \ve{r}^{\,0}_A\left(s_1\right)\right|} 
+ \frac{\pi}{2} \right) \bigg].  
\label{Term_rho_8_1_D}
\end{eqnarray}

\noindent
The upper limit of the absolute value is given by
\begin{eqnarray}
\rho_8\left(s_1,s_1\right) \le \frac{15}{4}\,\pi\,\frac{m_A^2}{P_A^2}\,,  
\label{Term_rho_8_2}
\end{eqnarray}

\noindent
which is less than $1\,{\rm nas}$ for all Solar System bodies and for the Sun at $45^{\circ}$. Hence (\ref{Term_rho_8_1_D}) is not taken into
account in the simplified transformation (\ref{Simplified_Transformation_k_to_sigma}).

\subsection{Estimation of $\rho_9$}

The term in the eleventh line of (\ref{Transformation_k_to_sigma}) is denoted as $\ve{\rho}_9$ and reads 
\begin{eqnarray}
\ve{\rho}_9\left(s_1,s_0\right) = - \frac{1}{4}\frac{m_A^2}{R}
\left(\frac{\ve{d}^k_A\left(s_1\right)}{\left(r_A^{\,1}\left(s_1\right)\right)^2} 
- \frac{\ve{d}^k_A\left(s_0\right)}{\left(r_A^{\,0}\left(s_0\right)\right)^2}\right)  
\label{Term_rho_9_1}
\\
\nonumber\\
\hspace{1.6cm} = \ve{\rho}_9\left(s_1,s_1\right) + {\cal O}\left(c^{-5}\right),  
\label{Term_rho_9_1_A}
\end{eqnarray}

\noindent
where 
\begin{eqnarray}
\ve{\rho}_9\left(s_1,s_1\right) = - \frac{1}{4}\frac{m_A^2}{R}
\left(\frac{\ve{d}^k_A\left(s_1\right)}{\left(r_A^{\,1}\left(s_1\right)\right)^2}
- \frac{\ve{d}^k_A\left(s_1\right)}{\left(r_A^{\,0}\left(s_1\right)\right)^2}\right).  
\label{Term_rho_9_1_B}
\end{eqnarray}

\noindent
The upper limit of the absolute value is  
\begin{eqnarray}
\rho_9\left(s_1,s_1\right) \le \frac{1}{4}\,\frac{m_A^2}{P_A^2}\,,  
\label{Term_rho_9_2}
\end{eqnarray}

\noindent
which is less than $1\,{\rm nas}$ for all Solar System bodies and for the Sun at $45^{\circ}$.  
Hence (\ref{Term_rho_9_1}) is not taken into
account in the simplified transformation (\ref{Simplified_Transformation_k_to_sigma}).

Numerical values for the upper limits $\rho_1$, $\dots$ , $\rho_9$ are presented in Table~\ref{Table2}.  
\begin{table}[h!]
\caption{\label{Table2}The numerical magnitude of the upper limits of Eqs.~(\ref{Estimation_rho1}), (\ref{Term_rho_2_2}), (\ref{Term_rho_3_3_A}),
(\ref{Estimation_rho_4_1}), (\ref{Term_rho_5_2}), (\ref{Term_rho_6_2}), (\ref{Term_rho_7_2}), (\ref{Term_rho_8_2}), (\ref{Term_rho_9_2}),
and (\ref{Appendix_estimation_epsilon_3}).
The parameters of the most massive bodies of the Solar System are given in Table~\ref{Table1}. 
The solar aspect angle (angle between the direction of the Sun and the satellite's spin axis as seen from the satellite) 
in the Hipparcos mission \cite{Hipparcos} and Gaia mission \cite{GAIA} is $43^{\circ}$ and $45^{\circ}$, respectively. Therefore, we 
consider only astrometric observations equal to or larger than $45^{\circ}$ angular radii from the Sun.  
All values are given in ${\rm nas}$. A blank means the value is less than $1\,{\rm nas}$.}  
\footnotesize
\begin{tabular}{@{}ccccccccccc}
\br
& $\rho_1$ & $\rho_2$ & $\rho_3$ & $\rho_4$ & $\rho_5$ & $\rho_6$ &$\rho_7$& $\rho_8 \dots \rho_9$& $\hat{\epsilon}_2$ \\ 
\mr
Sun at $45^{\circ}$ & $1.2 \cdot 10^7$ & $-$     & $-$ & $-$     & $-$ & $0.95$     & $0.95$ & $-$ & $-$\\
Mercury             & $0.8 \cdot 10^5$ & $6.5$   & $-$ & $13.1$  & $-$ & $2.8$     & $2.8$ & $-$ & $-$\\
Venus               & $0.5 \cdot 10^6$ & $28.8$  & $-$ & $57.7$  & $-$ & $50.2$   & $50.2$ & $-$ & $-$ \\
Earth               & $0.6 \cdot 10^6$ & $28.4$  & $-$ & $56.8$  & $-$ & $-$ & $-$ & $-$ & $-$\\
Mars                & $0.1 \cdot 10^6$ & $4.6$   & $-$ & $9.3$   & $-$ & $7.6$    & $7.6$ & $-$ & $-$\\
Jupiter             & $1.6 \cdot 10^7$ & $358.0$ & $-$ & $716.0$ & $-$ & $1.6 \cdot 10^4$ & $1.6 \cdot 10^4$ & $-$ & $-$\\
Saturn              & $0.6 \cdot 10^7$ & $92.4$  & $-$ & $184.9$ & $-$ & $4.4 \cdot 10^3$ & $4.4 \cdot 10^3$ & $-$ & $-$\\
Uranus              & $0.2 \cdot 10^7$ & $23.8$  & $-$ & $47.5$  & $-$ & $2.5 \cdot 10^3$ & $2.5 \cdot 10^3$ & $-$ & $-$\\
Neptune             & $0.2 \cdot 10^7$ & $22.8$  & $-$ & $45.6$  & $-$ & $5.8 \cdot 10^3$ & $5.8 \cdot 10^3$ & $-$ & $-$\\
\br
\end{tabular}\\
\end{table}

\section{Estimation of the terms in the transformation $\ve{\sigma}$ to $\ve{n}$}\label{Appendix_EstimationB}

The transformation $\ve{\sigma}$ to $\ve{n}$ is given by Eq.~(\ref{Transformation_sigma_to_n}).
In what follows an upper limit of each individual term of this transformation is given.
The approach is the same as described and used in the previous sections. 
If the terms depend solely on the retarded $s_1$ then a series expansion is not necessary. 
The estimations are straightforward and they are just given.

\subsection{Estimation of $\varphi_1$}

The term in the second line of (\ref{Transformation_sigma_to_n}) is denoted as $\ve{\varphi}_1$ and reads 
\begin{eqnarray}
\ve{\varphi}_1\left(s_1\right) 
= 2\,\frac{m_A}{r^{\,1}_A\left(s_1\right)}\,\left|\frac{\ve{d}_A^k\left(s_1\right)}{k \cdot r^{\,1}_A\left(s_1\right)}\right|,  
\label{Appendix_Estimation_phi_1_A}
\\
\nonumber\\
\varphi_1\left(s_1\right) \le 4\,\frac{m_A}{P_A}\,,  
\label{Appendix_Estimation_phi_1}
\end{eqnarray}

\noindent 
which has to be taken into account in the simplified transformation (\ref{Simplified_Transformation_sigma_to_n}).

\subsection{Estimation of $\varphi_2$}

The term in the third line of (\ref{Transformation_sigma_to_n}) is denoted as $\ve{\varphi}_2$ and reads 
\begin{eqnarray}
\ve{\varphi}_2\left(s_1\right) = - 4\,\frac{m_A}{r^{\,1}_A\left(s_1\right)}\,\ve{k}\,\frac{\ve{k} \cdot \ve{v}_A\left(s_1\right)}{c}\,, 
\label{Appendix_Estimation_phi_2_A}
\\
\nonumber\\
\varphi_2\left(s_1\right) \le 4\,\frac{m_A}{r^{\,1}_A\left(s_1\right)}\,\frac{v_A\left(s_1\right)}{c}\,,   
\label{Appendix_Estimation_phi_2}
\end{eqnarray}

\noindent
which is less than $1\,{\rm nas}$ for all Solar System bodies. Furthermore, (\ref{Appendix_Estimation_phi_2_A}) is 
a {\it scaling term}.

\subsection{Estimation of $\varphi_3$}

The term in the fourth line of (\ref{Transformation_sigma_to_n}) is denoted as $\ve{\varphi}_3$ and reads 
\begin{eqnarray}
\ve{\varphi}_3\left(s_1\right) = - 2\,\frac{m_A}{r^{\,1}_A\left(s_1\right)}\,\frac{\ve{d}^k_A\left(s_1\right)}{k \cdot r^{\,1}_A\left(s_1\right)}\,
\frac{\ve{k}\cdot \ve{v}_A\left(s_1\right)}{c}\,,  
\label{Appendix_Estimation_phi_3_A}
\\
\nonumber\\
\varphi_3\left(s_1\right) \le 4\,\frac{m_A}{P_A}\,\frac{v_A\left(s_1\right)}{c}\,,   
\label{Appendix_Estimation_phi_3}
\end{eqnarray}

\noindent
which has to be taken into account in the simplified transformation (\ref{Simplified_Transformation_sigma_to_n}).

\subsection{Estimation of $\varphi_4$}

The term in the fifth line of (\ref{Transformation_sigma_to_n}) is denoted as $\ve{\varphi}_4$ and reads 
\begin{eqnarray}
\ve{\varphi}_4\left(s_1\right) = 4\,\frac{m_A}{r^{\,1}_A\left(s_1\right)}\,\frac{\ve{v}_A\left(s_1\right)}{c} 
+ 2\,\frac{m_A}{\left(r^{\,1}_A\left(s_1\right)\right)^2}\,\ve{d}^k_A\left(s_1\right)\,
\frac{\ve{k} \cdot \ve{v}_A\left(s_1\right)}{c}\,, 
\label{Appendix_Estimation_phi_4_A}  
\\
\nonumber\\
\varphi_4\left(s_1\right) \le 6\,\frac{m_A}{r^{\,1}_A\left(s_1\right)}\,\frac{v_A\left(s_1\right)}{c}\,,  
\label{Appendix_Estimation_phi_4}  
\end{eqnarray}

\noindent
which is less than $1\,{\rm nas}$ for all Solar System bodies, hence (\ref{Appendix_Estimation_phi_4_A}) is not taken into
account in the simplified transformation (\ref{Simplified_Transformation_sigma_to_n}).

\subsection{Estimation of $\varphi_5$}

The term in the sixth line of (\ref{Transformation_sigma_to_n}) is denoted as $\ve{\varphi}_5$ and reads 
\begin{eqnarray}
\ve{\varphi}_5\left(s_1\right) 
= 2\,\frac{m_A}{\left(r^{\,1}_A\left(s_1\right)\right)^2}\,\frac{\ve{d}^k_A\left(s_1\right)}{k \cdot r^{\,1}_A\left(s_1\right)}\,
\frac{\ve{d}^k_A\left(s_1\right) \cdot \ve{v}_A\left(s_1\right)}{c}\,, 
\label{Appendix_Estimation_phi_5_A} 
\\
\nonumber\\  
\varphi_5\left(s_1\right) \le 4\,\frac{m_A}{r^{\,1}_A\left(s_1\right)}\,\frac{v_A\left(s_1\right)}{c}\,,   
\label{Appendix_Estimation_phi_5}
\end{eqnarray}

\noindent
which is less than $1\,{\rm nas}$ for all Solar System bodies, hence (\ref{Appendix_Estimation_phi_5_A}) is not taken into
account in the simplified transformation (\ref{Simplified_Transformation_sigma_to_n}).

\subsection{Estimation of $\varphi_6$}

The term in the seventh line of (\ref{Transformation_sigma_to_n}) is denoted as $\ve{\varphi}_6$ and reads 
\begin{eqnarray}
\ve{\varphi}_6\left(s_1\right) = - 2\,\ve{k}\,\frac{m_A^2}{\left(r^{\,1}_A\left(s_1\right)\right)^2}\,
\frac{\ve{d}^k_A\left(s_1\right) \cdot \ve{d}^k_A\left(s_1\right)}{\left(k \cdot r^{\,1}_A\left(s_1\right)\right)^2}\,, 
\label{Appendix_Estimation_phi_6_A}
\\
\nonumber\\ 
\varphi_6\left(s_1\right) \le 8\,\frac{m_A^2}{P_A^2}\,,   
\label{Appendix_Estimation_phi_6}
\end{eqnarray}

\noindent
which is less than $1\,{\rm nas}$ for all Solar System bodies. Furthermore, (\ref{Appendix_Estimation_phi_6_A}) is 
a {\it scaling term}.

\subsection{Estimation of $\varphi_7$}

The term in the eighth line of (\ref{Transformation_sigma_to_n}) is denoted as $\ve{\varphi}_7$ and reads 
\begin{eqnarray}
\fl \ve{\varphi}_7\left(s_1,s_0\right) = 4\,\frac{m_A^2}{r^{\,1}_A\left(s_1\right)}\,\frac{1}{k \cdot r^{\,1}_A\left(s_1\right)}\, 
\frac{\ve{k}}{R}\, \left(\frac{\ve{d}^k_A\left(s_1\right) \cdot \ve{d}^k_A\left(s_1\right)}{k \cdot r^{\,1}_A\left(s_1\right)}
- \frac{\ve{d}^k_A\left(s_0\right) \cdot \ve{d}^k_A\left(s_1\right)}{k \cdot r_A^{\,0}\left(s_0\right)}\right)  
\label{Appendix_Estimation_phi_7_1}
\\
\nonumber\\
\fl \hspace{1.6cm} = \ve{\varphi}_7\left(s_1,s_1\right) + {\cal O}\left(c^{-5}\right),  
\end{eqnarray}

\noindent
where  
\begin{eqnarray}
\fl \ve{\varphi}_7\left(s_1,s_1\right) 
= 4\,\frac{m_A^2}{r^{\,1}_A\left(s_1\right)}\,\frac{1}{k \cdot r^{\,1}_A\left(s_1\right)}\,
\frac{\ve{k}}{R}\, \left(\frac{\ve{d}^k_A\left(s_1\right) \cdot \ve{d}^k_A\left(s_1\right)}{k \cdot r^{\,1}_A\left(s_1\right)}
- \frac{\ve{d}^k_A\left(s_1\right) \cdot \ve{d}^k_A\left(s_1\right)}{k \cdot r_A^{\,0}\left(s_1\right)}\right).  
\label{Appendix_Estimation_phi_7_1_B}
\end{eqnarray}
 
\noindent
For the upper limit one finds 
\begin{eqnarray}
\varphi_7\left(s_1,s_1\right) \le 16\,\frac{m_A^2}{P_A^2}\,,  
\label{Appendix_Estimation_phi_7_2}
\end{eqnarray}

\noindent
which is less than $1.3\,{\rm nas}$ for all Solar System bodies. Furthermore, (\ref{Appendix_Estimation_phi_7_1}) is 
a {\it scaling term}.

\subsection{Estimation of $\varphi_8$}

The term in the ninth line of (\ref{Transformation_sigma_to_n}) is denoted as $\ve{\varphi}_8$ and reads 
\begin{eqnarray}
\ve{\varphi}_8\left(s_1\right) 
= 4\,\frac{m_A^2}{r^{\,1}_A\left(s_1\right)}\,\frac{\ve{d}_A^k\left(s_1\right)}{\left(k \cdot r^{\,1}_A\left(s_1\right)\right)^2}\,, 
\label{Appendix_Estimation_phi_8_1}
\\
\nonumber\\ 
\varphi_8\left(s_1\right) \le 16\,\frac{m_A^2}{P_A^2}\,\frac{r^{\,1}_A\left(s_1\right)}{P_A}\,,   
\label{Appendix_Estimation_phi_8_2}
\end{eqnarray}

\noindent
which is an {\it enhanced term} because of the large factor $r^{\,1}_A\left(s_1\right)/P_A$, hence (\ref{Appendix_Estimation_phi_8_1}) 
must necessarily to be taken into account in the simplified transformation (\ref{Simplified_Transformation_sigma_to_n}).

\subsection{Estimation of $\varphi_9$}

The term in the tenth line of (\ref{Transformation_sigma_to_n}) is denoted as $\ve{\varphi}_9$ and reads 
\begin{eqnarray}
\fl \ve{\varphi}_9 \left(s_1,s_0\right) = 4\,\frac{m_A^2}{r^{\,1}_A\left(s_1\right)}\,\frac{1}{R} 
\frac{\ve{d}_A^k\left(s_1\right)}{\left(k \cdot r^{\,1}_A\left(s_1\right)\right)^2}
\left(\frac{\ve{d}^k_A\left(s_1\right) \cdot \ve{d}^k_A\left(s_1\right)}{k \cdot r^{\,1}_A\left(s_1\right)}
- \frac{\ve{d}^k_A\left(s_0\right) \cdot \ve{d}^k_A\left(s_1\right)}{k \cdot r_A^{\,0}\left(s_0\right)} \right)  
\nonumber\\ 
\label{Appendix_Estimation_phi_9_1}
\\
\fl \hspace{1.6cm} = \ve{\varphi}_9 \left(s_1,s_1\right) + {\cal O}\left(c^{-5}\right), 
\label{Appendix_Estimation_phi_9_1_A}
\end{eqnarray}

\noindent
where 
\begin{eqnarray}
\fl \ve{\varphi}_9\left(s_1,s_1\right) = 4\,\frac{m_A^2}{r^{\,1}_A\left(s_1\right)}\,\frac{1}{R}
\frac{\ve{d}_A^k\left(s_1\right)}{\left(k \cdot r^{\,1}_A\left(s_1\right)\right)^2}
\left(\frac{\ve{d}^k_A\left(s_1\right) \cdot \ve{d}^k_A\left(s_1\right)}{k \cdot r^{\,1}_A\left(s_1\right)}
- \frac{\ve{d}^k_A\left(s_1\right) \cdot \ve{d}^k_A\left(s_1\right)}{k \cdot r_A^{\,0}\left(s_1\right)} \right).  
\nonumber\\
\label{Appendix_Estimation_phi_9_1_B}
\end{eqnarray}

\noindent
For the upper limit one finds
\begin{eqnarray}
\varphi_9\left(s_1,s_1\right) \le 32\,\frac{m_A^2}{P_A^2}\,\frac{r^{\,1}_A\left(s_1\right)}{P_A}\,,  
\label{Appendix_Estimation_phi_9_2}
\end{eqnarray}

\noindent
which contains the large factor $r^{\,1}_A\left(s_1\right)/P_A$ hence is an {\it enhanced term}, so that 
(\ref{Appendix_Estimation_phi_9_1_B}) must be taken into account in the simplified transformation (\ref{Simplified_Transformation_sigma_to_n}).

\subsection{Estimation of $\varphi_{10}$}

The term in the eleventh line of (\ref{Transformation_sigma_to_n}) is denoted as $\ve{\varphi}_{10}$ and reads 
\begin{eqnarray}
\fl \ve{\varphi}_{10}\left(s_1,s_0\right) = 4\,\frac{m_A^2}{r^{\,1}_A\left(s_1\right)}\,\frac{1}{R}\,
\frac{\ve{k} \cdot \ve{r}^{\,1}_A\left(s_1\right)}{k \cdot r^{\,1}_A\left(s_1\right)}
\left(\frac{\ve{d}^k_A\left(s_1\right)}{k \cdot r^{\,1}_A\left(s_1\right)}
- \frac{\ve{d}^k_A\left(s_0\right)}{k \cdot r_A^{\,0}\left(s_0\right)} \right) 
\label{Appendix_Estimation_phi_10_1}
\\
\nonumber\\
\fl \hspace{1.75cm} = \ve{\varphi}_{10}\left(s_1,s_1\right) + {\cal O}\left(c^{-5}\right),   
\label{Appendix_Estimation_phi_10_1_A}
\end{eqnarray}

\noindent
where 
\begin{eqnarray}
\fl \ve{\varphi}_{10}\left(s_1,s_1\right) = 4\,\frac{m_A^2}{r^{\,1}_A\left(s_1\right)}\,\frac{1}{R}\,
\frac{\ve{k} \cdot \ve{r}^{\,1}_A\left(s_1\right)}{k \cdot r^{\,1}_A\left(s_1\right)}
\left(\frac{\ve{d}^k_A\left(s_1\right)}{k \cdot r^{\,1}_A\left(s_1\right)}
- \frac{\ve{d}^k_A\left(s_1\right)}{k \cdot r_A^{\,0}\left(s_1\right)} \right)\,. 
\label{Appendix_Estimation_phi_10_1_B}
\end{eqnarray}

\noindent
For the upper limit one finds  
\begin{eqnarray}
\varphi_{10}\left(s_1,s_1\right) \le 16\,\frac{m_A^2}{P_A^2}\,\frac{r^{\,1}_A\left(s_1\right)}{P_A}\,,  
\label{Appendix_Estimation_phi_10_2}
\end{eqnarray}

\noindent
which contains the large factor $r^{\,1}_A\left(s_1\right)/P_A$ hence is an {\it enhanced term}, so that
(\ref{Appendix_Estimation_phi_10_1_B}) must be taken into account in the simplified transformation (\ref{Simplified_Transformation_sigma_to_n}).

\subsection{Estimation of $\varphi_{11}$}

The term in the twelfth line of (\ref{Transformation_sigma_to_n}) is denoted as $\ve{\varphi}_{11}$ and reads 
\begin{eqnarray}
\ve{\varphi}_{11}\left(s_1\right) = - 4\,\frac{m_A^2}{\left(r_A^{\,1}\left(s_1\right)\right)^2}\,
\frac{\ve{d}_A^k\left(s_1\right)}{k \cdot r^{\,1}_A\left(s_1\right)}\,,  
\label{Appendix_Estimation_phi_11_1}
\\
\nonumber\\
\varphi_{11}\left(s_1\right) \le 8\,\frac{m_A^2}{P_A\,r_A^{\,1}\left(s_1\right)}\,,   
\label{Appendix_Estimation_phi_11_2}
\end{eqnarray}

\noindent
which contributes less than $1\,{\rm nas}$ for all Solar System bodies, hence (\ref{Appendix_Estimation_phi_11_1}) is not taken into
account in the simplified transformation (\ref{Simplified_Transformation_sigma_to_n}).

\subsection{Estimation of $\varphi_{12}$}

The term in the thirteenth line of (\ref{Transformation_sigma_to_n}) is denoted as $\ve{\varphi}_{12}$ and reads 
\begin{eqnarray}
\fl \ve{\varphi}_{12}\left(s_1\right) 
= - \frac{m_A^2}{2}\,\ve{d}_A^k\left(s_1\right)\frac{\ve{k} \cdot \ve{r}^{\,1}_A\left(s_1\right)}{\left(r^{\,1}_A\left(s_1\right)\right)^4}
- \frac{15}{4}\,\frac{m_A^2}{\left(r_A^{\,1}\left(s_1\right)\right)^2}\,\ve{d}^k_A\left(s_1\right)
\frac{\ve{k} \cdot \ve{r}^{\,1}_A\left(s_1\right)}{\left|\ve{k} \times \ve{r}^{\,1}_A\left(s_1\right)\right|^2}\,,  
\label{Appendix_Estimation_phi_12_1}
\\
\nonumber\\
\fl \varphi_{12}\left(s_1\right)  \le \frac{1}{2}\,\frac{m_A^2}{\left(r^{\,1}_A\left(s_1\right)\right)^2} 
+ \frac{15}{4}\,\frac{m_A^2}{P_A\,r^{\,1}_A\left(s_1\right)}\,,  
\label{Appendix_Estimation_phi_12_2}
\end{eqnarray}

\noindent
which is less than $1\,{\rm nas}$ for all Solar System bodies, hence (\ref{Appendix_Estimation_phi_12_1}) is not taken into
account in the simplified transformation (\ref{Simplified_Transformation_sigma_to_n}).

\subsection{Estimation of $\varphi_{13}$}

The term in the fourteenth line of (\ref{Transformation_sigma_to_n}) is denoted as $\ve{\varphi}_{13}$ and reads 
\begin{eqnarray}
\fl \ve{\varphi}_{13}\left(s_1\right) = - \frac{15}{4}\,m_A^2\,  
\frac{\ve{d}^k_A\left(s_1\right)}{\left|\ve{k} \times \ve{r}^{\,1}_A\left(s_1\right)\right|^3}
\left(\arctan \frac{\ve{k} \cdot \ve{r}^{\,1}_A\left(s_1\right)}{\left|\ve{k} \times \ve{r}^{\,1}_A\left(s_1\right)\right|} + \frac{\pi}{2} \right), 
\label{Appendix_Estimation_phi_13_1}
\\
\nonumber\\
\fl \varphi_{13}\left(s_1\right)  \le \frac{15}{4}\,\pi\,\frac{m_A^2}{P_A^2}\,, 
\label{Appendix_Estimation_phi_13_2}
\end{eqnarray}

\noindent
which is less than $1\,{\rm nas}$ for all Solar System bodies, hence (\ref{Appendix_Estimation_phi_13_1}) is not taken into
account in the simplified transformation (\ref{Simplified_Transformation_sigma_to_n}).

\noindent 
The numerical values for the upper limits are presented in Table~\ref{Table3}.
\begin{table}[h!]
\caption{\label{Table3}The numerical magnitude of the upper limits of (\ref{Appendix_Estimation_phi_1}), (\ref{Appendix_Estimation_phi_2}),
(\ref{Appendix_Estimation_phi_3}), (\ref{Appendix_Estimation_phi_4}), (\ref{Appendix_Estimation_phi_5}), (\ref{Appendix_Estimation_phi_6}),
(\ref{Appendix_Estimation_phi_7_2}), (\ref{Appendix_Estimation_phi_8_2}), (\ref{Appendix_Estimation_phi_9_2}), (\ref{Appendix_Estimation_phi_10_2}),
(\ref{Appendix_Estimation_phi_11_2}), (\ref{Appendix_Estimation_phi_12_2}), (\ref{Appendix_Estimation_phi_13_2}),
and (\ref{Appendix_estimation_epsilon_1}).  
The parameters of the most massive bodies of the Solar System are given in Table~\ref{Table1}. 
Like in Table~\ref{Table2}, we consider only astrometric observations larger or equal than $45^{\circ}$ angular radii from the Sun 
in accordance with the solar aspect angle in the Gaia mission \cite{GAIA}.
All values are given in ${\rm nas}$. A blank means the value is less than $1\,{\rm nas}$.}
\footnotesize
\begin{tabular}{@{}cccccccccccc}
\br
& $\varphi_1$ & $\varphi_2$ & $\varphi_3$ & $\varphi_4 \dots \varphi_6$ & $\varphi_7$ & $\varphi_8$ & $\varphi_9$ & $\varphi_{10}$ & $\varphi_{11} \dots \varphi_{13}$ & $\hat{\epsilon}_1$ \\
\mr
Sun at $45^{\circ}$ & $1.2 \cdot 10^7$ & $-$ & $-$ & $-$ & $-$ & $-$ & $1.8$ & $-$     & $-$ & $-$ \\
Mercury & $\!0.8 \cdot 10^5\!$&$-$&$\!13.1\!$& $-$& $-$& $2.8$             & $\!5.7\!$   & $\!2.8\!$  & $-$ & $-$ \\
Venus   & $\!0.5 \cdot 10^6\!$&$-$&$\!57.7\!$& $-$& $-$& $50.2$            & $\!100.4\!$ & $\!50.2\!$ & $-$ & $-$ \\
Earth   & $\!0.6 \cdot 10^6\!$&$-$&$\!56.8\!$& $-$& $-$& $-$ & $-$         & $-$     & $-$    & $-$ \\
Mars    & $\!0.1 \cdot 10^6\!$&$-$&$\!9.3\!$&$-$ &$-$& $7.7$ & $15.3$  & $7.7$  & $-$ & $-$ \\
Jupiter & $\!1.6 \cdot 10^7\!$&$-$&$\!716.0\!$&$-$&$1.3$ & $\!1.6 \cdot 10^4\!$ & $\!3.2 \cdot 10^4\!$ & $\!1.6 \cdot 10^4\!$ & $-$ & $-$ \\
Saturn  & $\!0.6 \cdot 10^7\!$&$-$&$\!184.9\!$&$-$&$-$& $\!4.4 \cdot 10^3\!$ & $\!8.8 \cdot 10^3\!$ & $\!4.4 \cdot 10^3\!$ & $-$ & $-$ \\
Uranus  & $\!0.2 \cdot 10^7\!$ & $-$ & $\!47.5\!$&$-$&$-$&$\!2.5 \cdot 10^3\!$ & $\!5.1 \cdot 10^3\!$ & $\!2.5 \cdot 10^3\!$ & $-$ & $-$ \\
Neptune & $\!0.2 \cdot 10^7\!$ & $-$ & $\!45.6\!$&$-$&$-$&$\!5.8 \cdot 10^3\!$ & $\!1.2 \cdot 10^4\!$ & $\!5.8 \cdot 10^3\!$ & $-$ & $-$ \\
\br
\end{tabular}\\
\end{table}
\normalsize

\section{The terms $\hat{\ve \epsilon}_1$ and $\hat{\ve \epsilon}_2$}\label{Appendix_epsilon} 

\subsection{The upper limit of the term $\hat{\ve \epsilon}_1$} 

The expression of the term $\hat{\ve \epsilon}_1$ in (\ref{Transformation_sigma_to_n_5}) and (\ref{Transformation_sigma_to_n}) reads 
\begin{eqnarray}
\hat{\ve \epsilon}_1\left(s_1\right)  
= m_A\,\ve{\sigma} \times \left(\ve{\epsilon}_1\left(\ve{r}^{\,1}_A\left(s_1\right),\ve{v}_A\left(s_1\right)\right) \times \ve{\sigma}\right),   
\label{epsilon1}
\end{eqnarray}

\noindent
where $\ve{\epsilon}_1$ is given by Eq.~(\ref{epsilon_1}).  
The vectorial term $\hat{\ve \epsilon}_1$ is of order ${\cal O}\left(c^{-4}\right)$. Due to  
$\ve{\sigma} = \ve{k} + {\cal O}\left(c^{-2}\right)$ we may replace the unit-vector $\ve{\sigma}$ in (\ref{epsilon1}) 
by the unit-vector $\ve{k}$, because such a replacement would cause an error of the order ${\cal O}\left(c^{-6}\right)$ which is 
beyond 2PN approximation. Hence, we get  
\begin{eqnarray}
\fl \hat{\ve \epsilon}_1\left(s_1\right) 
= m_A\,\ve{k} \times \left(\ve{\epsilon}_1\left(\ve{r}^{\,1}_A\left(s_1\right),\ve{v}_A\left(s_1\right)\right) \times \ve{k}\right) + {\cal O}\left(c^{-6}\right)  
\nonumber\\
\nonumber\\
\fl \hspace{0.25cm} = + 4\,\frac{m_A}{r^{\,1}_A\left(s_1\right)}\,\frac{\ve{k} \times \left(\ve{v}_A\left(s_1\right) \times \ve{k}\right)}{c}
\frac{\ve{r}^{\,1}_A\left(s_1\right) \cdot \ve{v}_A\left(s_1\right)}{r^{\,1}_A\left(s_1\right)\,c}
\nonumber\\
\nonumber\\
\fl \hspace{0.5cm} - 4\,\frac{m_A}{r^{\,1}_A\left(s_1\right)}\,\frac{\ve{k} \times \left(\ve{v}_A\left(s_1\right) \times \ve{k}\right)}{c}\,
\frac{\ve{k} \cdot \ve{v}_A\left(s_1\right)}{c}
\nonumber\\
\nonumber\\
\fl \hspace{0.5cm} - \frac{m_A}{r^{\,1}_A\left(s_1\right)} \frac{\ve{d}_A^k\left(s_1\right)}{k \cdot r^{\,1}_A\left(s_1\right)}
\left[\frac{v^2_A\left(s_1\right)}{c^2}
+ 2 \left(\frac{\ve{r}^{\,1}_A\left(s_1\right) \cdot \ve{v}_A\left(s_1\right)}{r^{\,1}_A\left(s_1\right)\,c}
+ \frac{\ve{k} \cdot \ve{v}_A\left(s_1\right)}{c}\right)^2\right] + {\cal O}\left(c^{-6}\right).
\nonumber\\
\label{Transformation_sigma_to_n_epsilon}
\end{eqnarray}

\noindent
The upper limit of the absolute value of $\hat{\epsilon}_1 = \left|\hat{\ve \epsilon}_1\right|$ is given by
\begin{eqnarray}
\hat{\epsilon}_1 = \left|\hat{\ve \epsilon}_1\left(s_1\right)\right| \le 8\,\frac{m_A}{r^{\,1}_A\left(s_1\right)}\,\frac{v_A^2\left(s_1\right)}{c^2} 
+ 18\,\frac{m_A}{P_A}\,\frac{v_A^2\left(s_1\right)}{c^2} + {\cal O}\left(c^{-6}\right).  
\label{Appendix_estimation_epsilon_1}
\end{eqnarray}

\subsection{The upper limit of the term $\hat{\ve \epsilon}_2$} 

The expression of the term $\hat{\ve \epsilon}_2$ in (\ref{Transformation_k_to_sigma_5}) and (\ref{Transformation_k_to_sigma}) reads 
as follows:  
\begin{eqnarray}
\hat{\ve \epsilon}_2\left(s_1,s_0\right) 
= \frac{m_A}{R}\,\Bigg(\ve{\sigma} \times \bigg[ \ve{\sigma} \times \ve{\epsilon}_2\left(s_1,s_0\right)\bigg]\Bigg),  
\label{epsilon2}
\end{eqnarray}

\noindent 
where $\ve{\epsilon}_2$ is given by Eq.~(\ref{epsilon_3}).  
Because the vectorial term $\hat{\ve \epsilon}_2$ is of order ${\cal O}\left(c^{-4}\right)$, we  
$\ve{\sigma} = \ve{k} + {\cal O}\left(c^{-2}\right)$ we may replace the unit-vector $\ve{\sigma}$ in (\ref{epsilon2})
by the unit-vector $\ve{k}$, because such a replacement would cause an error of the order ${\cal O}\left(c^{-6}\right)$ which is 
beyond 2PN approximation. So, we obtain  
\begin{eqnarray}
\fl \hat{\ve \epsilon}_2\left(s_1,s_0\right)  
= \frac{m_A}{R}\,\Bigg(\ve{k} \times \bigg[ \ve{k} \times \ve{\epsilon}_2\left(s_1,s_0\right)\bigg]\Bigg) + {\cal O}\left(c^{-6}\right)  
\nonumber\\
\nonumber\\
\fl \hspace{0.5cm} = - \frac{m_A}{R}\,\frac{v_A^2\left(s_1\right)}{c^2}\,
\frac{\ve{d}^k_A\left(s_1\right)}{k \cdot r^{\,1}_A\left(s_1\right)}
+ \frac{m_A}{R}\,\frac{v_A^2\left(s_0\right)}{c^2}\,
\frac{\ve{d}^k_A\left(s_0\right)}{k \cdot r_A^{\,0}\left(s_0\right)}
\nonumber\\
\nonumber\\
\fl \hspace{0.75cm} - 2\,\frac{m_A}{R}\,\ve{d}^k_A\left(s_0\right) \frac{\ve{k} \cdot \ve{a}_A\left(s_1\right)}{c^2}
\ln \frac{k \cdot r^{\,1}_A\left(s_1\right)}{k \cdot r_A^{\,0}\left(s_0\right)}
\nonumber\\
\nonumber\\
\fl \hspace{0.75cm} + 2\,\frac{m_A}{R} \frac{\ve{k} \times \left(\ve{a}_A\left(s_1\right) \times \ve{k}\right)}{c^2}
\left[k \cdot r^{\,1}_A\left(s_1\right) - k \cdot r_A^{\,0}\left(s_0\right)\right]
\nonumber\\
\nonumber\\
\fl \hspace{0.75cm} - 2\,\frac{m_A}{R}\,\frac{\ve{k} \times \left(\ve{a}_A\left(s_1\right) \times \ve{k} \right)}{c^2}
\left(k \cdot r^{\,1}_A\left(s_1\right) \right)
\ln \frac{k \cdot r^{\,1}_A\left(s_1\right)}{k \cdot r_A^{\,0}\left(s_0\right)} + {\cal O}\left(c^{-6}\right), 
\label{Transformation_k_to_sigma_epsilon}
\end{eqnarray}

\noindent
where all the acceleration terms carry the same argument because of 
$\ve{a}_A\left(s_0\right) = \ve{a}_A\left(s_1\right) + {\cal O}\left(c^{-1}\right)$. In this respect we recall that  
$\ve{d}_A^k\left(s_0\right) = \ve{d}_A^k\left(s_1\right) + {\cal O}\left(c^{-1}\right)$ and  
$\ve{v}_A\left(s_0\right) = \ve{v}_A\left(s_1\right) + {\cal O}\left(c^{-1}\right)$, hence also the impact vectors 
and velocities in (\ref{Transformation_k_to_sigma_epsilon}) may actually be written such that they carry the same argument.  
Here we also notice that the origin of last term in (\ref{Transformation_k_to_sigma_epsilon}) is just the combination 
of the last both terms in (\ref{epsilon_3b}).  
The series expansion of (\ref{Transformation_k_to_sigma_epsilon}) around $s_1$ reads 
\begin{eqnarray}
\hat{\ve{\epsilon}}_2\left(s_1,s_0\right) = \hat{\ve{\epsilon}}_2\left(s_1,s_1\right) + {\cal O}\left(c^{-5}\right).   
\label{Appendix_series_expansion_epsilon_2}
\end{eqnarray}

\noindent 
For the upper limit of the absolute value of (\ref{Appendix_series_expansion_epsilon_2}) one finds  
\begin{eqnarray}
\hat{\epsilon}_2 = \left|\hat{\ve \epsilon}_2\left(s_1,s_1\right)\right|  
\le 2\,\frac{m_A}{P_A}\,\frac{v_A^2\left(s_1\right)}{c^2}  
+ 10\,m_A\,\frac{a_A\left(s_1\right)}{c^2}\,.  
\label{Appendix_estimation_epsilon_3}
\end{eqnarray}

\section{Proof of inequality (\ref{Inequality_A})}\label{Appendix_Inequality_A} 

We will show the inequality (\ref{Inequality_A}), which reads 
\begin{eqnarray}
\fl \left|\Delta \ve{\rho}_1^A\left(s_1,s_1\right) + \Delta \ve{\rho}_1^B\left(s_1,s_1\right) 
+ \ve{\varphi}_4\left(s_1\right) + \ve{\varphi}_5\left(s_1\right)\right|
\le 10\,\frac{m_A}{r^{\,1}_A\left(s_1\right)}\,\frac{v_A\left(s_1\right)}{c}\,,
\label{Appendix_Inequality_A_5}
\end{eqnarray}

\noindent
where $\Delta \ve{\rho}_1^A$, $\Delta \ve{\rho}_1^B$, $\ve{\varphi}_4$, and $\ve{\varphi}_5$ are given by Eqs.~(\ref{series_expansion_rho_1_B_1}),  
(\ref{series_expansion_rho_1_B_2}), (\ref{Appendix_Estimation_phi_4_A}), and (\ref{Appendix_Estimation_phi_5_A}).  
From (\ref{Impact_Vector_k1}) follows the relation 
\begin{eqnarray}
\fl \hspace{1.5cm} \frac{\ve{r}^{\,0}_A\left(s_1\right)}{r^{\,0}_A\left(s_1\right)} \cdot \frac{\ve{v}_A\left(s_1\right)}{c} 
= \frac{\ve{d}^k_A\left(s_1\right)}{r^{\,0}_A\left(s_1\right)} \cdot \frac{\ve{v}_A\left(s_1\right)}{c} 
+ \left(\frac{\ve{k} \cdot \ve{v}_A\left(s_1\right)}{c}\right) \frac{\ve{k} \cdot \ve{r}^{\,0}_A\left(s_1\right)}{r^{\,0}_A\left(s_1\right)}\,,  
\label{Appendix_Inequality_A_10}
\end{eqnarray}

\noindent  
which allows to rewrite the term $\Delta \ve{\rho}_1^A$ in (\ref{series_expansion_rho_1_B_1}) in the form  
\begin{eqnarray}
\fl \Delta \ve{\rho}_1^A\left(s_1,s_1\right) = - 2\,\frac{m_A}{R}\,  
\frac{\ve{d}^k_A\left(s_1\right)}{r^{\,0}_A\left(s_1\right)}
\left(\frac{\ve{k} \cdot \ve{v}_A\left(s_1\right)}{c}\right) 
\frac{k \cdot r^{\,0}_A\left(s_1\right) - k \cdot r^{\,1}_A\left(s_1\right)}{k \cdot r^{\,0}_A\left(s_1\right)}
\nonumber\\ 
\nonumber\\ 
\fl \hspace{2.35cm} - 2\,\frac{m_A}{R}\, \frac{\ve{d}^k_A\left(s_1\right)}{k \cdot r^{\,0}_A\left(s_1\right)} 
\left(\frac{\ve{d}^k_A\left(s_1\right)}{r^{\,0}_A\left(s_1\right)} \cdot \frac{\ve{v}_A\left(s_1\right)}{c} \right) 
\frac{k \cdot r^{\,0}_A\left(s_1\right) - k \cdot r^{\,1}_A\left(s_1\right)}{k \cdot r^{\,0}_A\left(s_1\right)}\,. 
\label{Appendix_Inequality_A_15}
\end{eqnarray}

\noindent
The term $\Delta \ve{\rho}_1^B$ in (\ref{series_expansion_rho_1_B_2}) is written as follows,  
\begin{eqnarray}
\fl \hspace{1.5cm} \Delta \ve{\rho}^B_1\left(s_1,s_1\right) = - 2\,\frac{m_A}{R}\, \frac{\ve{v}_A\left(s_1\right)}{c} 
\frac{k \cdot r^{\,0}_A\left(s_1\right) - k \cdot r^{\,1}_A\left(s_1\right)}{k \cdot r^{\,0}_A\left(s_1\right)}\,, 
\label{Appendix_Inequality_A_20}
\end{eqnarray}

\noindent
while the term proportional to three-vector $\ve{k}$ is omitted because it does not contribute to the light deflection. 
Using the expressions (\ref{Appendix_Inequality_A_15}) - (\ref{Appendix_Inequality_A_20}) for $\Delta \ve{\rho}^A_1$ and $\Delta \ve{\rho}^B_1$  
as well as Eqs.~(\ref{Appendix_Estimation_phi_4_A}) and (\ref{Appendix_Estimation_phi_5_A}) for $\ve{\varphi}_4$ and $\ve{\varphi}_5$ 
we get 
\begin{eqnarray}
\fl \left|\Delta \ve{\rho}_1^A\left(s_1,s_1\right) + \Delta \ve{\rho}_1^B\left(s_1,s_1\right)
+ \ve{\varphi}_4\left(s_1\right) + \ve{\varphi}_5\left(s_1\right)\right| = \left| \ve{T}_1 + \ve{T}_2 + \ve{T}_3 \right|,  
\label{Appendix_Inequality_A_25}
\end{eqnarray}

\noindent
where the terms of same algebraic structure are grouped together,  
\begin{eqnarray}
\fl \ve{T}_1 = + 4\,\frac{m_A}{r^{\,1}_A\left(s_1\right)}\,\frac{\ve{v}_A\left(s_1\right)}{c} 
- 2\,\frac{m_A}{R}\, \frac{\ve{v}_A\left(s_1\right)}{c}
\frac{k \cdot r^{\,0}_A\left(s_1\right) - k \cdot r^{\,1}_A\left(s_1\right)}{k \cdot r^{\,0}_A\left(s_1\right)}\,,  
\label{Appendix_Inequality_A_30}
\\
\nonumber\\
\fl \ve{T}_2 = + \frac{2\,m_A}{r^{\,1}_A\left(s_1\right)}\,\frac{\ve{d}^k_A\left(s_1\right)}{r^{\,1}_A\left(s_1\right)}\,
\frac{\ve{k} \cdot \ve{v}_A\left(s_1\right)}{c} - 2\,\frac{m_A}{R}\,
\frac{\ve{d}^k_A\left(s_1\right)}{r^{\,0}_A\left(s_1\right)}\, 
\frac{\ve{k} \cdot \ve{v}_A\left(s_1\right)}{c}\, 
\frac{k \cdot r^{\,0}_A\left(s_1\right) - k \cdot r^{\,1}_A\left(s_1\right)}{k \cdot r^{\,0}_A\left(s_1\right)}\,,  
\nonumber\\
\label{Appendix_Inequality_A_35}
\\
\fl \ve{T}_3 = + 2\,\frac{m_A}{\left(r^{\,1}_A\left(s_1\right)\right)^2}\,\frac{\ve{d}^k_A\left(s_1\right)}{k \cdot r^{\,1}_A\left(s_1\right)}\,
\frac{\ve{d}^k_A\left(s_1\right) \cdot \ve{v}_A\left(s_1\right)}{c} 
\nonumber\\
\fl \hspace{0.9cm} - 2\,\frac{m_A}{r^{\,0}_A\left(s_1\right)}\, \frac{\ve{d}^k_A\left(s_1\right)}{k \cdot r^{\,0}_A\left(s_1\right)}\,\frac{1}{R}\, 
\frac{\ve{d}^k_A\left(s_1\right) \cdot \ve{v}_A\left(s_1\right)}{c}\, 
\frac{k \cdot r^{\,0}_A\left(s_1\right) - k \cdot r^{\,1}_A\left(s_1\right)}{k \cdot r^{\,0}_A\left(s_1\right)}\,. 
\label{Appendix_Inequality_A_40}
\end{eqnarray}
 
\noindent
Then, using the approach described above (items 2. - 5. in \ref{Approach_Appendix}) one may demonstrate that the upper limits are 
\begin{eqnarray}
\left|\ve{T}_1\right| \le 4\,\frac{m_A}{r^{\,1}_A\left(s_1\right)}\,\frac{v_A\left(s_1\right)}{c}\,,  
\label{Appendix_Inequality_A_45}
\\
\left|\ve{T}_2\right| \le 2\,\frac{m_A}{r^{\,1}_A\left(s_1\right)}\,\frac{v_A\left(s_1\right)}{c}\,,
\label{Appendix_Inequality_A_50}
\\
\left|\ve{T}_3\right| \le 4\,\frac{m_A}{r^{\,1}_A\left(s_1\right)}\,\frac{v_A\left(s_1\right)}{c}\,,
\label{Appendix_Inequality_A_55}
\end{eqnarray}

\noindent
while their total sum confirms the asserted inequality (\ref{Appendix_Inequality_A_5}), that is (\ref{Inequality_A}).

\newpage



\section*{References}

\begin{thebibliography}{99}


\bibitem{History_Astrometry1} 
E. H{\o}g, 
{\it Astrometry Lost and Regained},  
Baltic Astronomy {\bf 20} (2011) 221. 

\bibitem{History_Astrometry2}
M. Perryman,
{\it The History of Astrometry},  
Eur. Phys. J. H {\bf 37} (2012) 745. 

\bibitem{Kovalevsky}
J. Kovalevsky,
{\it Modern Astrometry},
Springer, $2^{\rm nd}$ ed., 2002.

\bibitem{Hipparcos}
{\it The Hipparcos and Tycho Catalogues}, ESA SP {\bf 1200} (1997) Noordwijk: ESA Publishing Division.

\bibitem{Hipparcos1}
E. H{\o}g, G. B\"assgen, U. Bastian, et al.,
{\it The Tycho Catalogue},
Astronomy \& Astrophysics {\bf 323} (1997) L57.

\bibitem{Hipparcos2}
E. H{\o}g,  C. Fabricius, V.V. Makarov, et al.,
{\it The Tycho-2 Catalogue of the 2.5 million brightest stars},
Astronomy \& Astrophysics {\bf 355} (2000) L27.

\bibitem{GAIA}
{\it The Three-Dimensional Universe with Gaia},
Observatoire de Paris-Meudon, France, 4-7 October 2004,
Editors: C. Turon, K.S. O'Flaherty, M.A.C. Perryman.

\bibitem{GAIA1}
T. Prusti, J.H. J. de Bruijne, A.G.A. Brown, et al.,
{\it The Gaia mission},
Astronomy \& Astrophysics {\bf 595} (2016) A1.

\bibitem{Theory_Experiment}
C.M. Will,
{\it The Confrontation between General Relativity and Experiment},
Living Rev. Relativity {\bf 17} (2014) 4.

\bibitem{Kopeikin_Efroimsky_Kaplan}
S. Kopeikin, M. Efroimsky, G. Kaplan,
{\it Relativistic Celestial Mechanics of the Solar System},
Wiley-VCH, Signapure (2012).

\bibitem{Klioner2003b}
S.A. Klioner,
{\it A practical relativistic model for microarcsecond astrometry in space},
The Astronomical Journal {\bf 125} (2003) 1580.

\bibitem{Book_Clifford_Will}
C.M. Will,
{\it Theory and Experiment in gravitational physics},
Cambridge University Press, Cambridge/UK, 1993.

\bibitem{GAIA_DR1_1}
A.G.A. Brown, A. Vallenari, T. Prusti, et al.,
{\it Data Release 1: Summary of the astrometric, photometric, and survey properties},
Astronomy \& Astrophysics {\bf 595} (2016) A2.

\bibitem{GAIA_DR1_2}
L. Lindegren, U. Lammers, U. Bastian, et al.,
{\it Gaia Data Release 1: Astrometry - one billion positions, two million proper motions and parallaxes},
Astronomy \& Astrophysics {\bf 595} (2016) A4.

\bibitem{GAIA_DR2_1} 
A.G.A. Brown, A. Vallenari, T. Prusti, et al.:
{\it Gaia Data Release 2: Summary of the contens and survey properties},  
Astronomy and Astrophysics {\bf 616} (2018) A1. 

\bibitem{Gaia_Archive} 
https://gea.esac.esa.int/archive/ 

\bibitem{Gaia_CRF2} 
F. Mignard, S. Klioner, L. Lindegren, et al.,  
{\it Gaia Data Release 2: The Celestial reference frame (Gaia-CRF2)}, 
Astronomy and Astrophysics {\bf 616} (2018) A14. 

\bibitem{ICRF3_A}  
IAU  Working  Group  on  the  Third  Realization  of  the  ICRF, 
{\it The Third Realization of the International Celestial Reference Frame}, 
IAU Resolution B2, Proceedings of the XXX General Assembly of IAU, Vienna/Austria, 
20 - 31 August 2018.
 
\bibitem{GAIA_DR2_2}  
X. Luri, A.G.A.  Brown, L.M.  Sarro, et al.   
{\it Gaia Data Release 2: Using Gaia parallaxes}, 
Astronomy and Astrophysics {\bf 616} (2018) A9.  

\bibitem{Jasmine1} 
N. Gouda, Y. Kobayashi, Y. Yamada, T. Yano, and JASMINE Working Group,  
{\it Infrared space astrometry project JASMINE}  
Proceedings of the International Astronomical Union {\bf 248} (2008) 248.  

\bibitem{article_sub_micro_1} 
E. H{\o}g,   
{\it Absolute astrometry in the next 50 years}, 
arXiv:astro-ph/1408.2190v7. 

\bibitem{article_sub_micro_2}
E. H{\o}g,   
{\it Astrometry for Dynamics}, 2013,  
Response to the call for White Papers for the definition of the L2 and L3 missions in the ESA Science Programme,
arXiv:astro-ph/arXiv:1408.3299v1. 

\bibitem{article_sub_micro_3}
A.G.A. Brown, 
{\it Space-Time Structure Explorer Sub-microarcsecond astrometry for the 2030s}, 2013,  
Response to the call for White Papers for the definition of the L2 and L3 missions in the ESA Science Programme,  
available at http://home.strw.leidenuniv.nl/~brown/docs/GAIA-CG-TN-LEI-AB-039.pdf  

\bibitem{article_sub_micro_4}
A. Vallenari, {\it The future of Astrometry in Space},
Front. Astron. Space Sci. {\bf 5} (2018) 11; doi: 10.3389/fspas.2018.00011.

\bibitem{Conference_Cambridge}
A.G.A. Brown (Leiden Observatory), T. Prusti (ESA), N. Walton (IoA, Cambridge), 
Conference {\it Next Steps Towards Future Space Astrometry Missions},
University of Cambridge, UK, Date: 6-8 July 2015.  

\bibitem{Theia}
The Theia Collaboration (2017)
{\it Theia: Faint objects in motion or the new astrometry frontier},
arXiv:astro-ph/1707.01348

\bibitem{Kopeikin_Gwinn}
S.M. Kopeikin, C.R. Gwinn,
{\it Sub-Microarcsecond Astrometry and New Horizons in Relativistic Gravitational Physics},
Proceeding of IAU Colloquium {\bf 180} "Towards Models and Constants for Sub-Microarcsecond Astrometry",
Washington DC, 26.03. - 02.04. 2000;
arXiv:gr-qc/0004064.

\bibitem{Gaia_NIR}
D. Hobbs, A. Brown, A. Mora, et al., (2017)
{\it GaiaNIR - Combining optical and Near-Infra-Red (NIR) capabilities with
Time-Delay-Integration (TDI) sensors for a future Gaia-like mission},
arXiv:astro-ph/1609.07325.

\bibitem{NEAT1}
{\it NEAT: Nearby Earth Astrometric Telescope} (2011) http://neat.obs.ujf-grenoble.fr/NEAT.html

\bibitem{NEAT2}
F. Malbet, A. L\'eger, R. Goullioud, et al. (2011)
{\it An Astrometric Telescope To Probe Planetary Systems Down To The Earth Mass Around Nearby
Solar-Type Stars},
arXiv:astro-ph/1108.4784.

\bibitem{NEAT3}
F. Malbet, A. L\'eger, M. Shao, et al. 
{\it High precision astrometry mission for the detection and characterization of nearby habitable planetary systems 
with the Nearby Earth Astrometric Telescope}  
Experimental Astronomy {\bf 34} (2012) 385.  

\bibitem{Brumberg1991}
V.A. Brumberg,
{\it Essential Relativistic Celestial Mechanics},
(1991) Bristol: Adam Hilder.

\bibitem{Zschocke1}
S. Zschocke,
{\it Light propagation in the gravitational field of N arbitrarily moving bodies in 1PN approximation for high-precision astrometry},
Phys. Rev. D {\bf 92} (2015) 063015 (Publishers Note: Phys. Rev. D {\bf 93} (2016) 069903(E)).

\bibitem{Zschocke2}
S. Zschocke,
{\it Light propagation in the gravitational field of N arbitrarily moving bodies in the 1.5PN approximation for high-precision astrometry},
Phys. Rev. D {\bf 93} (2016) 103010 (Erratum: Phys. Rev. D {\bf 94} (2016) 029902(E)).

\bibitem{KK1992}
S.A. Klioner, S.M. Kopeikin,
{\it Microarcsecond astrometry in space: relativistic effects and reduction of observations},
Astron. J. {\bf 104} (1992) 897.

\bibitem{Kopeikin1997} S.M. Kopeikin,
{\it Propagation of light in the stationary field of multipole gravitational lens},
J. Math. Phys. {\bf 38} (1997) 2587.

\bibitem{KS1999}
S.M. Kopeikin, G. Sch\"afer,
{\it Lorentz covariant theory of light propagation in gravitational fields of arbitrary-moving bodies},
Phys. Rev. D {\bf 60} (1999) 124002.

\bibitem{KSGE}
S.M. Kopeikin, G. Sch\"afer, C.R. Gwinn, T.M. Eubanks,
{\it Astrometric and timing effects of gravitational waves from localized sources},
Phys. Rev. D {\bf 59} (1999) 084023.

\bibitem{Klioner2003a}
S.A. Klioner,
{\it Light propagation in the gravitational field of moving bodies by means of Lorentz transformation:
I. Mass monopoles moving with constant velocities},
Astronomy \& Astrophysics {\bf 404} (2003) 783.

\bibitem{KopeikinMashhoon2002}
S.M. Kopeikin, B. Mashhoon,
{\it Gravitomagnetic-effects in the propagation of electromagnetic waves in variable
gravitational fields of arbitrary-moving and spinning bodies},
Phys. Rev. D {\bf 65} (2002) 064025.

\bibitem{Einstein1}
A. Einstein, {\it Die Feldgleichungen der Gravitation},
Sitzungsberichte der K\"oniglich-Preussischen Akademie der Wissenschaften zu Berlin {\bf 2} (1915) 844.

\bibitem{Einstein2}
A. Einstein, {\it Die Grundlage der allgemeinen Relativit\"atstheorie},
Annalen der Physik (Ser. 4) {\bf 49} (1916) 769.

\bibitem{MTW}
C.W. Misner, K.S. Thorne, J.A. Wheeler,
{\it Gravitation}, Palgrave Macmillan (1973), W.H. Freeman, Oxford/UK.  

\bibitem{Landau_Lifschitz}
L.D. Landau, E.M. Lifshitz,
{\it The Classical Theory of Fields}, Vol. 2, $3^{\rm rd}$ ed., 1971, Pergamon Press.

\bibitem{Fock}
V. Fock,
{\it The Theory of Space, Time and Gravitation},  
$2^{\rm nd}$ ed., Pergamon Press, Oxford, 1964.

\bibitem{Observables} 
P.G. Bergmann,  
{\it Observables in General Relativity},  
Rev. Mod. Phys. {\bf 33} (1961) 510.  

\bibitem{Brumberg2010} 
V.A. Brumberg, 
{\it Relativistic Celestial Mechanics on the verge of its 100 year anniversary}, 
Celest. Mech. Dyn. Astr. {\bf 106} (2010) 209. 

\bibitem{IAU_Resolution1}
M. Soffel, S.A. Klioner, G. Petit, et al.,
{\it The IAU 2000 resolutions for astrometry, celestial mechanics and
metrology in the relativistic framework: explanatory supplement},
Astron. J. {\bf 126} (2003) 2687.

\bibitem{Donder}
T. de Donder,
{\it La Gravifique Einteinienne}, Gauthier-Villars, Paris, 1921.

\bibitem{Lanczos}
K. Lanczos,
{\it Ein vereinfachendes Koordinatensystem f\"ur die Einsteinschen Gravitationsgleichungen},
Physikalische Zeitschrift {\bf 23} (1923) 537.

\bibitem{Thorne} 
K. Thorne, 
{\it Multipole expansions of gravitational radiation}, 
Rev. Mod. Phys. {\bf 52} (1980) 299.  

\bibitem{Poisson_Lecture_Notes}
E. Poisson,
{\it Post-Newtonian theory for the common reader},
Lecture Notes (July 2007),
Department of Physics University of Guelph, Canada,
http://www.physics.uoguelph.ca/poisson/research/postN.pdf.

\bibitem{Poisson_Will}
E. Poisson, C.M. Will,
{\it Gravity - Newtonian, Post-Newtonian, Relativistic},
Cambridge University Press (2014).

\bibitem{Sommerfeld1}
A. Sommerfeld,
{\it Electrodynamics: Lectures on Theoretical Physics, Vol. III},
Academic Press Inc., $1^{\rm st}$ ed., New York, 1952.

\bibitem{Sommerfeld2} 
A. Sommerfeld, 
{\it Partial Differential Equations in Physics: Lectures on Theoretical Physics, Vol. VI}, 
Academic Press Inc., $1^{\rm st}$ ed., New York, 1949.  

\bibitem{Whittaker} 
E.T. Whittaker, 
{\it Note on the law that light-rays are the null geodecics of a gravitational field},  
Proceedings of the Cambridge Philosophical Society {\bf 24} (1928) 32.  

\bibitem{Article_Charaketristiken}
H. Handrek,
{\it \"Uber die Differentialgleichungen in der allgemeinen Relativit\"atstheorie},
Zeitschrift f\"ur Physik {\bf 50} (1928) 397.

\bibitem{Sexl_Urbantke}
R.U. Sexl, H.K. Urbantke,
{\it Gravitation und Kosmologie}, $4^{\rm th}$ ed.,
Spektrum Akademischer Verlag, Heidelberg, Berlin, Oxford, 2002.

\bibitem{Iverno}
R. d'Iverno,
{\it Introducing Einstein's Relativity}, $1^{\rm th}$ ed.,
Oxford University Press, New York, 1992.

\bibitem{Ligo1}
B.P. Abbott et al. (LIGO Scientific Collaboration and Virgo Collaboration),
{\it GW170817: Observation of Gravitational Waves from a Binary Neutron Star Inspiral},
Phys. Rev. Lett. {\bf 119} (2017) 161101.

\bibitem{Ligo2}
B.P. Abbott et al. (LIGO Scientific Collaboration and Virgo Collaboration),
{\it Multi-messenger Observations of a Binary Neutron Star Merger},
Astrophys. J. Lett. {\bf 848} (2017) L12.

\bibitem{Ligo3}
B.P. Abbott et al. (LIGO Scientific Collaboration and Virgo Collaboration),
{\it Gravitational waves and Gamma-rays from a binary neutron star merger: GW170817 and GRB 170817A},
Astrophys. J. Lett. {\bf 848} (2017) L13.

\bibitem{Kopeikin_A} 
S.M. Kopeikin,  
{\it Testing the Relativistic Effect of the Propagation of Gravity by Very Long Baseline Interferometry},
The Astrophysical Journal {\bf 556} (2001) L1.  

\bibitem{Will_2003}
C.M. Will,
{\it Propagation Speed of Gravity and the Relativistic Time Delay},
The Astrophysical Journal {\bf 590} (2003) 683.

\bibitem{Kopeikin_E}
S.M. Kopeikin, E.B. Fomalont,
{\it Aberration and the Speed of Gravity in the Jovian Deflection Experiment},
Foundations of Phys. {\bf 36} (2006) 1244.

\bibitem{Kopeikin_B} 
S.M. Kopeikin, E.B. Fomalont, 
{\it General relativistic model for experimental measurement of the speed of propagation of gravity by VLBI},  
Proceedings of the 6th European VLBI Network Symposium, Ros, E., Porcas, R.W., Zensus, J.A. (eds.), 
25 June - 28 June 2002, Bonn, Germany; 
arXiv:gr-qc/0206022. 

\bibitem{Samuel_1} 
S. Samuel, 
{\it On the Speed of Gravity and the v/c Corrections to the Shapiro Time Delay}, 
Phys. Rev. Lett. {\bf 90} (2003) 231101.  

\bibitem{Faber} 
J.A. Faber, 
{\it The speed of gravity has not been measured from time delays},  
arXiv:astro-ph/0303346 

\bibitem{Asada1} 
H. Asada, 
{\it Comments on "Measuring the Gravity Speed by VLBI"}, 
Proc. of "Physical Cosmology", the XVth Rencontres de Blois, 15-20 June 2003, 
arXiv:astro-ph/0308343. 

\bibitem{Asada2}
H. Asada, 
{\it Remarks on "Comments on 'On the speed of gravity and the Jupiter/Quasar measurement' " by S. Samuel},
Int. J. Mod. Phys. {\bf 15} (2009) 289. 

\bibitem{Carlip} 
S. Carlip, 
{\it Model-Dependence of Shapiro Time Delay and the "Speed of Gravity/Speed of Light" Controversy}, 
Class. Quantum Gravit. {\bf 21} (2004) 3803.  

\bibitem{Pascual} 
J.-F. Pascual-Sanchez,  
{\it Speed of gravity and gravitomagnetism}, 
Int. J. Mod. Phys. D {\bf 13} (2004) 2345.  

\bibitem{Samuel_2} 
S. Samuel, 
{\it On the Speed of Gravity and the Jupiter/Quasar Measurement}, 
Int. J. Mod. Phys. D {\bf 13} (2004) 1753.  

\bibitem{Kopeikin_I} 
S.M. Kopeikin, 
{\it The post-Newtonian treatment of the VLBI experiment on September 8, 2002}, 
Phys. Lett. A {\bf 312} (2003) 147. 
 
\bibitem{Kopeikin_CQG} 
S.M. Kopeikin, 
{\it Speed of Gravity in General Relativity and Theoretical Interpretation of the Jovian Deflection Experiment}, 
Class. Quantum Gravit. {\bf 21} (2004) 3251.  

\bibitem{Kopeikin_D} 
S.M. Kopeikin, E.B. Fomalont, 
{\it On the Speed of Gravity and Relativistic v/c Corrections to the Shapiro Time Delay}, 
Phys. Lett. A {\bf 355} (2006) 163.  

\bibitem{Kopeikin_F} 
S.M. Kopeikin, E.B. Fomalont,  
{\it Comments on "On the speed of gravity and the Jupiter/Quasar measurement" by S. Samuel},  
Int. J. Mod. Phys. D {\bf 15} (2006) 273.  

\bibitem{Kopeikin_G} 
S.M. Kopeikin, E.B. Fomalont,  
{\it Gravitomagnetism and the Speed of Gravity}, 
Int. J. Mod. Phys. D {\bf 15} (2006) 305.  

\bibitem{Kopeikin_H} 
S.M. Kopeikin, E.B. Fomalont,
{\it Comment on "Model-dependence of Shapiro time delay and the 'speed of gravity/speed of light' controversy"}, 
Class. Quantum Grav. {\bf 22} (2005) 5181.  

\bibitem{Frittelli} 
S. Frittelli, 
{\it Aberration by gravitational lenses in motion}, 
Mon. Not. R. Astron. Soc. {\bf 344} (2003) L85.  

\bibitem{Mignard_Crosta} 
M.T. Crosta, F. Mignard, 
{\it Micro-arcsecond light bending by Jupiter}, 
Class. Quantum Grav. {\bf 23} (2006) 4853.  

\bibitem{Malkin} 
Z.M. Malkin, V.N. L'vov, S.D. Tsekmeister,  
{\it Forthcoming Close Angular Approaches of Planets to Radio Sources and Possibilities to Use Them as GR Tests},  
Solar System Research {\bf 43} (2009) 313.  

\bibitem{Zhu} 
Y. Zhu, 
{\it Measurement of the speed of gravity}, 
Chinese Phys. Lett. {\bf 28} (2011) 070401.  

\bibitem{Expansion_2PN}
L. Blanchet,
{\it Gravitational radiation from post-Newtonian sources and inspiralling compact binaries},
Living Rev. Relativity {\bf 9} (2006) 4.

\bibitem{Zschocke3}
S. Zschocke,
{\it Light propagation in the field of one arbitrarily moving pointlike body in the 2PN approximation},
Phys. Rev. D {\bf 94} (2016) 124007 (Publishers Note: Phys. Rev. D {\bf 95} (2017) 069905(E);
Erratum: Physical Phys. Rev. D {\bf 96} (2017) 049906).

\bibitem{Zschocke4}
S. Zschocke,
{\it Light propagation in 2PN approximation in the field of one moving monopole I. Initial value problem},
Class. Quant. Grav. {\bf 35} (2018) 055013.

\bibitem{Bruegmann2005}
M.H. Br\"ugmann,
{\it Light deflection in the post-linear gravitational field of bounded point-like masses},
Phys. Rev. D {\bf 72} (2005) 024012.

\bibitem{EpsteinShapiro}
R. Epstein, I.I. Shapiro,
{\it Post-post-Newtonian deflection of light by the Sun},
Phys. Rev. D {\bf 22} (1980) 2947.

\bibitem{FischbachFreeman}
E. Fischbach, B.S. Freeman,
{\it Second-oder contribution to the gravitational deflection of light},
Phys. Rev. D {\bf 22} (1980) 2950.

\bibitem{RM1}
G.W. Richter, A. Matzner,
{\it Second-order contributions to gravitational deflection of light in the parameterized post-Newtonian formalism},
Phys. Rev. D {\bf 26} (1982) 1219.

\bibitem{RM2}
G.W. Richter, A. Matzner,
{\it Second-order contributions to gravitational deflection of light in the parameterized post-Newtonian formalism. II. Photon orbits and deflections in three dimensions},
Phys. Rev. D {\bf 26} (1982) 2549.

\bibitem{RM3}
G.W. Richter, A. Matzner,
{\it Second-order contributions to relativistic time delay in the parameterized post-Newtonian formalism},
Phys. Rev. D {\bf 28} (1983) 3007.

\bibitem{Cowling} S.A. Cowling,
{\it Gravitational light deflection in the Solar System},
Mon.Not.R.Astr.Soc. {\bf 209} (1984) 415.

\bibitem{BodennerWill2003}
J. Bodenner, C.M. Will,
{\it Deflection of light to second order: A tool for illustrating principles of general relativity},
Am. J. Phys. {\bf 71} (2003) 770.

\bibitem{Moving_Kerr_Black_Hole2}
G. He, W. Lin,
{\it Second-order time delay by a radially moving Kerr-Newman black hole},
Phys. Rev. D {\bf 94} (2016) 063011.

\bibitem{Brumberg1987}
V.A. Brumberg,
{\it Post-post-Newtonian propagation of light in the Schwarzschild field},
Kinematika Fis Nebesnykh Tel {\bf 3} (1987) 8, in Russian.

\bibitem{Deng_Xie}
X.M. Deng, Yi Xie,
{\it Two-post-Newtonian light propagation in the scalar-tensor theory: An $N$-point mass case},
Phys. Rev. D {\bf 86} (2012) 044007.

\bibitem{Deng_2015}
X-M. Deng,
{\it The second post-Newtonian light propagation and its astrometric measurement in the solar system},
Int. J. Mod. Phys. D {\bf 24} (2015) 1550056.

\bibitem{Minazzoli2}
O. Minazzoli, B. Chauvineau,
{\it Scalar-tensor propagation of light in the inner solar system including relevant $c^{-4}$ contributions for ranging and time transfer},
Class. Quantum. Grav. {\bf 28} (2011) 085010.

\bibitem{Article_Zschocke1}
S.A. Klioner, S. Zschocke,
{\it Numerical versus analytical accuracy of the formulas for light propagation},
Class. Quantum Grav. {\bf 27} (2010) 075015.

\bibitem{LLT2004}
Chr. Le Poncin-Lafitte, B. Linet, P. Teyssandier,
{\it World function and time transfer: general post-Minkowskian expansions},
Class. Quantum Grav. {\bf 21} (2004) 4463.

\bibitem{TL2008}
P. Teyssandier, Chr. Le Poncin-Lafitte,
{\it General post-Minkowskian expansion of time transfer functions},
Class. Quantum Grav. {\bf 25} (2008) 145020.

\bibitem{Teyssandier}
P. Teyssandier,
{\it Direction of light propagation to order $G^2$ in static, spherically symmetric spacetimes: a new derivation},
Class. Quant. Grav. {\bf 29} (2012) 245010.

\bibitem{HBL2014b}
A. Hees, S. Bertone, C. Le Poncin-Lafitte,
{\it Relativistic formulation of coordinate light time, Doppler and astrometric observables up to the second post-Minkowskian order},
Phys. Rev. D {\bf 89} (2014) 064045.

\bibitem{AshbyBertotti2010}
N. Ashby, B. Bertotti,
{\it Accurate light-time correction due to a gravitating mass},
Class. Quantum Grav. {\bf 27} (2010) 145013.

\bibitem{Moving_Kerr_Black_Hole1}
G. He, W. Lin,
{\it Second order Kerr-Newman time delay},
Phys. Rev. D {\bf 93} (2016) 023004.

\bibitem{Xu_Wu}
C. Xu, X. Wu,
{\it Extending the first-order post-Newtonian scheme in
multiple systems to the second-order contributions to light propagation},
Chin. Phys. Lett. {\bf 20} (2003) 195.

\bibitem{Xu_Gong_Wu_Soffel_Klioner}
C. Xu, Y. Gong, X. Wu, M. Soffel, S.A. Klioner (2005),
{\it Second order post-Newtonian Equations of light propagation in multiple systems},
arXiv:gr-qc/0510074.

\bibitem{Minazzoli1}
O. Minazzoli, B. Chauvineau,
{\it Post-Newtonian metric of general relativity including all the $c^{-4}$ terms in the continuity of the IAU2000 resolutions},
Phys. Rev. D {\bf 79} (2009) 084027.

\bibitem{2PN_Light_PropagationA}
G. Yan-Xiang, Wu Xiao-Mei,
{\it Post-post-Newtonian deflection of light ray in multiple systems with PPN parameters},
Chin. Phys. Lett. {\bf 20} (2011) 020403.

\bibitem{Xie_Huang}
Yi Xie, T.Y. Huang,
{\it Second post-Newtonian approximation of Einstein-aether theory},
Phys. Rev. D {\bf 77} (2008) 124049.

\bibitem{Zschocke_Soffel}
S. Zschocke, M.H. Soffel,
{\it Gravitational field of one uniformly moving extended body and $N$ arbitrarily
moving pointlike bodies in post-Minkowskian approximation},
Class. Quantum Grav. {\bf 31} (2014) 175001.

\bibitem{JPL}
{\it Highly accurate ephemerides for solar system objects}, provided by
Jet Propulsion Laboratory (JPL) at http://ssd.jpl.nasa.gov/

\bibitem{Mathematical_Methods}
G.B. Arfken, H.J. Weber, {\it Mathematical Methods for Physicists},
Academic Press Inc., $4^{\rm th}$ ed., New York, 1995.

\bibitem{KopeikinMakarov2007}
S.M. Kopeikin, V.V. Makarov,
{\it Gravitational bending of light by planetary multipoles and its measurement with micro-arcsecond astronomical interferometers},
Phys. Rev. D {\bf 75} (2007) 062002.

\bibitem{Zschocke_Lense_Equation}
S. Zschocke,
{\it A generalized lens equation for light-deflection in weak gravitational fields},
Class. Quantum Grav. {\bf 28} (2011) 125016.

\bibitem{LinetTeyssandier2013_a}
B. Linet, P. Teyssandier,
{\it New method for determining the light travel time in static, spherically symmetric
spacetimes. Calculation of the terms of order $G^3$},
Class. Quantum Grav. {\bf 30} (2013) 175008.

\bibitem{LinetTeyssandier2013_b}
P. Teyssandier, B. Linet,  
{\it Enhanced term of order $G^3$ in the light travel time: discussion for some solar system experiments},  
Proceedings of the Conference {\it Journ\'ees 2013 Syst\`emes de r\'ef\'erence spatio-temporels 
(scientific developments from highly accurate space-time reference systems)}, 16-18 September 2013, Observatoire de Paris, France. 

\bibitem{nas_telescopes}
Guyon, E.A. Bendek, T.D. Milster, et al.,
{\it High-precision astrometry with a diffractive pupil telescope},
The Astrophysical Journal Supplement {\bf 200} (2012) 11.

\bibitem{KopeikinKorobkovPolnarev2006}
S. Kopeikin, P. Korobkov, A. Polnarev,
{\it Propagation of light in the field of stationary and radiative gravitational multipoles},
Class. Quantum Grav. {\bf 23} (2006) 4299.

\bibitem{KopeikinKorobkov2005}
S. Kopeikin, P. Korobkov,
{\it General Relativistic Theory of Light Propagation in the Field of Radiative Gravitational Multipoles}
(2005) arXiv: gr-qc/0510084v2.

\bibitem{moving_axisymmetric_body}
A. Hees, S. Bertone, C. Le Poncin-Lafitte,
{\it Light propagation in the field of a moving axisymmetric body: theory and application to JUNO},
Phys. Rev. D {\bf 90} (2014) 084020.

\bibitem{Soffel_Han}
M.H. Soffel, Wen-Biao Han,
{\it The gravitational time delay in the field of a slowly moving body with arbitrary multipoles},
Phys. Lett. A {\bf 379} (2015) 233.

\bibitem{Abramowitz_Stegun} 
M. Abramowitz, I.A. Stegun, 
{\it Handbook of Mathematical Functions}, 
Dover Publications, Inc., New York, Ninth Printing, 1970.  

\bibitem{Maple}
F. Garvan, {\it The MAPLE Book}, Chapman \& Hall (2002).


\end{thebibliography}
\end{document}